\documentclass[aps,prl,twocolumn,amsmath,amssymb,nofootinbib,superscriptaddress,floatfix,reprint,longbibliography]{revtex4-2}

\usepackage{graphicx}
\usepackage{latexsym}
\usepackage{amsmath}
\usepackage{amsfonts}
\usepackage{amssymb}
\usepackage{bm}
\usepackage{color}
\usepackage{txfonts}
\usepackage{braket}
\usepackage{url}
\usepackage{CJKutf8}
\usepackage{braket} 
\usepackage{svg}
\usepackage{xcolor}
\usepackage[colorlinks=true, allcolors=blue]{hyperref}
\usepackage{ulem}
\usepackage{cleveref}
\usepackage{bm,color,amsmath,amssymb,mathrsfs,latexsym,graphicx,psfrag,mathtools}

\newcommand{\downdivider}{%
  \par\noindent
  \hbox{%
    \rule{\textwidth}{0.6pt}%
    \kern-0.6pt%
    \raisebox{-2.5mm}{\rule{0.6pt}{2.5mm}}%
  }%
  \par\vspace{0.8ex}%
}

\newcommand{\updivider}{%
  \par\noindent
  \hbox{%
    \rule{\textwidth}{0.6pt}%
    \kern-0.6pt%
    \raisebox{0pt}{\rule{0.6pt}{2.5mm}}
  }%
  \par\vspace{0.8ex}%
}

\crefname{Supplement}{Supp.\,}{Supps.\,}

\creflabelformat{Supplement}{[#2#1#3]}

\newcommand{\prlheading}[1]{%
  \par\indent\textcolor{blue}{\textit{#1}---}\ignorespaces
}
\newcommand{\tmt}{\text{$t$MoTe$_2$}}
\begin{document}
	\newcommand{\fig}[2]{\includegraphics[width=#1]{#2}}
	\def\gsim{~\rlap{$>$}{\lower 1.0ex\hbox{$\sim$}}}
	\setlength{\unitlength}{1mm}
	\newcommand{\bea}{\begin{equation} \begin{aligned}}
    \newcommand{\eea}{\end{aligned} \end{equation} }
    \def\abs#1{\left|{#1}\right|}          			
    \def\bra#1{\left<{#1}\right|}					
    \def\ket#1{\left|{#1}\right>}					

\title{Integer and Fractional Chern Plateaus Shaped by Layer-Dependent Disorder}

\author{Ziyue Qi}
\affiliation{Beijing National Laboratory for Condensed Matter Physics and,\\ Institute of Physics, Chinese Academy of Sciences, Beijing 100190, China}
\affiliation{School of Physical Sciences, University of Chinese Academy of Sciences, Beijing 100190, China}
 
\author{Zhan Wang}
\email{zhan.wang@iphy.ac.cn}
\affiliation{Beijing National Laboratory for Condensed Matter Physics and,\\ Institute of Physics, Chinese Academy of Sciences, Beijing 100190, China}

\author{Hongming Weng}
\email{hmweng@iphy.ac.cn}
\affiliation{Beijing National Laboratory for Condensed Matter Physics and,\\ Institute of Physics, Chinese Academy of Sciences, Beijing 100190, China}
\affiliation{School of Physical Sciences, University of Chinese Academy of Sciences, Beijing 100190, China}

\author{Kun Jiang}
\email{jiangkun@iphy.ac.cn}
\affiliation{Beijing National Laboratory for Condensed Matter Physics and,\\ Institute of Physics, Chinese Academy of Sciences, Beijing 100190, China}
\affiliation{School of Physical Sciences, University of Chinese Academy of Sciences, Beijing 100190, China}

\date{\today}

\begin{abstract}
The realization of integer and fractional Chern insulators in moir\'e systems provides a lattice setting for exploring how the structure of Bloch wave functions influences the interplay between topology and disorder. Here we investigate how layer-dependent disorder shapes Hall plateau widths and transitions. Using a realistic three-orbital model of twisted bilayer MoTe$_2$, we compare Hall and bulk transport under onsite-potential, displacement-field, and twist-angle disorder. At equal bare disorder strength, onsite-potential disorder produces broad integer plateaus with nearly direct transitions, whereas displacement-field disorder yields narrow plateaus separated by conducting regions with pronounced Hall-resistance dips. These contrasting behaviors arise from the layer composition of the Bloch states: contributions from different layer under displacement-field disorder partially cancel, weakening the bulk localization. Composite-fermion mean-field theory further yields fractional plateaus at $\nu_h=2/3$ and $3/5$ with the same disorder dependence of plateau widths. These findings identify layer-dependent disorder as an important mechanism shaping Chern plateau stability in moir\'e systems beyond bare disorder strength alone.
\end{abstract}
\maketitle

\prlheading{Introduction}
The discovery of the integer and fractional quantum Hall effects (IQHE and FQHE) in two-dimensional electron gases established a remarkable interplay between topology, electron correlations, and disorder~\cite{Klitzing1980,Huckestein1995,Tsui1982,Laughlin1983}. While topology underlies the quantization of Hall conductivity, disorder localizes bulk electronic states and stabilizes quantized Hall plateaus over finite ranges of carrier density~\cite{Laughlin1981,AokiAndo1981,Halperin1982,Laughlin1983}. The realization of integer and fractional Chern insulators (CIs and FCIs) extends these phenomena to lattice systems without external magnetic fields~\cite{Haldane1988,Tang2011,Sun2011,Neupert2011,Sheng2011,RegnaultBernevig2011,2025Shan}. Unlike conventional Landau levels, Chern bands possess nontrivial lattice and orbital structures, introducing additional degrees of freedom that can substantially influence electronic localization and transport.

\begin{figure} [htbp!]
    \centering
    \includegraphics[width=\linewidth]{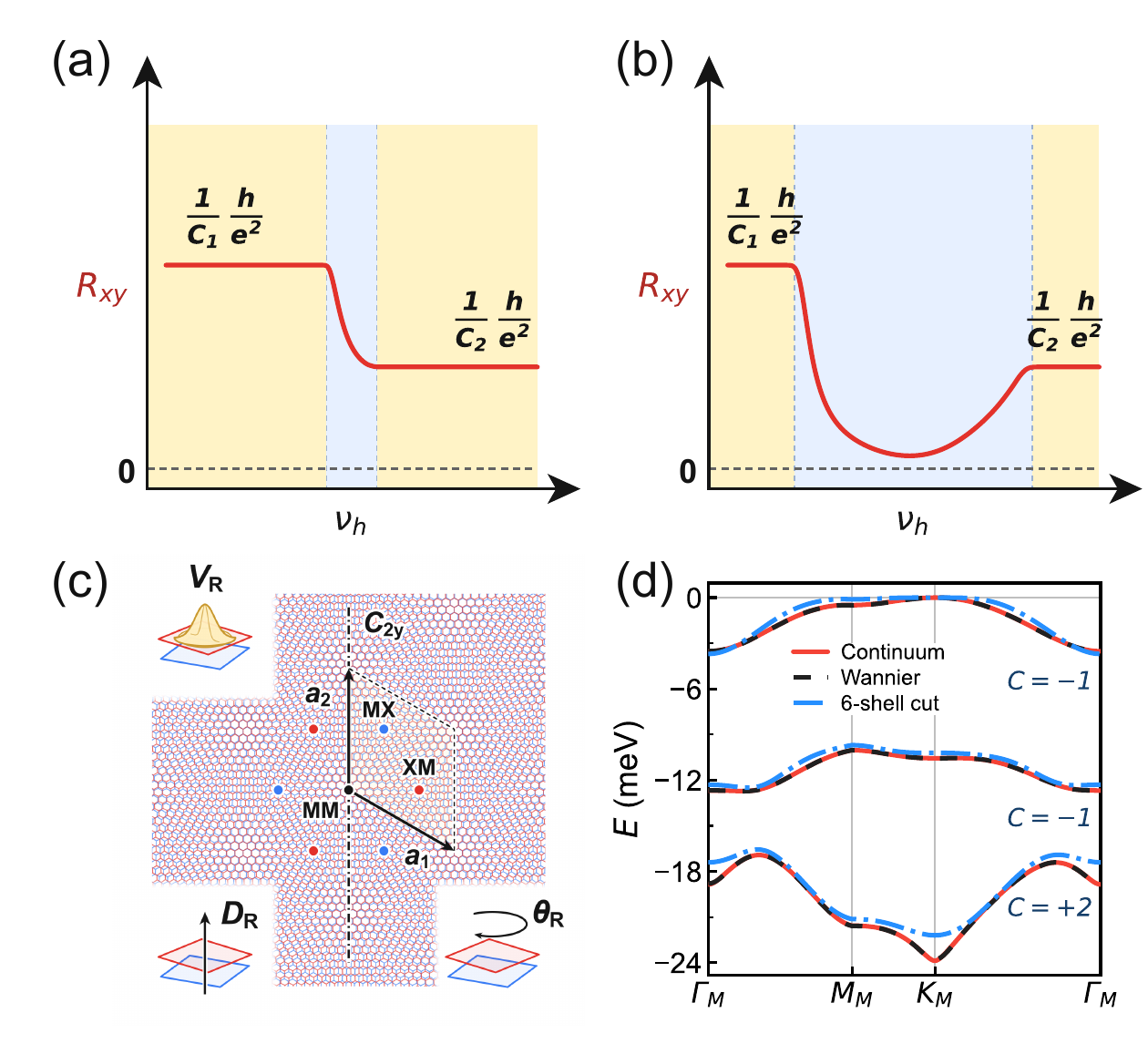}
    \caption{\textbf{Hall-plateau transitions and moir\'e model.} (a),(b) Schematic Hall resistance $R_{xy}$ versus hole filling $\nu_h$: broad plateaus separated by a narrow, direct transition in (a), and narrow plateaus separated by a broad transition with a Hall dip toward zero in (b). Yellow and blue shading highlights plateau and transition regions, respectively. (c) Moir\'e lattice of \tmt{} with MM (black), MX (blue), and XM (red) stacking regions, primitive vectors $\bm{a}_1$ and $\bm{a}_2$, and the $C_{2y}$ rotation axis. Insets illustrate local potential $V_R$, perpendicular displacement field $D_R$, and twist-angle $\theta_R$ disorder. (d) The three topmost $K$-valley valence bands at $\theta=2.45^\circ$ from the continuum model (red solid), Wannier TB model (black dashed), and TB model with hoppings retained through the sixth neighbor shell (blue dash-dotted). Band Chern numbers are labeled.}
    \label{fig:Figure1}
\end{figure}

Moir\'e materials provide a highly tunable platform for exploring these effects, with integer and fractional Chern states observed in twisted transition-metal dichalcogenides~\cite{Cai2023,Zeng2023,Park2023,Xu2023} and rhombohedral graphene superlattices~\cite{LuGraphene2024,LuGraphene2025}. Transport experiments have revealed strikingly different behaviors of Hall plateaus and their transitions~\cite{Park2023,Xu2023,LuGraphene2024,LuGraphene2025,Park2025Dissipationless,Xu2025Superconductivity,Park2025HigherBand,Xu2025SecondBand,DingGraphene2025Chirality,XieGraphene2025,LiGraphene2026HighChern,ButlerGraphene2026OneThird}. In some systems, broad quantized plateaus are connected by narrow, nearly direct transitions~\cite{LuGraphene2025,Xu2025SecondBand}, resembling conventional quantum Hall behavior shown in Fig.~\ref{fig:Figure1}(a). In others, narrow plateaus are separated by extended conducting regions where the Hall resistance is strongly suppressed, sometimes approaching zero before recovering toward the next quantized value~\cite{Park2023,Xu2023,LuGraphene2024,Park2025Dissipationless,Xu2025Superconductivity,Park2025HigherBand,DingGraphene2025Chirality,XieGraphene2025,LiGraphene2026HighChern,ButlerGraphene2026OneThird}, as shown in Fig.~\ref{fig:Figure1}(b). These contrasting behaviors suggest that the microscopic disorder landscape plays an important role beyond conventional plateau broadening. In layered moir\'e systems, different disorder sources can act differently on the layer and orbital degrees of freedom~\cite{Redekop2024,Thompson2025,ResendizVazquez2026,Uri2020TwistDisorder,Wilson2020TwistDisorder}, potentially modifying the localization of bulk states. This raises a central question: How does the layer-dependent nature of disorder govern the stability of integer and fractional Chern plateaus and the transport behavior between them?

In this work, we identify the layer-dependent disorder sensitivity as a common mechanism in shaping the Chern plateau stability. Using a realistic three-orbital Wannier model of twisted MoTe$_2$ (\tmt{}) as an example, we investigate onsite-potential, displacement-field, and twist-angle disorder, as shown in Fig.~\ref{fig:Figure1}(c). Their distinct orbital structures control bulk localization through different effective disorder strengths, leading to markedly different Hall plateau and transition behaviors sketched in Fig.~\ref{fig:Figure1}(a,b). We elucidate this mechanism through orbital form factors and disorder-related charge responses, and extend the analysis to fractional Chern plateaus within composite-fermion (CF) mean-field theory~\cite{Jain1989, LopezFradkin1991}.

\begin{figure*} 
    \centering
    \includegraphics[width=\linewidth]{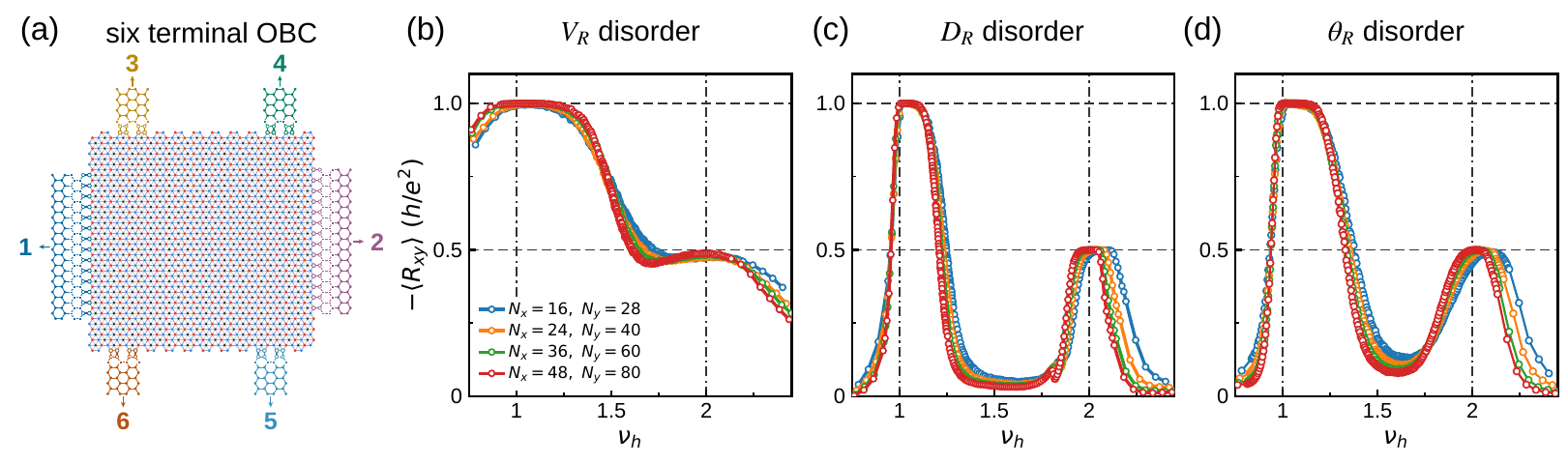}
    \caption{\textbf{Hall transport under orbital-selective disorder.} (a) Six-terminal geometry with open boundary conditions (OBC). (b)--(d) Disorder-averaged Hall resistance $-\langle R_{xy}\rangle$ versus hole filling $\nu_h$ for $V_R$, $D_R$, and $\theta_R$ disorder, respectively. Curve colors indicate the system sizes listed in (b). Horizontal dashed lines mark $h/e^2$ and $h/(2e^2)$ and vertical lines mark $\nu_h=1,2$, repectively. Here $N_x$ and $N_y$ count lattice repeats of length $\sqrt{3}a_M$ along $x$ and $a_M$ along $y$, respectively, where $a_M$ is the moir\'e lattice constant. Parameters: $\theta=2.45^\circ$, $\lambda=2$, and $\gamma_a=3\,\mathrm{meV}$, corresponding to $W_\theta\simeq1.0^\circ$ for twist-angle disorder (see SM.~\ref{sm:disorder}). Throughout this work, each transport data point is averaged over 500 disorder realizations.}
    \label{fig:Figure2}
\end{figure*}

\prlheading{Model construction}
The moir\'e superlattice of \tmt{} is triangular, with MM, MX, and XM stacking regions corresponding to Mo--Mo, Mo--Te, and Te--Mo alignments between the top and bottom layers, respectively, as shown in Fig.~\ref{fig:Figure1}(c). The low-energy valence states originate from the monolayer $K$ and $K'$ valleys, where strong spin--orbit coupling produces spin--valley locking. Opposite spin-polarizations in the two valleys suppress intervalley scattering under smooth moir\'e potentials~\cite{Wu2019Topological,Devakul2021Magic,Reddy2023FQAH,Wang2024FCI,Jia2024}. For the valley-polarized Chern states considered here, we focus on the single-$K$-valley sector and construct a three-band tight-binding (TB) model. To this end, we start from a continuum model extracted directly from first-principles calculations without parameter fitting~\cite{Zhang2024NoFit}, whose $K$-valley bands at $\theta=2.45^\circ$ are shown in red solid lines in Fig.~\ref{fig:Figure1}(d). Throughout the twist angle range $\theta=2.45^\circ$--$3.89^\circ$, the three highest valence bands have zero total Chern number, although their individual Chern numbers vary with twist angle. For example, at $\theta=2.45^\circ$, they carry Chern numbers $-1$, $-1$, and $2$, respectively, as labeled in Fig.~\ref{fig:Figure1}(d). 

The vanishing total Chern number allows us to construct three maximally localized Wannier functions (MLWFs) using \textsc{WANNIER90}~\cite{MOSTOFI2008685}. These Wannier functions, hereafter referred to as local \textit{orbitals}, are centered at the MX, XM, and MM stacking regions, respectively. Importantly, they exhibit distinct layer polarizations: the MX and XM orbitals are predominantly localized in opposite layers, whereas the MM orbital has approximately equal weights in both layers~\cite{Devakul2021Magic,Qiu2023ThreeOrbital,Xu2024Wannier,Crepel2024ThreeOrbital,Qi2026Multivalley}. This pattern is consistent with $C_{2y}\mathcal{T}$ symmetry~\cite{Jia2024}, where $\mathcal{T}$ denotes time reversal and $C_{2y}$ is a $180^\circ$ rotation about the in-plane $y$ axis that exchanges the layers and MX/XM centers while leaving MM fixed, as shown in Fig.~\ref{fig:Figure1}(c). The resulting three-orbital TB model accurately reproduces the continuum band structure, as shown in black dashed lines Fig.~\ref{fig:Figure1}(d). For real-space transport, we truncate hoppings beyond the sixth-neighbor shell, preserving the overall band dispersion and the individual band Chern numbers, as shown in blue dash-dotted lines in Fig.~\ref{fig:Figure1}(d). Further details of the model construction are provided in Supplementary Material \ref{sm:model} (SM.~\ref{sm:model}).

\begin{figure}
    \centering
    \includegraphics[width=\linewidth]{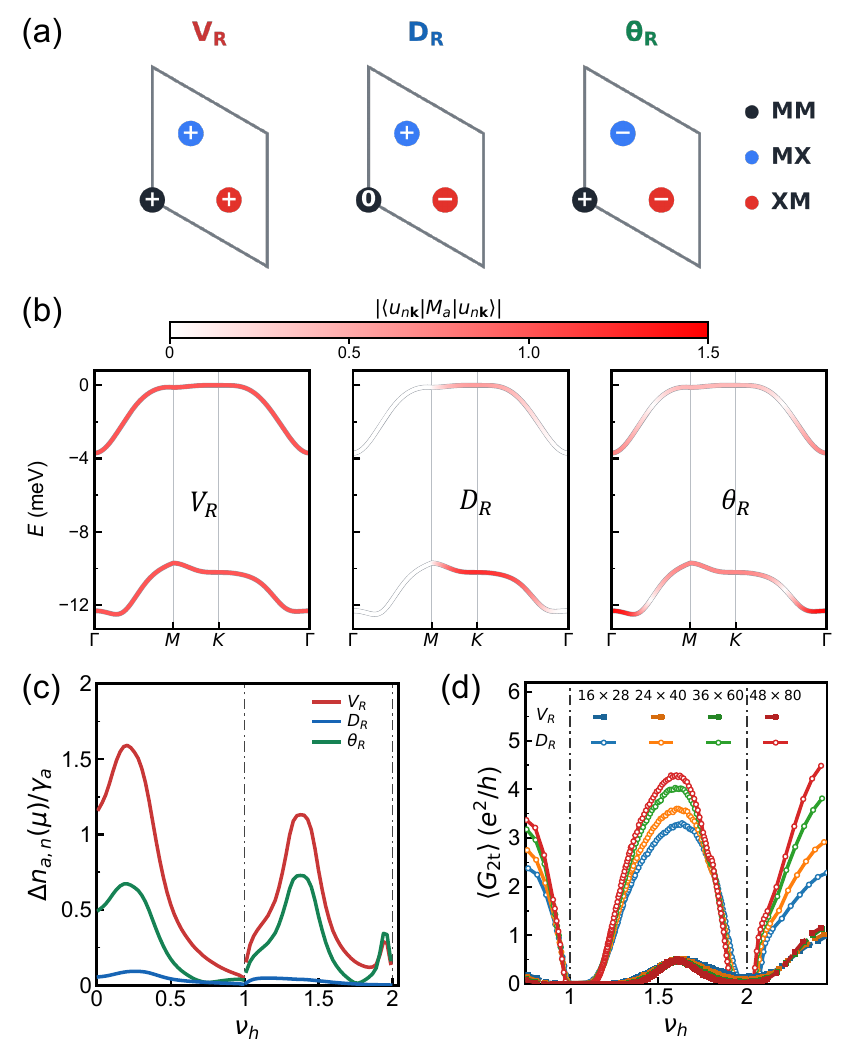}
    \caption{\textbf{Orbital projections, charge response, and bulk transport.} (a) Wannier orbital structure $M_a$ for $V_R$, $D_R$, and $\theta_R$ disorder (left to right), with black, blue, and red dots denote MM, MX, and XM orbitals, respectively. (b) The two top valence bands colored by the coupling magnitude $|F_a^{(n)}(\bm k,0)|=|\langle u_{n\bm k}|M_a|u_{n\bm k}\rangle|$ for the corresponding disorder channels. (c) Charge response amplitudes per unit disorder strength, $\Delta n_{a,n}(\mu)/\gamma_a$, versus hole filling $\nu_h$ for the top ($0<\nu_h<1$) and second ($1<\nu_h<2$) valence bands. Red, blue, and green curves correspond to $V_R$, $D_R$, and $\theta_R$ disorder, respectively. (d) $\langle G_{2t}\rangle$ under $V_R$ (filled squares) and $D_R$ (open circles) disorder. The parameters are $\theta=2.45^\circ$ in (a-d), and $\lambda=2$, $\gamma_a=3\,\mathrm{meV}$ in (c,d).}
    \label{fig:Figure3}
\end{figure}

We next introduce three types of disorder commonly encountered in moir\'e systems~\cite{Uri2020,Foutty2024,Redekop2024,Uri2020TwistDisorder,Wilson2020TwistDisorder}: local onsite potential variations $V_R$, perpendicular displacement-field variations $D_R$, and twist-angle variations $\theta_R=\theta+\delta\theta_R$, defined on each moir\'e unit cell $R$, as illustrated in Fig.~\ref{fig:Figure1}(c). These disorder channels act differently on the local Wannier orbitals. Potential disorder shifts all three orbital energies equally, whereas displacement-field disorder produces opposite shifts of the layer-polarized MX and XM orbitals, with a negligible contribution to the layer-balanced MM orbital. Twist-angle disorder primarily changes the relative onsite energies of the MX/XM and MM orbitals, reflecting their different responses to local variations of the moir\'e structure. Although twist-angle disorder also modifies hopping, we find that its onsite contribution dominates the transport behavior, with comparatively small corrections from hopping variations (see SM.~\ref{sm:disorder}). We therefore retain the onsite contributions of all three disorder channels and obtain the Hamiltonian,
\begin{equation}
    \begin{aligned}
        H &= H_0+H_{\mathrm{dis}},\\
        H_{\mathrm{dis}} = \sum_R w_a(R)&c_R^\dagger M_a c_R,
        \qquad a=V,D,\theta.
    \end{aligned}
    \label{eq:orbital_disorder}
\end{equation}
Here $H_0$ is the clean TB Hamiltonian and $c_R=(c_{R,\mathrm{MX}},c_{R,\mathrm{XM}},c_{R,\mathrm{MM}})^T$ contains the annihilation operators of the three local Wannier orbitals. Different disorder channels are labeled by the index $a$ and are considered separately. The scalar field $w_a(R)$ specifies the disorder amplitude in cell $R$, while the dimensionless diagonal matrix $M_a$ describes its orbital structure, as illustrated in Fig.~\ref{fig:Figure3}(a). Specifically, $M_V=I_3$, $M_D=\sqrt{3/2}\operatorname{diag}(1,-1,0)$, and $M_\theta=\sqrt{1/2}\operatorname{diag}(-1,-1,2)$. The prefactors ensure a common normalization, $\operatorname{Tr}(M_a^2)/3=1$, allowing the three disorder channels to be compared at equal bare disorder strength. Their effective coupling to the electronic bands depends on the orbital composition of the Bloch states, as discussed below.

We model $w_a(R)$ as a zero-mean Gaussian disorder field with variance $\gamma_a^2$, characterizing the bare disorder strength. We consider both spatially uncorrelated disorder ($\lambda=0$) and smoothly correlated disorder with dimensionless correlation length $\lambda$, measured in units of the moir\'e lattice constant $a_M$. The latter captures smooth inhomogeneity on moir\'e length scales~\cite{Uri2020,Foutty2024,Redekop2024,Uri2020TwistDisorder,Wilson2020TwistDisorder}. The disorder field is shared by the three orbitals within each unit cell, with their distinct onsite shifts determined by $M_a$. Details of the disorder construction are provided in SM.~\ref{sm:disorder}.

\prlheading{Integer Chern plateau transitions}
To investigate the effects of different disorder channels on Chern plateau stability and transitions, we calculate the six-terminal Hall resistance $R_{xy}$ using the Landauer--B\"uttiker formalism~\cite{Buttiker1986} implemented in \textsc{KWANT}~\cite{Groth2014}. In the Hall-bar geometry with open boundaries, as shown in Fig.~\ref{fig:Figure2}(a), current $I$ flows from lead 1 to lead 2, while leads 3--6 serve as voltage probes carrying zero net current. The Hall resistance is defined as
\begin{equation}
    R_{xy}=\frac{(V_3-V_6)+(V_4-V_5)}{2I},
    \label{eq:hall_resistance}
\end{equation}
where $V_i$ is the voltage at lead $i$.

Figs.~\ref{fig:Figure2}(b--d) compare the disorder-averaged Hall resistance under the three disorder channels at $\theta=2.45^\circ$, with identical bare disorder strength $\gamma_a=3\,\mathrm{meV}$ and correlation length $\lambda=2$. Throughout this work, $\nu_h$ denotes the hole filling of the three valence bands shown in Fig.~\ref{fig:Figure1}(d), with $\nu_h=0$ corresponding to the chemical potential at the top of the highest valence band. Potential disorder $V_R$ produces broad plateaus at $-\langle R_{xy}\rangle=h/e^2$ and $h/(2e^2)$ over $\nu_h\simeq0.85$--$1.35$ and $1.85$--$2.1$, corresponding to widths $\Delta\nu_h\simeq0.50$ and $0.25$, respectively, as shown in Fig.~\ref{fig:Figure2}(b). These quantized values correspond to cumulative Chern numbers $C=1$ and $2$, consistent with the two highest Chern bands in Fig.~\ref{fig:Figure1}(d). The plateaus are connected by a narrow, nearly direct transition, resembling conventional quantum Hall behavior illustrated in Fig.~\ref{fig:Figure1}(a). In contrast, $D_R$ disorder yields much narrower plateaus at the same quantized values, with widths $\Delta\nu_h\simeq0.15$ and $0.13$, respectively, as shown in Fig.~\ref{fig:Figure2}(c). Between these plateaus, $-\langle R_{xy}\rangle$ exhibits a pronounced dip toward zero before recovering to the second quantized value, characteristic of the broad transition region illustrated in Fig.~\ref{fig:Figure1}(b). Twist-angle disorder produces intermediate behavior, as shown in Fig.~\ref{fig:Figure2}(d): the first plateau has a width $\Delta\nu_h\simeq0.25$, between those of $V_R$ and $D_R$, while the second remains narrow ($\Delta\nu_h\simeq0.15$). The intervening Hall-resistance dip is also less pronounced than for $D_R$ disorder. Similar trends in plateau widths and transition profiles are found at $\theta=3.89^\circ$, with more detailed results presented in SM.~\ref{sm:transport}. As discussed below, these contrasting behaviors originate from the interplay between the Bloch wave function structures of the Chern bands and the layer-dependent nature of disorder.

\prlheading{Layer-dependent disorder and localization}The three disorder channels act differently on the local Wannier orbitals through their onsite matrices $M_a$, as illustrated in Fig.~\ref{fig:Figure3}(a). Their effects on quasiparticles in the Chern bands are determined by the orbital composition of the Bloch states $|u_{n\bm k}\rangle$. At momentum transfer $\bm q$, the disorder potential projected onto the $n$-th band is characterized by the orbital form factor $F_a^{(n)}(\bm k,\bm q)=\langle u_{n,\bm k+\bm q}|M_a|u_{n\bm k}\rangle$, multiplied by the Fourier component $w_a(\bm q)$ of the disorder field. Fig.~\ref{fig:Figure3}(b) shows the magnitude of the zero-momentum form factor $|F_a^{(n)}(\bm k,0)|$ for the two highest valence bands. For potential disorder $V_R$, equal coupling to all three orbitals gives $F_V^{(n)}(\bm k,0)=1$. For displacement-field disorder $D_R$, the opposite contributions from MX and XM can largely cancel, while MM does not contribute. For twist-angle disorder $\theta_R$, cancellation instead occurs between the MM and MX/XM contributions. Consequently, the diagonal form factors with $\bm q=0$ are substantially suppressed for $D_R$ and $\theta_R$ compared with $V_R$. This illustrates how the layer-dependent orbital structure weakens the effective disorder coupling to the Chern bands.

To further characterize the effects of different disorder channels, we calculate the total charge-density response to an orbital-dependent perturbation. The intraband density response is $\delta n_{a,n}(\bm q;\mu)=-\chi_{\rho,a}^{(n)}(\bm q;\mu)w_a(\bm q)$, where
\begin{equation}
    \chi_{\rho,a}^{(n)}(\bm q;\mu) =\frac{1}{N_k}\sum_{\bm k} \frac{f_{n\bm k}-f_{n,\bm k+\bm q}} {\varepsilon_{n,\bm k+\bm q}-\varepsilon_{n\bm k}} \langle u_{n\bm k}|u_{n,\bm k+\bm q}\rangle F_a^{(n)}(\bm k,\bm q).
\label{eq:density_response_kernel}
\end{equation}
Here $f_{n\bm k}$ is the Fermi--Dirac distribution, and the form factor represents the perturbation vertex. We characterize the typical charge rearrangement by the root-mean-square response amplitude
\begin{equation}
    \Delta n_{a,n}(\mu) =\left[ \frac{1}{N_q}\sum_{\bm q} |\chi_{\rho,a}^{(n)}(\bm q;\mu)|^2 \overline{|w_a(\bm q)|^2} \right]^{1/2},
    \label{eq:charge_response}
\end{equation}
where $\overline{|w_a(\bm q)|^2}=\gamma_a^2 S_{a,\lambda}(\bm q)$ incorporates the disorder correlations controlled by $\lambda$ (see SM.~\ref{sm:disorder}). At equal bare disorder strength, $\Delta n_{a,n}/\gamma_a$ quantifies the relative total-charge response to different disorder channels. Fig.~\ref{fig:Figure3}(c) shows that the response is generally strongest for $V_R$ and weakest for $D_R$, with $\theta_R$ intermediate, across the two highest valence bands. This hierarchy is consistent with the projected disorder couplings in Fig.~\ref{fig:Figure3}(b), indicating that displacement-field disorder induces substantially weaker total-charge redistribution than potential disorder. Further details of the charge-response analysis are provided in SM.~\ref{sm:response}.

To examine the corresponding bulk transport behavior, we calculate the two-terminal conductance under $V_R$ and $D_R$ disorder. We retain leads 1 and 2 of the geometry in Fig.~\ref{fig:Figure2}(a) and impose periodic boundary conditions along the $y$ direction to eliminate edge conduction. After correcting for the Sharvin contact resistance~\cite{Schep1997}, the conductance reads
\begin{equation}
    G_{2t}=\frac{e^2}{h} \left[ \frac{1}{T}-\frac{1}{2} \left(\frac{1}{N_1}+\frac{1}{N_2}\right) \right]^{-1},
\label{eq:bulk_conductance}
\end{equation}
where $T=\operatorname{Tr}(t^\dagger t)$ is the total transmission, $t$ is the transmission matrix, and $N_1,N_2$ denote the numbers of propagating modes in the two leads. Fig.~\ref{fig:Figure3}(d) shows that the disorder-averaged bulk conductance $\langle G_{2t}\rangle$ approaches zero within the Hall plateau regions for both disorder channels, consistent with bulk localization. Between plateaus, however, $V_R$ disorder produces a small, nearly size-independent conductance peak of $\sim0.5\,e^2/h$, whereas the peak under $D_R$ disorder increases from $\sim3.3$ to $4.3\,e^2/h$ as $N_x$ increases from 16 to 48. This demonstrates substantially stronger bulk conduction under $D_R$ disorder. 

In the isotropic bulk approximation, the Hall resistance satisfies $|R_{xy}|\simeq|\sigma_{xy}|/(\sigma_{xx}^2+\sigma_{xy}^2)$. On quantized plateaus, $\sigma_{xx}\rightarrow0$ and $|R_{xy}|\simeq1/|\sigma_{xy}|$. Between plateaus, stronger localization under $V_R$ disorder suppresses bulk conduction, yielding nearly direct Hall transitions resembling those in conventional quantum Hall systems. By contrast, the larger longitudinal conduction under $D_R$ disorder suppresses $|R_{xy}|$ relative to $1/|\sigma_{xy}|$, producing the pronounced Hall-resistance dip observed in Fig.~\ref{fig:Figure2}(c). 
Furthermore, $R_{xy}$ in Fig.~\ref{fig:Figure2} exhibits asymmetric broadening, particularly under $D_R$ and $\theta_R$ disorder around $\nu_h=1$, as shown in Fig.~\ref{fig:Figure2}(c,d). This asymmetry is qualitatively accounted for by the unequal effective disorder coupling at the band edges on either side of $\nu_h=1$, as reflected in the projected disorder matrix elements and charge-response amplitudes in Fig.~\ref{fig:Figure3}(b,c). The stronger projected coupling and larger charge response on the $\nu_h>1$ side suggest greater sensitivity to disorder, supporting more effective bulk localization and greater plateau broadening. Conversely, the weaker coupling on the $\nu_h<1$ side is consistent with the more rapid loss of Hall quantization. Similar bulk transport trends are found at $\theta=3.89^\circ$, with further details provided in SM.~\ref{sm:transport}.

\prlheading{Fractional Chern plateau transport}
Motivated by the fractional Chern plateaus observed in~\cite{Cai2023,Zeng2023,Park2023,Xu2023,Park2025Dissipationless}, we extend the disorder analysis to $\theta=3.89^\circ$, a representative twisting angle in the experimentally relevant regime. We retain the full three-band Wannier Hamiltonian with MX, XM, and MM orbitals, as detailed in SM.~\ref{sm:model}, while focusing on hole doping of the top valence band over $0\leq\nu_h\leq1$. Here $\nu_h=1$ corresponds to an empty top electron band. This band carries electronic Chern number $C_e=-1$, corresponding to $C_h=+1$ in the hole representation.

\begin{figure} 
    \centering
    \includegraphics[width=\linewidth]{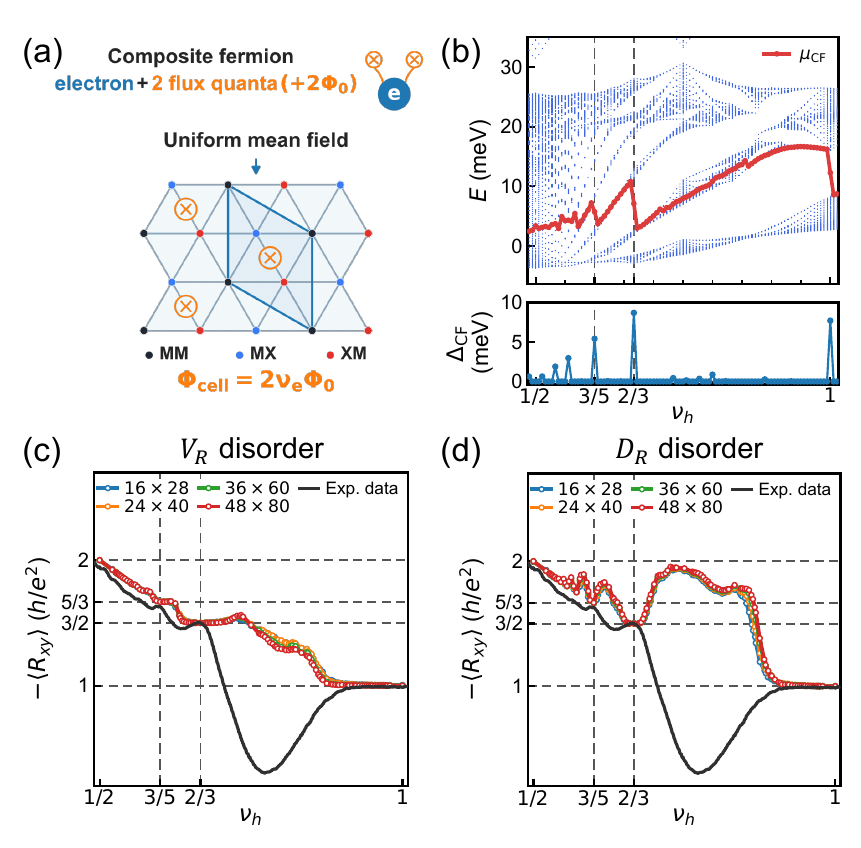}
    \caption{\textbf{CF spectrum and fractional plateau transport.} (a) Schematic of the CF construction and its uniform mean-field representation, with $2\Phi_0$ attached to each electron and a flux $\Phi_{\mathrm{cell}}=2\nu_e\Phi_0$ uniformly distributed within the moir\'e cell. (b) Upper panel: Clean CF Hofstadter spectrum (blue) and chemical potential $\mu_{\mathrm{CF}}$ (red) versus hole filling $\nu_h$. Lower panel: The corresponding spectral gap $\Delta_{\mathrm{CF}}$ at $\mu_{\mathrm{CF}}$. (c),(d) Disorder-averaged physical Hall resistance $-\langle R_{xy}\rangle$ of disordered CF model with $V_R$ and $D_R$ disorder, respectively. Colored curves correspond to different system sizes. Black curves show experimental data obtained from Ref.~\cite{Park2025Dissipationless}. The calculations  in (c,d) use $\theta=3.89^\circ$, with $\gamma_a=3\,\mathrm{meV}$ and $\lambda=2$.}
    \label{fig:Figure4}
\end{figure}

We adopt a CF mean-field description~\cite{Jain1989,LopezFradkin1991} previously employed in studies of moir\'e FCIs~\cite{SternFu2023,LuSantos2024,HuXiaoRan2024,LuWuSantos2025,HuRanXiao2026,JinShi2026,PatriDongSenthil2024,DingMillisZeng2026}. We implement flux attachment in the electron representation, assigning two statistical flux quanta $+2\Phi_0$ to each electron of charge $-e$, with $\Phi_0=h/e$. The electron filling counts all three bands, so $\nu_e=3-\nu_h$. In the spatially uniform mean-field approximation, the statistical flux per moir\'e cell is therefore $\Phi_{\mathrm{cell}}=2\nu_e\Phi_0=(6-2\nu_h)\Phi_0$, as illustrated in Fig.~\ref{fig:Figure4}(a). The CF Hamiltonian expressed in hole representation reads
\begin{equation}
    \begin{aligned}
        H_{\mathrm{CF}}
        =-\sum_{ij}[h^e_{ij}]^*e^{i\varphi_{ij}}f_i^\dagger f_j,\quad
        \varphi_{ij}
        =\frac{2\pi}{\Phi_0}
        \int_{\bm{r}_j}^{\bm{r}_i}\bm{A}_{\mathrm{CS}}\cdot d\bm{\ell}.
    \end{aligned}
    \label{eq:cf_hamiltonian}
\end{equation}
Here $f_i$ annihilates a CF in the hole representation, $-[h^e]^*$ is the hole TB matrix, and $\bm A_{\mathrm{CS}}$ generates the statistical flux defined above. To obtain the physical transport response, we use the Chern--Simons (CS) relation between the physical and CF Hall resistances in the ideal Hall-bar approximation~\cite{SternFu2023,LuSantos2024},
\begin{equation}
    \langle R_{xy}\rangle
    \simeq\langle R_{xy}^{\mathrm{CF}}\rangle-2\frac{h}{e^2}.
    \label{eq:cf_hall_resistance}
\end{equation}
Here $\langle...\rangle$ denote disorder averaging.

We first examine the clean CF spectrum. Varying $\nu_h$ changes both the hole density and the statistical flux, producing the Hofstadter spectrum shown in Fig.~\ref{fig:Figure4}(b). The chemical potential $\mu_{\mathrm{CF}}$, determined by the hole density, is shown in red, and exhibits jumps at $\nu_h=3/5$ and $2/3$ associated with pronounced spectral gaps $\Delta_{\mathrm{CF}}$, as depicted in the lower panel. These gaps identify incompressible states within the CF description. The occupied CF bands carry total Chern numbers $C_{\mathrm{CF}}=-3$ and $-2$, respectively, which yield fractionally quantized Hall resistances $-R_{xy}=5h/(3e^2)$ and $3h/(2e^2)$ through Eq.~\eqref{eq:cf_hall_resistance}. A gap also appears at $\nu_h=1$ with $C_{\mathrm{CF}}=-1$, corresponding to the integer Chern plateau with $-R_{xy}=h/e^2$.

We next include $V_R$ and $D_R$ disorder with the same bare strength $\gamma_a=3\,\mathrm{meV}$ and correlation length $\lambda=2$. Using the six-terminal geometry in Fig.~\ref{fig:Figure2}(a), we calculate $\langle R_{xy}^{\mathrm{CF}}\rangle$ and convert it to the physical Hall resistance using Eq.~\eqref{eq:cf_hall_resistance}. As shown in Fig.~\ref{fig:Figure4}(c,d), both disorder channels produce fractional plateaus near $\nu_h=3/5$ and $2/3$, together with an integer plateau near $\nu_h=1$, in agreement with the experimental plateaus~\cite{Park2025Dissipationless} shown in black. The fractional plateaus are broader under $V_R$ than under $D_R$, following the same disorder dependence as the integer plateaus in Fig.~\ref{fig:Figure2}(b,c). This trend is consistent with less effective localization of CF bulk states under $D_R$ at equal disorder strength. The narrower experimental fractional plateaus more closely resemble the $D_R$ results.

On the other hand, we note that the calculated transport behavior between fractional plateaus deviates from experiment, highlighting the limitations of the CF mean-field treatment in describing the plateau transitions. Under $D_R$ disorder, the weaker effective disorder coupling allows stronger CF bulk conduction, suppressing $\langle R_{xy}^{\mathrm{CF}}\rangle$ and leaving the CS contribution in Eq.~\eqref{eq:cf_hall_resistance} dominant. Consequently, $-\langle R_{xy}\rangle$ approaches $2h/e^2$, as shown in Fig.~\ref{fig:Figure4}(d), in contrast to the experimentally observed Hall-resistance dips shown by the black solid curve.

\prlheading{Summary}
Our results show that Chern plateau stability depends not only on the bare disorder strength but also on how disorder acts on the layer components of Bloch states. Taking \tmt{} as an example, we construct three-orbital Wannier-based TB models at different twist angles and compare Hall and bulk transport under onsite-potential, displacement-field, and twist-angle disorder. At equal bare disorder strength, their layer-dependent effects on localization yield contrasting integer plateau profiles: broad plateaus with nearly direct transitions under $V_R$, versus narrow plateaus separated by Hall dips under $D_R$, qualitatively consistent with experiment. CF mean-field theory also reproduces the $2/3$ and $3/5$ plateaus and captures their disorder-dependent widths. These findings suggest a common mechanism, rooted in layer-dependent disorder, for plateau stability in Chern systems and motivate further studies in other moir\'e platforms, including rhombohedral graphene/hBN superlattices~\cite{LuGraphene2024,LuGraphene2025}.

\prlheading{Acknowledgments}
This work was supported by the National Natural Science Foundation of China (Nos. 12494594 and 12574150), the Science Center of the NSFC (Grant Nos. 12188101 and 12621008), the National Key R\&D Program of China (Grant No. 2022YFA1403800), and the Chinese Academy of Sciences Project for Young Scientists in Basic Research (Grant No. 2022YSBR-048). H.W. also acknowledges support from the New Cornerstone Science Foundation through the XPLORER PRIZE.

\bibliography{reference}

\clearpage
\onecolumngrid

\pagenumbering{arabic}
\renewcommand{\thepage}{S\arabic{page}}

\setcounter{section}{0}
\setcounter{subsection}{0}
\setcounter{equation}{0}
\setcounter{figure}{0}
\setcounter{table}{0}
\setcounter{secnumdepth}{2}

\renewcommand{\thesection}{\Roman{section}}
\renewcommand{\thesubsection}{\Alph{subsection}}
\renewcommand{\theequation}{S\arabic{equation}}
\renewcommand{\thefigure}{S\arabic{figure}}
\renewcommand{\thetable}{S\arabic{table}}

\renewcommand{\theHsection}{SM.\arabic{section}}
\renewcommand{\theHequation}{SM.\arabic{equation}}
\renewcommand{\theHfigure}{SM.\arabic{figure}}
\renewcommand{\theHtable}{SM.\arabic{table}}

\begin{center}
  {\large\bfseries Supplemental Material\par}
  \vspace{0.5em}
  {\bfseries
  Integer and Fractional Chern Plateaus Shaped by
  Orbital-Selective Disorder}
\end{center}

\vspace{1em}

\newcommand{\SMsection}[1]{%
  \section{#1}%
  \addcontentsline{smtoc}{section}
    {\protect\numberline{\thesection}#1}%
}

\newcommand{\SMsubsection}[1]{%
  \subsection{#1}%
  \addcontentsline{smtoc}{subsection}
    {\protect\numberline{\thesubsection}#1}%
}

\begin{center}
  \textbf{CONTENTS}
\end{center}

\makeatletter
\@starttoc{smtoc}
\makeatother

\vspace{1em}

\SMsection{Moir\'e model construction}
\label{sm:model}

This section describes the construction of the three-orbital tight-binding (TB) model of \tmt{} used in our transport calculations. We first introduce the single-$K$-valley continuum model extracted directly from first-principles calculations without parameter fitting~\cite{Zhang2024NoFit}. We then construct maximally localized Wannier functions (MLWFs) for its three highest valence bands using \textsc{WANNIER90} package~\cite{MOSTOFI2008685}, whose Chern numbers sum to zero, and obtain the corresponding TB Hamiltonian.
\SMsubsection{Accurate continuum model without fitting}
\label{sm:continuum_model}

Unlike earlier continuum models fitted to density-functional-theory (DFT) bands~\cite{Wu2019Topological,Devakul2021Magic,Reddy2023FQAH,Wang2024FCI,Jia2024}, we adopt a model extracted directly from first-principles Hamiltonians without fitting, which accurately reproduces the low-energy DFT bands and wavefunctions~\cite{Zhang2024NoFit}.
The underlying DFT calculations use OpenMX software~\cite{ozaki_Variationally_2003,ozaki_Numerical_2004,ozaki_Efficient_2005} with the standard pseudo-atomic-orbital basis \texttt{Mo7.0-s3p2d2} and \texttt{Te7.0-s3p2d2f1}, with a radial cutoff of $7$ Bohr.
We use the resulting models at six twist angles, $\theta=2.45^\circ$, $2.65^\circ$, $2.88^\circ$, $3.15^\circ$, $3.48^\circ$, and $3.89^\circ$.

The extraction begins by projecting the DFT Hamiltonian onto atomic plane waves near the $K$ valley using the truncated atomic plane wave (TAPW) method~\cite{Miao2023TAPW}.
Integrating out remote states then leaves one spin-polarized valence state per layer and plane wave.
The resulting low-energy Hamiltonian $H_{CM}$ is expanded in symmetry-allowed continuum terms, whose coefficients are determined by projection~\cite{Zhang2024NoFit}.
Grouping the associated annihilation operators by layer, we write
\begin{equation}
\Psi_{\bm k}=
\begin{pmatrix}
\psi_{\bm k,\bm Q_1^{(t)}}\\
\vdots\\
\psi_{\bm k,\bm Q_{N_t}^{(t)}}\\
\psi_{\bm k,\bm Q_1^{(b)}}\\
\vdots\\
\psi_{\bm k,\bm Q_{N_b}^{(b)}}
\end{pmatrix},
\qquad
\begin{aligned}
H_{\mathrm{CM}}
&=\sum_{\bm k}\Psi_{\bm k}^\dagger
h(\bm k)\Psi_{\bm k},\\
h_{ij}(\bm k)
&=\langle \bm k-\bm Q_i,\ell_i|
\hat h_K|\bm k-\bm Q_j,\ell_j\rangle .
\end{aligned}
\label{eq:sm_continuum_basis}
\end{equation}
Here $N_\ell$ counts the retained plane waves in layer $\ell=t,b$.
The index $i=1,\ldots,N_t+N_b$ runs over both layers in the order shown in $\Psi_{\bm k}$, with each entry specified by its wave vector $\bm Q_i$ and layer $\ell_i$.
The basis state $|\bm k-\bm Q_i,\ell_i\rangle$ is a plane wave with envelope momentum $\bm k-\bm Q_i$ in layer $\ell_i$.
The corresponding component of $\Psi_{\bm k}$ annihilates a particle in this state.
The wave vectors satisfy $\bm Q_i=\bm\kappa_{\ell_i}+\bm G_i$, where $\bm G_i$ is a moir\'e reciprocal vector and $\bm\kappa_\ell$ locates the layer's $K$ point relative to a common origin; $\bm k$ lies in the moir\'e Brillouin zone.
Here $\hat h_K$ is the single-particle $K$-valley Hamiltonian.

The continuum Hamiltonian combines the valence-band dispersion with moir\'e couplings within and between the layers:
\begin{equation}
h(\bm k)=
\begin{pmatrix}
E_t(\bm k)+V_t(\bm k) & T(\bm k)\\
T^\dagger(\bm k) & E_b(\bm k)+V_b(\bm k)
\end{pmatrix},
\label{eq:sm_continuum_blocks}
\end{equation}
where $E_\ell$, $V_\ell$, and $T$ describe kinetic energy, intralayer moir\'e coupling, and interlayer tunneling, respectively.
These contributions include both higher moir\'e harmonics and momentum-dependent couplings.
To write their general form, label each basis state by $(\ell_i,\bm Q_i)$ and define $\bm p_i=\bm k-\bm Q_i$ and $\bm q_{ij}=\bm Q_i-\bm Q_j$.
The matrix elements take the form
\begin{equation}
h_{ij}(\bm k)
=\sum_{m,n\geq0}p_{i,x}^{m}p_{i,y}^{n}
\begin{cases}
a_{\ell_i}^{mn},
& i=j,\\[2pt]
V_{\ell_i}^{mn}(\bm q_{ij}),
& i\neq j,\ \ell_i=\ell_j,\\[2pt]
w_{\ell_i\ell_j}^{mn}(\bm q_{ij}),
& \ell_i\neq\ell_j .
\end{cases}
\label{eq:sm_continuum_expansion}
\end{equation}
The three lines give the kinetic, intralayer, and interlayer terms.
The quadratic kinetic coefficients satisfy $a_\ell^{20}=a_\ell^{02}=-\hbar^2/(2m^*)$, with $m^*>0$ the magnitude of the valence-band effective mass.
The momentum transfer $\bm q_{ij}$ labels the moir\'e harmonics, while $m,n$ label the momentum powers.
We retain the terms specified in the source model~\cite{Zhang2024NoFit,zhang_continuum_2024}, subject to Hermiticity and crystal symmetries.
The resulting bands and Bloch states provide the starting point for the Wannier construction below.

\begin{figure*}[htbp]
    \centering
    \includegraphics[width=\linewidth]{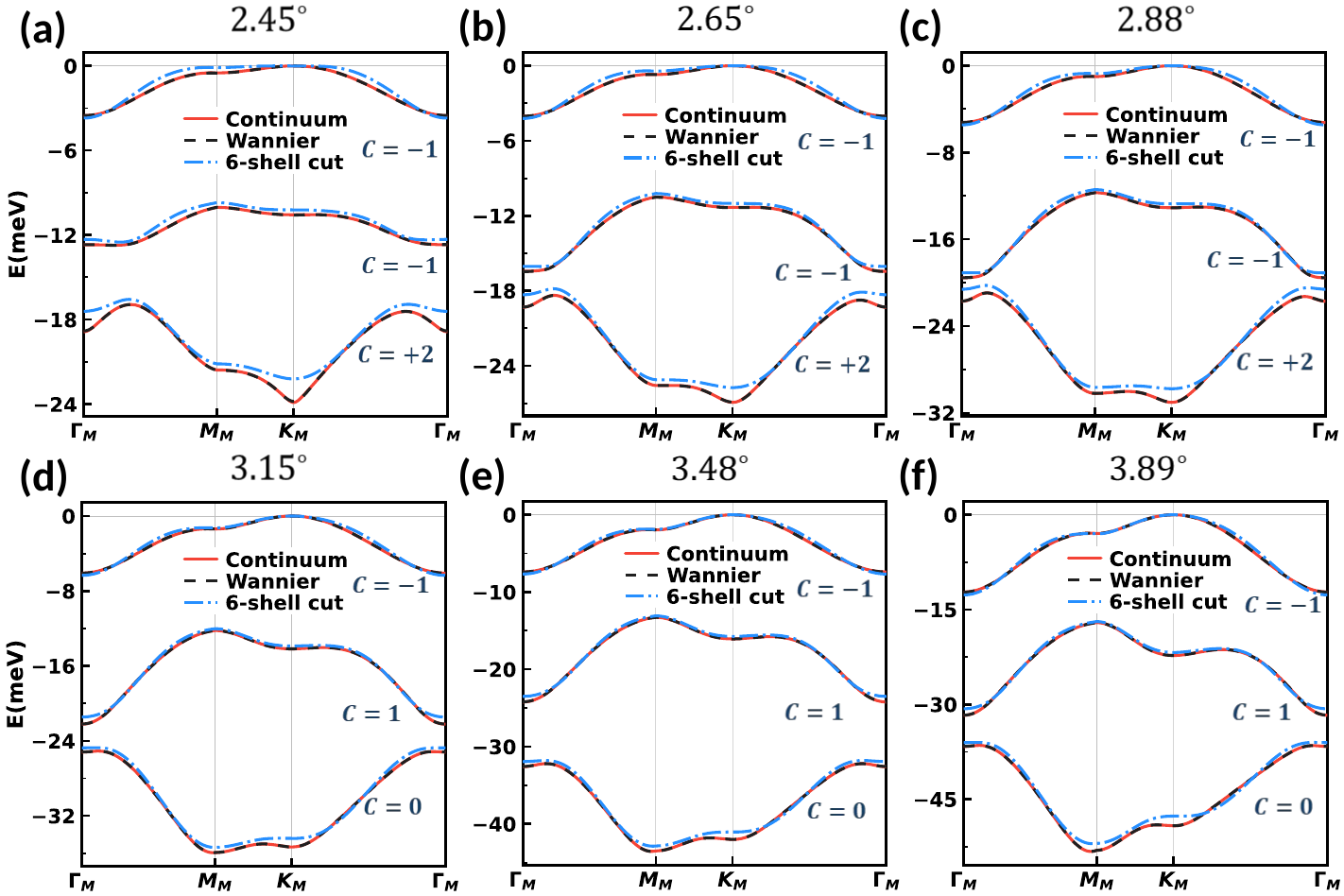}
    \caption{Three highest $K$-valley valence bands of \tmt{} at
(a) $2.45^\circ$, (b) $2.65^\circ$, (c) $2.88^\circ$,
(d) $3.15^\circ$, (e) $3.48^\circ$, and (f) $3.89^\circ$.
Red solid, black dashed, and blue dash-dotted lines correspond
to the continuum model, the Wannier TB model, and the TB model
with hoppings retained through the sixth neighbor shell,
respectively.
The band Chern numbers are indicated in each panel. }
    \label{fig:SM1_Continuum-TB-model}
\end{figure*}

Figure~\ref{fig:SM1_Continuum-TB-model} shows the $K$-valley continuum bands (red) at the six twist angles from $2.45^\circ$ to $3.89^\circ$.
The Chern numbers of the three highest valence bands, ordered from top to bottom, are $(-1,-1,2)$ at $\theta=2.45^\circ$, $2.65^\circ$, and $2.88^\circ$, and $(-1,1,0)$ at $\theta=3.15^\circ$, $3.48^\circ$, and $3.89^\circ$, as labeled in Fig.~\ref{fig:SM1_Continuum-TB-model}.

\SMsubsection{Wannierization and tight-binding model}
\label{sm:wannierization}

\begin{figure*}[htbp]
    \centering
    \includegraphics[width=\linewidth]{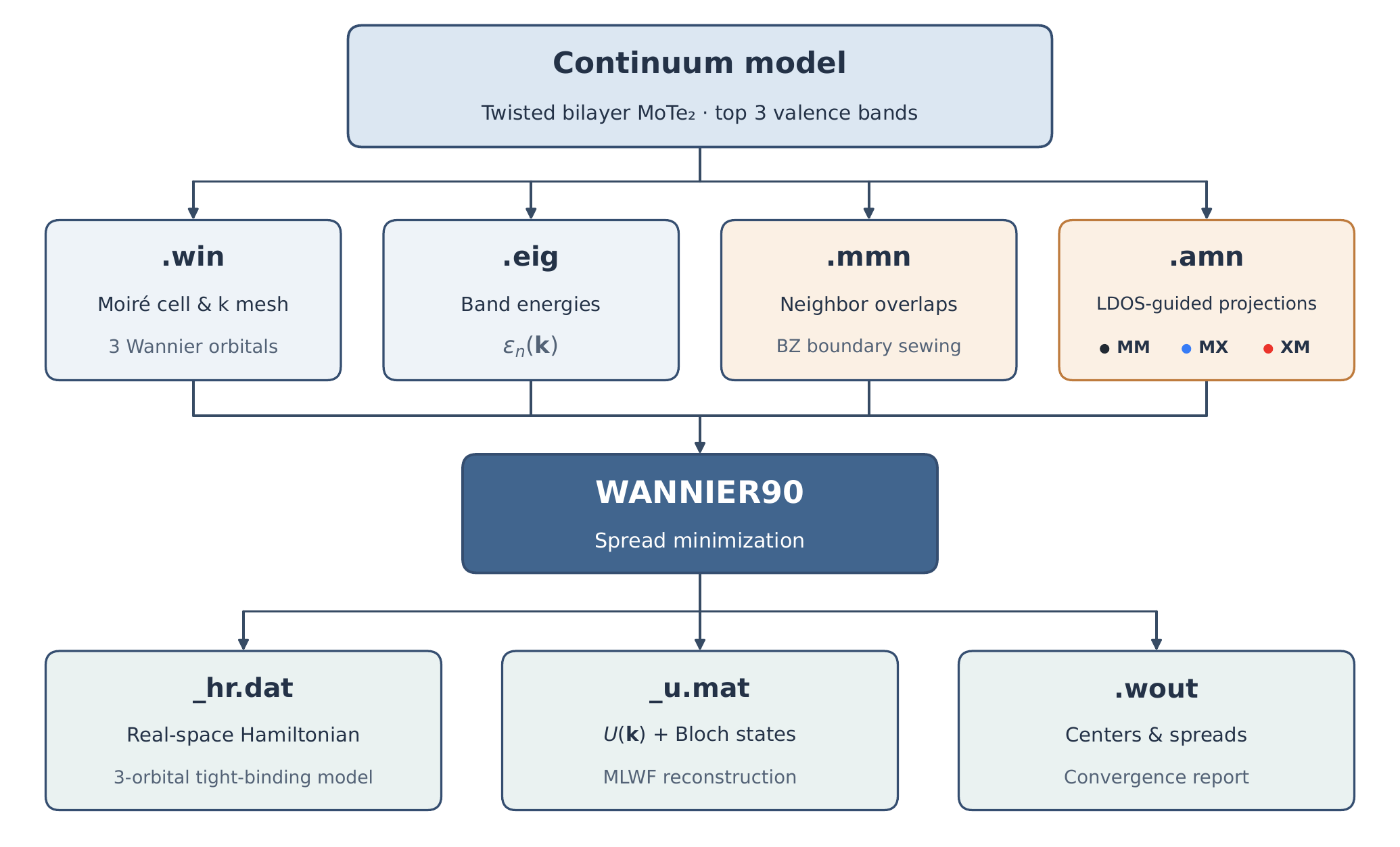}
    \caption{ Workflow for constructing a three-orbital TB model and MLWFs
from the three highest $K$-valley continuum bands of \tmt{}.
The four Wannier90 inputs contain the moir\'e cell and momentum
mesh (\texttt{.win}), band energies (\texttt{.eig}), neighbor
overlaps with Brillouin-zone boundary sewing (\texttt{.mmn}),
and LDOS-guided initial projections (\texttt{.amn}).
Spread minimization yields the real-space TB Hamiltonian
(\texttt{\_hr.dat}), unitary rotations for MLWF reconstruction
from the continuum Bloch states (\texttt{\_u.mat}), and
Wannier centers and spreads (\texttt{.wout}).}
    \label{fig:SM1_wannier-workflow-overview}
\end{figure*}

\begin{figure*}[htbp]
    \centering
    \includegraphics[width=\linewidth]{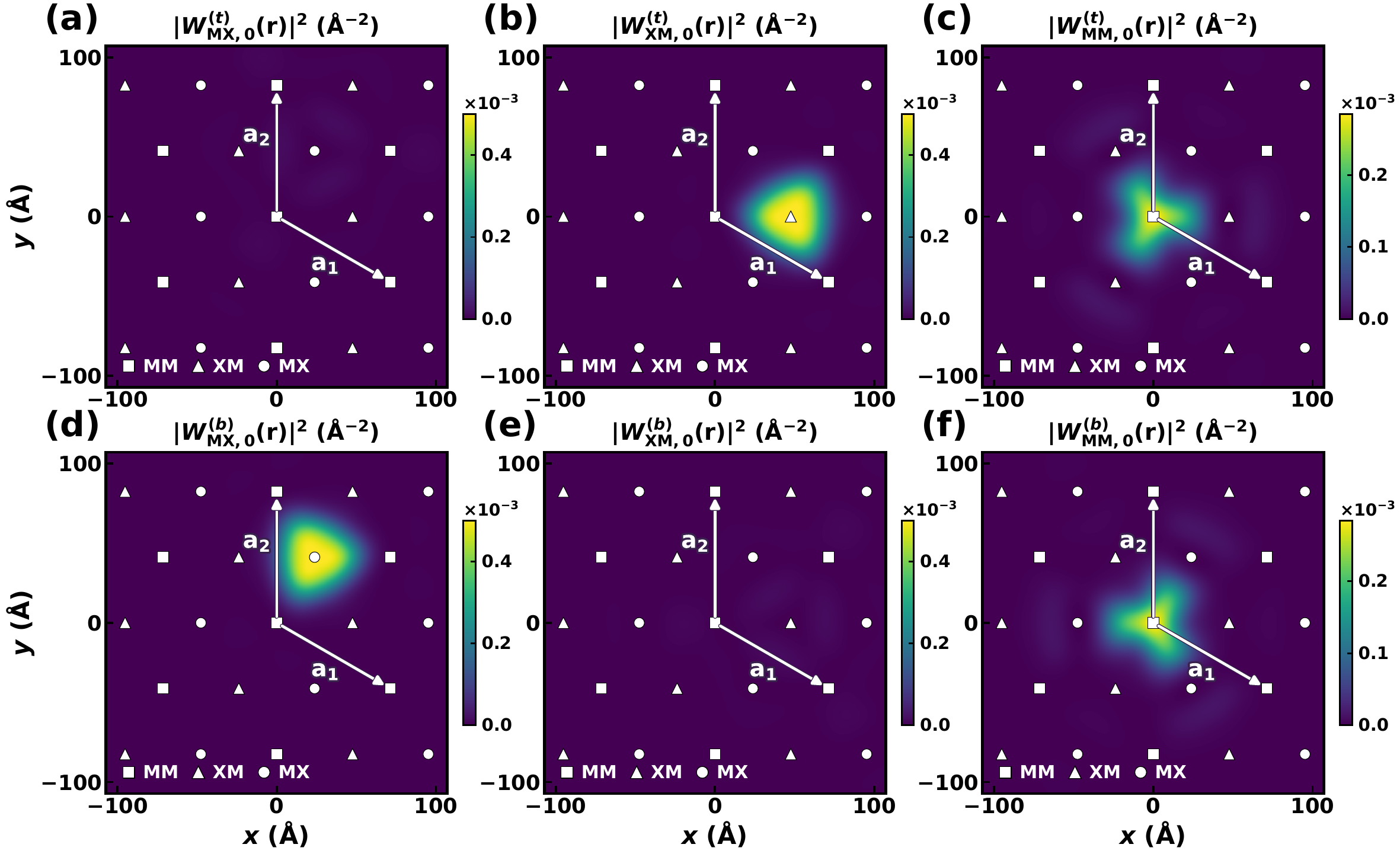}
    \caption{Layer-resolved Wannier function densities
$|W_{\alpha,0}^{(\ell)}(\bm r)|^2$ of \tmt{} at
$\theta=2.45^\circ$.
(a)--(c) Top-layer densities of the MX, XM, and MM orbitals;
(d)--(f) the corresponding bottom-layer densities.
Squares, triangles, and circles mark MM, XM, and MX sites,
respectively; arrows indicate the moir\'e lattice vectors
$\bm a_1$ and $\bm a_2$.
The two layers of each orbital share a color scale in \AA$^{-2}$. }
    \label{fig:SM1_wannier-density-2.45deg}
\end{figure*}

\begin{figure*}[htbp]
    \centering
    \includegraphics[width=\linewidth]{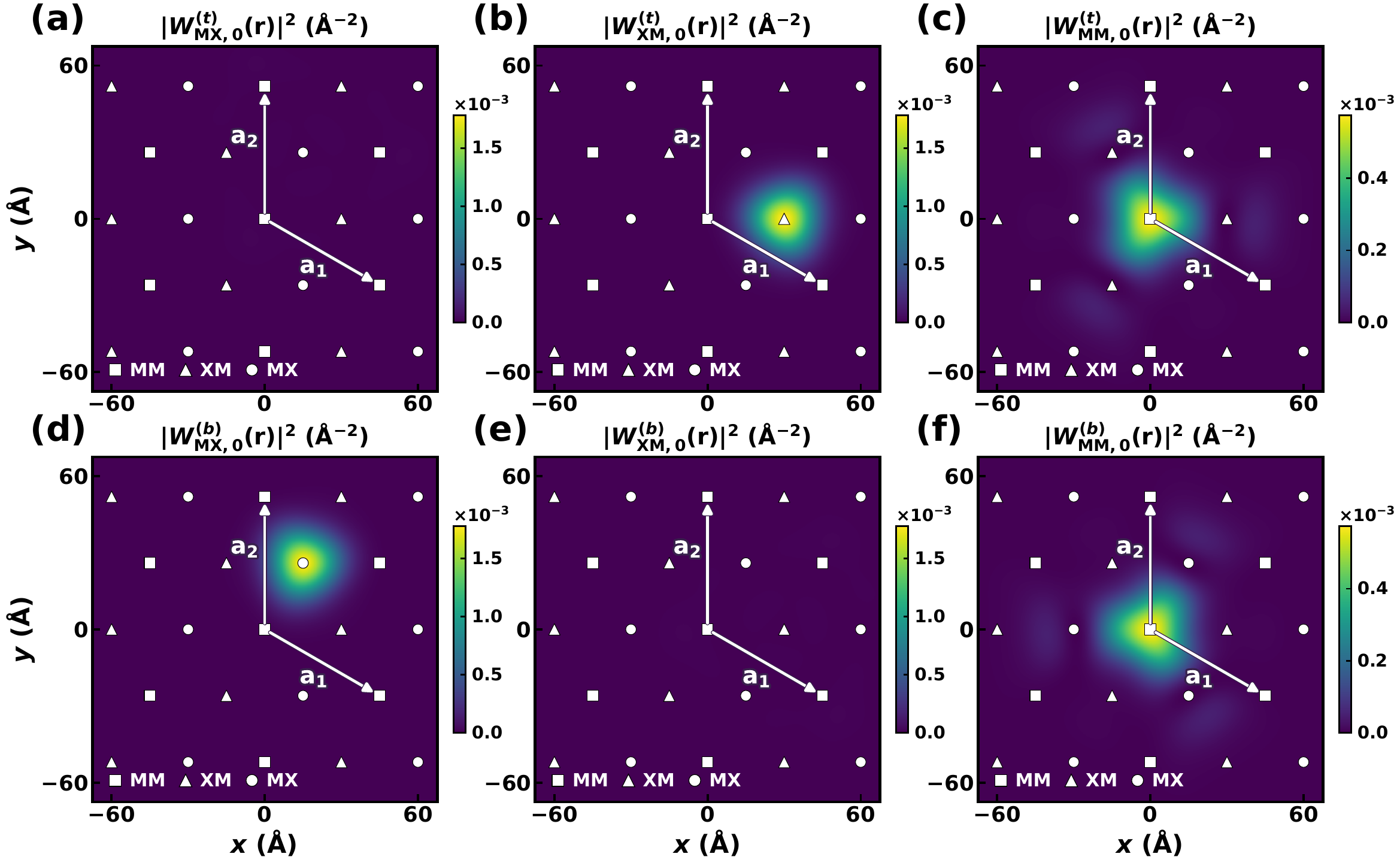}
    \caption{ Layer-resolved Wannier function densities
$|W_{\alpha,0}^{(\ell)}(\bm r)|^2$ of \tmt{} at
$\theta=3.89^\circ$.
(a)--(c) Top-layer densities of the MX, XM, and MM orbitals;
(d)--(f) the corresponding bottom-layer densities.
Squares, triangles, and circles mark MM, XM, and MX sites,
respectively; arrows indicate the moir\'e lattice vectors
$\bm a_1$ and $\bm a_2$.
The two layers of each orbital share a color scale in \AA$^{-2}$.}
    \label{fig:SM1_wannier-density-3.89deg}
\end{figure*}

For real-space transport calculations, we need to extract a lattice TB model from the low-energy bands of the continuum model.
The low-energy valence states of \tmt{} originate from the $K$ and $K'$ valleys, where strong spin--orbit coupling locks opposite spins to the two valleys and suppresses intervalley mixing~\cite{Wu2019Topological,Devakul2021Magic,Reddy2023FQAH,Wang2024FCI,Jia2024}.
We therefore work with the single-$K$-valley continuum models constructed above at the six twist angles from $2.45^\circ$ to $3.89^\circ$.
Their three highest valence bands form an isolated group with Chern numbers that sum to zero, allowing us to Wannierize them by using \textsc{WANNIER90} package~\cite{MOSTOFI2008685} to obtain MLWFs and a three-orbital TB model.

Different from the standard DFT--WANNIER90 workflow, we generate the four input files (\texttt{.win}, \texttt{.eig}, \texttt{.mmn}, \texttt{.amn}) directly from the continuum-model data.
This requires particular care with the plane-wave basis at the moir\'e Brillouin-zone boundary and with the choice of initial trial orbitals.
This continuum-model-based Wannierization approach is broadly applicable and has also been used to construct effective low-energy TB models for other moir\'e systems~\cite{song_1,Yu_2023_PhysRevB,Qi2026Multivalley}.
The workflow of this approach is summarized in Fig.~\ref{fig:SM1_wannier-workflow-overview}, with each step described below:
\begin{enumerate}

\item \textit{Lattice and band information
(\texttt{.win}, \texttt{.eig}).}
The \texttt{.win} file specifies the moir\'e cell, a
$151\times151$ momentum mesh, and
\texttt{num\_bands=num\_wann=3}.
The \texttt{.eig} file stores the corresponding continuum
eigenvalues $\varepsilon_s(\bm k)$ in eV, where $s=1,2,3$
is the band index.
All input files use the same momentum and band ordering.
No disentanglement is required for this isolated three-band group.

\item \textit{Neighbor overlaps (\texttt{.mmn}).}
The \texttt{.mmn} file stores the $3\times3$ overlap matrices
$M_{ss'}(\bm k,\bm b)$ between the cell-periodic parts of
Bloch states at neighboring momenta $\bm k$ and $\bm k+\bm b$.
To evaluate these overlaps, we expand the continuum eigenstates
in the layer-resolved plane-wave basis defined above:
\begin{equation}
|\Phi_{s\bm k}\rangle
=\sum_i C_{is}(\bm k)
|\bm k-\bm Q_i,\ell_i\rangle.
\label{eq:sm_cm_eigenstates}
\end{equation}
Here $C_{is}(\bm k)$ are the plane-wave coefficients, and
$|u_{s\bm k}\rangle
=e^{-i\bm k\cdot\bm r}|\Phi_{s\bm k}\rangle$
is the cell-periodic part.

The neighbor pairs and reciprocal translations are specified
in the \texttt{.nnkp} file generated by Wannier90 preprocessing.
When a neighbor crosses the moir\'e Brillouin-zone boundary,
we write $\bm k+\bm b=\bm k_b+\bm G$, with $\bm k_b$ on the
stored mesh and $\bm G$ a moir\'e reciprocal vector.
We introduce the sewing matrix $S_{\bm G}$ to express the
eigenvectors stored at $\bm k_b$ in the plane-wave basis at
$\bm k+\bm b$.
It shifts the $\bm Q$ labels by $\bm G$ without mixing layers,
giving
\begin{equation}
\begin{aligned}
M_{ss'}(\bm k,\bm b)
&=\langle u_{s\bm k}|u_{s',\bm k+\bm b}\rangle
=\big[C^\dagger(\bm k)S_{\bm G}C(\bm k_b)\big]_{ss'},\\
(S_{\bm G})_{ij}
&=
\begin{cases}
1,& \ell_i=\ell_j,\quad \bm Q_i=\bm Q_j+\bm G,\\
0,& \text{otherwise}.
\end{cases}
\end{aligned}
\label{eq:sm_wannier_sewing}
\end{equation}
Each nonzero element connects states in the same layer
with the same plane-wave momentum, since
$\bm k+\bm b-\bm Q_i=\bm k_b-\bm Q_j$.
We use $S_{\bm 0}=I$ and retain only matches within the
plane-wave cutoff.

\item \textit{Initial projections (\texttt{.amn}).}
The \texttt{.amn} file stores the initial projection matrices
$A_{s\alpha}(\bm k)$ for MLWF construction, defined by the overlaps
between continuum Bloch states and localized trial orbitals.
Guided by the layer-resolved local density of states (LDOS)
calculated from the continuum model, we choose Gaussian trial
orbitals centered at MX, XM, and MM, which take the form
\begin{equation}
g_\alpha^{(\ell)}(\bm r)
=\frac{\chi_{\ell\alpha}}{\sqrt{\pi}\sigma_\alpha}
\exp\!\left[
-\frac{|\bm r-\bm\tau_\alpha|^2}{2\sigma_\alpha^2}
\right],
\qquad
\alpha=\mathrm{MX},\mathrm{XM},\mathrm{MM}.
\label{eq:sm_wannier_trials}
\end{equation}
Here $\sigma_\alpha$ is the real-space width and
$\chi_{\ell\alpha}$ is a normalized layer spinor.
For moir\'e primitive vectors $\bm a_1,\bm a_2$, the centers are
$\bm\tau_{\mathrm{MX}}=(\bm a_1+2\bm a_2)/3$,
$\bm\tau_{\mathrm{XM}}=(2\bm a_1+\bm a_2)/3$, and
$\bm\tau_{\mathrm{MM}}=\bm 0$ (see Fig.\ref{fig:Figure1}).
The MX and XM trials occupy opposite layers, while the MM
trial has equal layer weights.
These layer weights respect the $C_{2y}\mathcal{T}$.
The $C_{2y}$ operation is a $180^\circ$ rotation about the
in-plane $y$ axis that flips the layers and exchanges MX/XM centers
while leaving MM fixed~\cite{Jia2024}, and $\mathcal{T}$ is time reversal symmetry.
The layer spinor also specifies the relative phase, which
is not determined by the density alone.
Fourier transformation gives the trial coefficients in
the continuum basis and hence the projection matrix:
\begin{equation}
\begin{aligned}
G_{i\alpha}(\bm k)
&=\mathcal N_\alpha\chi_{\ell_i\alpha}
e^{-\sigma_\alpha^2|\bm p_i|^2/2}
e^{-i\bm p_i\cdot\bm\tau_\alpha},
\qquad \bm p_i=\bm k-\bm Q_i,\\
A_{s\alpha}(\bm k)
&=\langle\Phi_{s\bm k}|g_\alpha\rangle
=\sum_i C_{is}^*(\bm k)G_{i\alpha}(\bm k).
\end{aligned}
\label{eq:sm_wannier_projections}
\end{equation}
Here $\mathcal N_\alpha$ is a momentum-independent
normalization constant.
The matrix $A(\bm k)$ must have full rank throughout the
mesh to initialize all three Wannier orbitals.
Their final layer weights are determined by the subsequent
localization.

\end{enumerate}

With these inputs, Wannier90 optimizes a $3\times3$ unitary
matrix $U(\bm k)$ to minimize the total quadratic spread
$\Omega=\sum_\alpha
(\langle r^2\rangle_\alpha-|\langle\bm r\rangle_\alpha|^2)$.
The resulting Wannier functions are
\begin{equation}
|W_{\alpha R}\rangle
=\frac{1}{\sqrt{N_k}}
\sum_{\bm k,s}e^{-i\bm k\cdot\bm R}
|\Phi_{s\bm k}\rangle U_{s\alpha}(\bm k),
\label{eq:sm_wannier_transform}
\end{equation}
where $N_k$ is the number of momentum points and $\bm R$
is a moir\'e lattice vector.
The expectation values in $\Omega$ refer to the home-cell
states $|W_{\alpha\bm 0}\rangle$.
The matrices $U(\bm k)$ are saved in \texttt{\_u.mat},
while \texttt{.wout} records the Wannier centers, spreads,
and minimization history.

The Wannier rotation and Fourier transformation give the clean
TB Hamiltonian $H_0$ in Eq.~\eqref{eq:orbital_disorder} of the main text:
\begin{equation}
\begin{aligned}
t_{\alpha\beta}(R'-R)
=\langle W_{\alpha R}|&\hat h_K|W_{\beta R'}\rangle=\frac{1}{N_k}\sum_{\bm k}e^{-i\bm k\cdot(R'-R)}
\big[U^\dagger(\bm k)\varepsilon(\bm k)U(\bm k)\big]_{\alpha\beta},\\
H_0&=\sum_{R,R'}c_R^\dagger t(R'-R)c_{R'}.
\end{aligned}
\label{eq:sm_wannier_hamiltonian}
\end{equation}
Here $R$ and $R'$ label moir\'e unit cells, and $t(R'-R)$ is the
$3\times3$ hopping matrix in the $(\mathrm{MX},\mathrm{XM},\mathrm{MM})$
orbital basis, with
$c_R=(c_{R,\mathrm{MX}},c_{R,\mathrm{XM}},c_{R,\mathrm{MM}})^T$
as in the main text.
The band energies form the diagonal matrix
$\varepsilon(\bm k)=\operatorname{diag}
[\varepsilon_1(\bm k),\varepsilon_2(\bm k),\varepsilon_3(\bm k)]$.
The matrix elements $t_{\alpha\beta}(R'-R)$ are stored in
\texttt{\_hr.dat}.
For the finite momentum mesh, the standard Wannier90
Wigner--Seitz weights and equivalent-cell images are retained
in the real-space interpolation.
The Wannier TB bands (black dashed lines) accurately reproduce
the continuum bands (red solid lines) at all six twist angles
[Fig.~\ref{fig:SM1_Continuum-TB-model}].
For real-space transport calculations, we truncate hoppings
beyond the sixth neighbor shell.
The truncated bands (blue dash-dotted lines) remain in good
agreement with the continuum dispersions and preserve the
Chern number of each band.

To visualize each Wannier orbital in real space, we Fourier transform the
rotated plane-wave coefficients $C(\bm k)U(\bm k)$ to obtain the layer-resolved Wannier function distribution.
For the home cell $R=0$, the Wannier function
density in layer $\ell=t,b$ are
\begin{equation}
\begin{aligned}
W_{\alpha,0}^{(\ell)}(\bm r)
=\frac{1}{N_k\sqrt{A_M}}&
\sum_{\bm k}\sum_{i:\ell_i=\ell}
[C(\bm k)U(\bm k)]_{i\alpha}
e^{i(\bm k-\bm Q_i)\cdot\bm r},\\
\rho_\alpha^{(\ell)}(\bm r)
&=\left|W_{\alpha,0}^{(\ell)}(\bm r)\right|^2.
\end{aligned}
\label{eq:sm_layer_wannier_density}
\end{equation}
Here $A_M$ is the moir\'e-cell area and $N_k$ is the total
number of momentum points.
Integrating $\rho_\alpha^{(\ell)}$ gives the weight in layer
$\ell$, with the two layer weights summing to unity.

Figures~\ref{fig:SM1_wannier-density-2.45deg} and
\ref{fig:SM1_wannier-density-3.89deg} show these Wannier function densities at
$\theta=2.45^\circ$ and $3.89^\circ$, respectively.
Although the individual band Chern numbers differ between
the two angles (see Fig.\ref{fig:SM1_Continuum-TB-model}), the Wannier centers remain at MX, XM, and MM.
MX and XM exhibit strong, opposite layer polarizations, with approximately $93$--$95\%$ of their weight in the bottom and top layers, respectively, whereas MM has nearly balanced layer components.
Moreover, the densities exhibit the expected $C_3$ symmetry about
each center and the layer-exchange relations of
$C_{2y}\mathcal T$.
All three orbitals are well localized, with spatial widths
$\sqrt{\langle r^2\rangle_\alpha-
|\langle\bm r\rangle_\alpha|^2}
\simeq(0.34\text{--}0.48)\cdot a_M$.

\begin{figure*}[htbp]
    \centering
    \includegraphics[width=\linewidth]{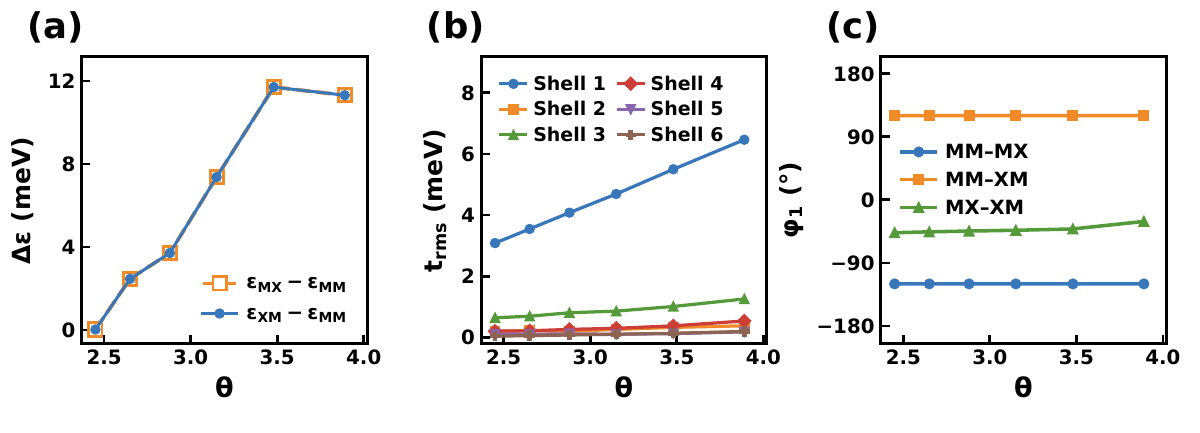}
    \caption{ Twist-angle dependence of the Wannier TB parameters.
(a) Onsite energy offsets $\epsilon_{\mathrm{MX}}-\epsilon_{\mathrm{MM}}$
and $\epsilon_{\mathrm{XM}}-\epsilon_{\mathrm{MM}}$.
(b) Root-mean-square hopping amplitudes (see Eq.~\eqref{eq:sm_shell_hopping}) for the first six neighbor
shells, ordered by the distance between Wannier centers.
(c) Phases of  intercell NN hoppings between
the indicated orbitals.
Results are shown for six twist angles from $2.45^\circ$ to $3.89^\circ$.}
    \label{fig:SM1_tb_parameters_vs_theta}
\end{figure*}

In this localized Wannier basis, we examine how the TB Hamiltonian parameters
vary with twist angle.
We focus on three aspects: onsite energies, hopping amplitudes and hopping phases, as shown in Fig.~\ref{fig:SM1_tb_parameters_vs_theta}.
The MX and XM onsite energies, defined as
$\epsilon_{MX}(\theta)$ and $\epsilon_{XM}(\theta)$,
coincide within numerical precision, as required by the
$C_{2y}\mathcal T$ symmetry that exchanges these two orbitals~\cite{Jia2024}.
Their common offset from MM rises from nearly zero at $2.45^\circ$
to $11.7\,\mathrm{meV}$ at $3.48^\circ$, followed by a slight decrease
at $3.89^\circ$ [Fig.~\ref{fig:SM1_tb_parameters_vs_theta}(a)].
For the hopping terms, we group them into shells according to the distance between
the connected Wannier centers: $j=1$ denotes nearest neighbors,
$j=2$ next-nearest neighbors, and so on.
The root-mean-square hopping amplitude in shell $j$ is
\begin{equation}
t_{\mathrm{rms}}^{(j)}(\theta)
=\left[
\frac{1}{N_j}
\sum_{(\alpha,\beta,R)\in\mathcal S_j}
|t_{\alpha\beta}(R;\theta)|^2
\right]^{1/2},
\label{eq:sm_shell_hopping}
\end{equation}
where $\mathcal S_j$ contains the $N_j$ bonds in that shell,
with each Hermitian pair counted once.
Figure~\ref{fig:SM1_tb_parameters_vs_theta}(b) shows that nearest-neighbor (shell 1)
hopping terms remain dominant throughout the studied range, increasing
from approximately $3.1$ to $6.5\,\mathrm{meV}$, while the more distant
shells remain much weaker, consistent with the well-localized character of the MLWFs.
Since first-shell hopping dominates, we focus on the corresponding
nearest-neighbor hopping phases.
To compare them across angles, we fix the Wannier gauge by making the
two intracell reference hoppings $t_{\mathrm{MX,MM}}(0;\theta)$ and
$t_{\mathrm{XM,MM}}(0;\theta)$ real and positive.
The phases in Fig.~\ref{fig:SM1_tb_parameters_vs_theta}(c) instead
refer to selected intercell nearest-neighbor bonds,
$\varphi_1^{\alpha\beta}(\theta)
=\arg t_{\alpha\beta}(R_{\alpha\beta}^{(1)};\theta)$,
with $R_{\alpha\beta}^{(1)}\ne0$; these hoppings can therefore
remain complex in the same gauge.
Over $\theta=2.45^\circ$--$3.89^\circ$, the phase pattern changes
little: the MM--MX and MM--XM phases stay nearly constant at
$-120^\circ$ and $+120^\circ$, respectively, while the MX--XM phase
shows a modest shift from approximately $-47^\circ$ to $-31^\circ$.
The systematic angle dependence of the onsite energy difference
between MM and MX/XM and of the nearest-neighbor hopping amplitudes
are used to construct the twist-angle disorder Hamiltonian below.

\clearpage
\SMsection{Disorder Hamiltonian construction}
\label{sm:disorder}
In this section, we introduce the
construction of the disorder Hamiltonians for three types of disorder discussed in the main text: potential disorder $V_R$,
displacement-field disorder $D_R$, and twist-angle disorder $\theta_R$. 
We first describe the construction of spatially uncorrelated disorder fields
and then introduce how to generate spatially correlated disorder from them.

\SMsubsection{Spatially uncorrelated disorder}
\label{sm:uncorrelated-disorder}

We consider the three disorder channels separately and first construct
spatially uncorrelated disorder by assigning an independent random value
to each moir\'e unit cell $R$.
For each disorder realization, we first generate $V_R$, $D_R$, or $\delta\theta_R$
and then construct the Hamiltonian as described below, reflecting how
local inhomogeneities modify the electronic structure of real moir\'e systems.

To compare their orbital effects, we first retain the onsite terms,
as in Eq.~\eqref{eq:orbital_disorder} of the main text:
\begin{equation}
H=H_0+\sum_R w_a(R)c_R^\dagger M_a c_R,
\qquad a=V,D,\theta,
\label{eq:sm_onsite_disorder}
\end{equation}
where $c_R=(c_{R,\mathrm{MX}},c_{R,\mathrm{XM}},c_{R,\mathrm{MM}})^T$.
In the common Gaussian convention, the spatial uncorrelated onsite disorder term $w_a(R)$ are
drawn independently for different cells, with the follow distribution:
\begin{equation}
\begin{aligned}
P_a(w)&=\frac{1}{\sqrt{2\pi}\gamma_a}
\exp\!\left(-\frac{w^2}{2\gamma_a^2}\right),\\
\overline{w_a(R)}&=0,\qquad
\overline{w_a(R)w_a(R')}=\gamma_a^2\delta_{RR'}.
\end{aligned}
\label{eq:sm_uncorrelated_field}
\end{equation}
All three orbitals in a cell share the same random variable, with
their orbital components set by $M_a$; independence refers to different
cells, not to different orbitals within a cell.
We impose $\operatorname{Tr}(M_a^2)/3=1$ so that $\gamma_a$ provides
a common measure of disorder strength for all three channels.
This allows us to compare the effects of orbital selectivity
at equal $\gamma_a$.

A local potential disorder $V_R$ shifts all three onsite energies equally, giving
$w_V(R)=V_R$ and $M_V=I_3$.
A perpendicular displacement field $D_R$ instead produces opposite
potential shifts in the two layers.
Since MX and XM are predominantly localized in opposite layers (see Fig.\ref{fig:SM1_wannier-density-2.45deg} and Fig.\ref{fig:SM1_wannier-density-3.89deg}),
their onsite shifts have opposite signs and nearly equal magnitudes,
whereas the shift of the layer-balanced MM orbital vanishes.
$D_R$ produces equal and opposite onsite shifts of MX and XM,
while the small shift of MM is neglected.
We therefore use $M_D=\sqrt{3/2}\operatorname{diag}(1,-1,0)$,
with $w_D(R)$ setting the energy scale.
The factor $\sqrt{3/2}$ ensures $\operatorname{Tr}(M_D^2)/3=1$,
so that $\gamma_D$ and $\gamma_V$ measure disorder strength
on the same scale.

For twist-angle disorder $\theta_R=\theta+\delta \theta_R$, we use the angle dependence of the Wannier
parameters in Fig.~\ref{fig:SM1_tb_parameters_vs_theta}.
At each clean angle, $C_{2y}\mathcal T$ symmetry relates MX and XM
and enforces equal onsite energies~\cite{Jia2024}.
Their energy relative to MM changes with angle
[Fig.~\ref{fig:SM1_tb_parameters_vs_theta}(a)].
We describe this change using
$\Delta_\epsilon(\theta)=\epsilon_{\mathrm{MM}}(\theta)
-[\epsilon_{\mathrm{MX}}(\theta)+\epsilon_{\mathrm{XM}}(\theta)]/2$,
the negative of the offset plotted in that panel.
For a local angle $\theta_R=\theta+\delta\theta_R$, let
$\delta\Delta_\epsilon(R)=\Delta_\epsilon(\theta_R)
-\Delta_\epsilon(\theta)$.
To isolate the layer-dependent onsite contribution of the twist angle disorder, we omit a shift
common to all three orbitals, which would enter the scalar-potential
channel.
The remaining onsite correction is
\begin{equation}
\begin{aligned}
\delta h_\theta^{\mathrm{on}}(R)
&=\frac{\delta\Delta_\epsilon(R)}{3}
\operatorname{diag}(-1,-1,2)
=w_\theta(R)M_\theta,\\
M_\theta&=\frac{1}{\sqrt{2}}\operatorname{diag}(-1,-1,2),\qquad
w_\theta(R)=\frac{\sqrt{2}}{3}\delta\Delta_\epsilon(R).
\end{aligned}
\label{eq:sm_twist_onsite}
\end{equation}
Thus, the signs in $M_\theta$ describe relative onsite shifts after
separating the common shift.
We obtain $\Delta_\epsilon(\theta)$ by piecewise-linear interpolation
of the six clean Wannier models, continuing the endpoint slopes
outside the tabulated angle range.

The relation to a twist angle disorder strength follows by drawing
$\delta\theta_R$ independently from $\mathcal N(0,\sigma_\theta^2)$.
Linearizing the onsite offset gives
\begin{equation}
\begin{aligned}
w_\theta(R)&\simeq
\frac{\sqrt{2}}{3}\Delta_\epsilon'(\theta)\delta\theta_R,\\
\gamma_\theta&\simeq
\frac{\sqrt{2}}{3}|\Delta_\epsilon'(\theta)|\sigma_\theta.
\end{aligned}
\label{eq:sm_twist_linear_strength}
\end{equation}
This connects Gaussian angle fluctuations to the common Gaussian
onsite disorder model.
For the full interpolated angle dependence, the disorder strength is
instead evaluated directly from the onsite changes:
\begin{equation}
\gamma_{\theta,\mathrm{eff}}
=\left[\frac{1}{3}
\overline{\operatorname{Tr}
\bigl[(\delta h_\theta^{\mathrm{on}}(R))^2\bigr]}\right]^{1/2}
=\frac{\sqrt{2}}{3}
\left[\overline{\delta\Delta_\epsilon(R)^2}\right]^{1/2}.
\label{eq:sm_twist_effective_strength}
\end{equation}
The angle-width label used in the transport figures is
$W_\theta=\sqrt{3}\sigma_\theta$.
At $\theta=2.45^\circ$, $W_\theta=0.2^\circ$, $0.6^\circ$, and
$1^\circ$ correspond to $\gamma_{\theta,\mathrm{eff}}\simeq0.65$,
$1.83$, and $3.03\,\mathrm{meV}$, respectively.
This mapping relates the twist-angle width $W_\theta$ to the onsite
disorder strength $\gamma_\theta$, enabling comparison with the other two channels.

Twist-angle fluctuations also modify the hopping amplitudes.
We restrict hopping disorder to the dominant nearest-neighbor (NN)
amplitudes, since longer-range hoppings are much weaker and the phases
vary little with angle [Fig.~\ref{fig:SM1_tb_parameters_vs_theta}(b,c)].
Let $A_{\alpha\beta}(\theta)$ denote the mean NN hopping magnitude
for an orbital-pair type: MM--MX and MM--XM share one response function since $A_{MM,XM}(\theta)=A_{MM,MX}(\theta)$ (protected by $C_{2y}\mathcal{T}$),
and MX--XM has a second, which reads  $A_{MX,XM}(\theta)$.
As for $\Delta_\epsilon$, we interpolate linearly between the tabulated
angles and extrapolate linearly beyond them.
For each NN bond, we use the average twist angle of its two endpoint
cells to determine the hopping change:
\begin{equation}
\begin{aligned}
\bar\theta_{RR'}&=\frac{\theta_R+\theta_{R'}}{2},\\
\delta t_{\alpha\beta}(R,R')
&=t_{\alpha\beta}(R'-R;\theta)
\left[\frac{A_{\alpha\beta}(\bar\theta_{RR'})}
{A_{\alpha\beta}(\theta)}-1\right].
\end{aligned}
\label{eq:sm_twist_hopping}
\end{equation}
Here $t_{\alpha\beta}(R'-R;\theta)$ is the clean hopping.
Only the real amplitude ratio is varied; if it becomes negative
under extrapolation, the hopping changes sign.
The lattice geometry and longer-range hoppings remain unchanged.

Combining the two contributions (onsite and NN hopping amplitude) gives the extended twist-disorder
Hamiltonian
\begin{equation}
\begin{aligned}
H^{(\theta)}={}&H_0+
\sum_R w_\theta(R)c_R^\dagger M_\theta c_R\\
&+\sum_{\langle R\alpha,R'\beta\rangle_1}
\left[\delta t_{\alpha\beta}(R,R')
c_{R,\alpha}^\dagger c_{R',\beta}+\mathrm{H.c.}\right],
\end{aligned}
\label{eq:sm_full_twist_disorder}
\end{equation}
where each NN bond is counted once.
Both contributions are generated from the same cell-angle field,
so no independent hopping noise is added.
The model in the main text (see Eq.~\eqref{eq:orbital_disorder}) isolates the layer-dependent onsite contribution of twist-angle disorder and adopts a unified strength convention for all three disorder types.
For twist anglke disorder, Sec.~\ref{sm:transport} compares transport with only the onsite correction, only
the NN correction, and both contributions retained.

\begin{figure*}[htbp]
    \centering
    \includegraphics[width=\linewidth]{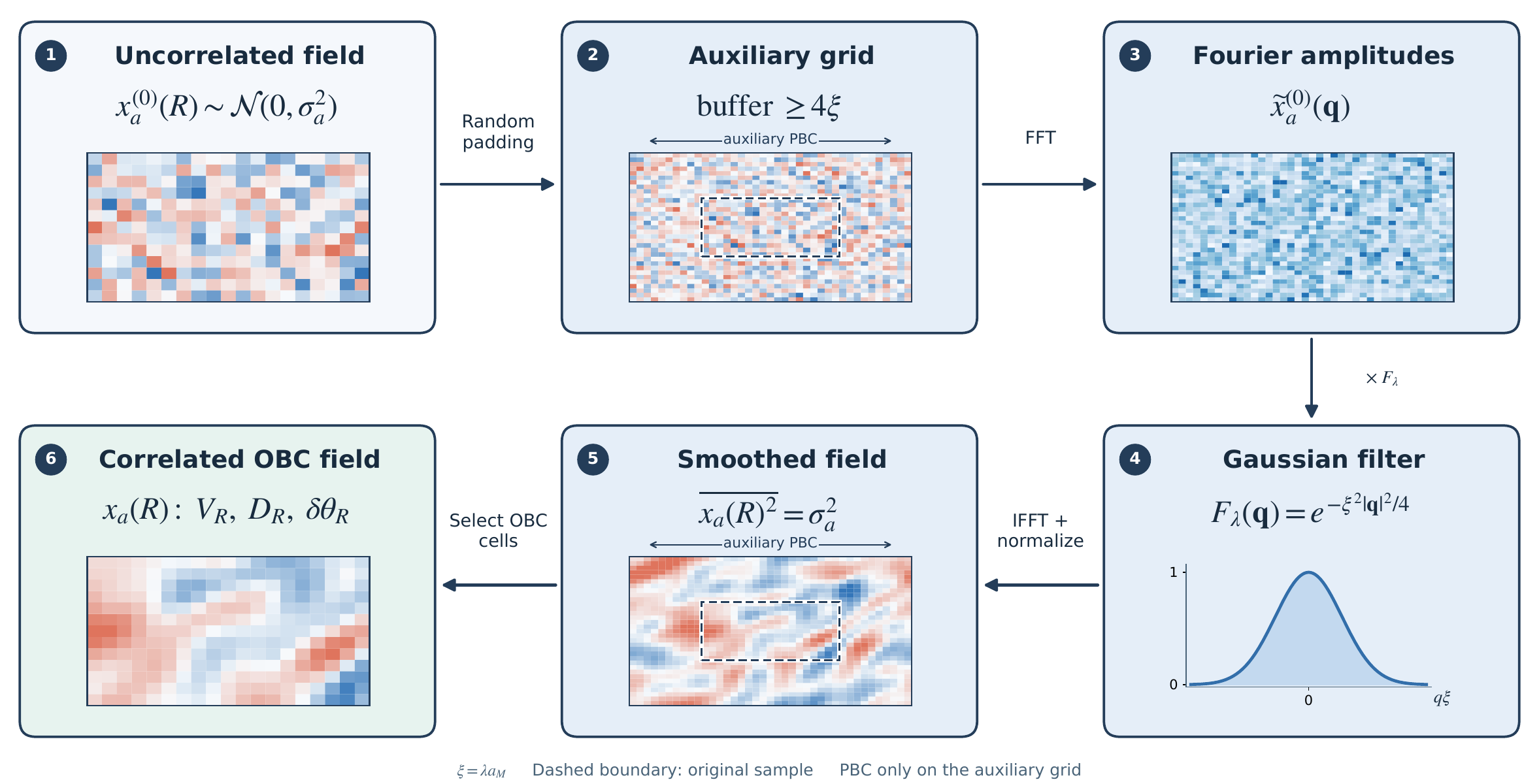}
    \caption{Workflow of the construction of spatially correlated disorder.
Independent Gaussian cell values are extended with random padding
on an auxiliary periodic grid.
The field is Fourier transformed, multiplied by
$F_\lambda(\bm q)=\exp[-\xi^2|\bm q|^2/4]$, and inverse transformed,
with $\xi=\lambda a_M$.
Normalization preserves the original ensemble variance $\sigma_a^2$.
Selecting the original OBC cells gives the correlated field $x_a(R)$,
representing $V_R$, $D_R$, or $\delta\theta_R$.
Dashed outlines mark the original sample. }
    \label{fig:SM2_correlated-disorder-workflow}
\end{figure*}

\SMsubsection{Spatially correlated disorder}

To construct spatially correlated disorder, we now introduce spatial correlations into the independent disorder
fields generated above.
We generate
spatially correlated disorder with correlation length $\lambda a_M$
($\lambda>0$), while keeping the Gaussian distribution at each cell
unchanged as Eq.~\eqref{eq:sm_uncorrelated_field}.
We use the notation $x_V(R)=V_R$, $x_D(R)=D_R$, and
$x_\theta(R)=\delta\theta_R$, with standard deviations $\sigma_a$ ($a=V,R,\theta$), to describe disorder field in the real space. 
The corresponding onsite disorder strengths are
$\gamma_V=\sigma_V$, $\gamma_D=\sigma_D$, and
 $\gamma_\theta\simeq\frac{\sqrt{2}}{3}
|\Delta_\epsilon'(\theta)|\sigma_\theta$ in the linearized limit, as discussed in Sec.~\ref{sm:uncorrelated-disorder}.

The workflow of the construction is summarized in Fig.~\ref{fig:SM2_correlated-disorder-workflow}, with each step described below:

\begin{enumerate}
\item \textit{Start from the uncorrelated field.}
We use the independent Gaussian field generated in the previous
subsection, denoted here by $x_a^{(0)}(R)$.
The superscript $(0)$ labels the input field, which has zero mean
and variance $\sigma_a^2$.

\item \textit{Add random padding.}
We surround the sample with a buffer at least $4\xi$ ($\xi=\lambda a_M$) thick, filled
with additional independent values from the same distribution.
The enlarged grid is treated as periodic for the fast Fourier transformation (FFT).
The buffer keeps the original sample far from these boundaries,
reducing artificial correlations caused by periodic repetition.
The sample itself retains open boundaries.

\item \textit{Transform to momentum space.}
The FFT gives the Fourier components
\begin{equation}
\widetilde{x}_a^{(0)}(\bm q)
=\sum_{R\in\mathrm{aux}}x_a^{(0)}(R)e^{-i\bm q\cdot R},
\label{eq:sm_disorder_fft}
\end{equation}
where the sum runs over the enlarged grid.
The wave vectors $\bm q$ are obtained from the moir\'e reciprocal vectors.

\item \textit{Smooth the field.}
We multiply each Fourier component by
\begin{equation}
F_\lambda(\bm q)=e^{-\xi^2|\bm q|^2/4}.
\label{eq:sm_disorder_filter}
\end{equation}
This suppresses short-wavelength variations and averages nearby cell
values in real space.
The filter smoothly suppresses large-$|\bm q|$ components,
smoothing the disorder field in real space.
Its squared amplitude, $|F_\lambda|^2=e^{-\xi^2|\bm q|^2/2}$,
gives a Gaussian spatial correlation of width $\xi=\lambda a_M$
in the smooth-field limit.

\item \textit{Transform back and preserve the variance.}
The inverse FFT gives
\begin{equation}
\begin{aligned}
x_a(R)
&=\frac{1}{N_p\mathcal N_\lambda}
\sum_{\bm q}F_\lambda(\bm q)\widetilde{x}_a^{(0)}(\bm q)
e^{i\bm q\cdot R},\\
\mathcal N_\lambda^2
&=\frac{1}{N_p}\sum_{\bm q}|F_\lambda(\bm q)|^2,
\end{aligned}
\label{eq:sm_disorder_ifft}
\end{equation}
where $N_p$ is the number of cells in the enlarged grid.
Dividing by $\mathcal N_\lambda$ restores the original variance,
$\overline{x_a(R)^2}=\sigma_a^2$.
We keep the $\bm q=0$ component and apply no further mean subtraction
or rescaling to individual samples.

\item \textit{Select the sample and construct its Hamiltonian.}
We keep only the values on the original OBC cells and discard the buffer.
The resulting $V_R$, $D_R$, or $\delta\theta_R$ is then used to construct
the sample Hamiltonian through the mappings in the previous subsection.
\end{enumerate}

The filtered fields remain Gaussian with zero ensemble mean and
variance $\sigma_a^2$.
Their spatial correlation on the finite grid is
\begin{equation}
\begin{aligned}
\overline{x_a(R)x_a(R')}
&=\sigma_a^2
\frac{\sum_{\bm q}|F_\lambda(\bm q)|^2
e^{i\bm q\cdot(R-R')}}
{\sum_{\bm q}|F_\lambda(\bm q)|^2}\\
&\simeq\sigma_a^2
\exp\!\left[-\frac{|R-R'|^2}{2\xi^2}\right].
\end{aligned}
\label{eq:sm_disorder_covariance}
\end{equation}
The last expression applies when the field varies smoothly over cells
and periodic copies have negligible influence.
For $V_R$, $D_R$, and the linearized twist-angle model, the mappings
to $w_a(R)$ preserve this correlation, with $\sigma_a^2$ replaced
by $\gamma_a^2$ as in the main text.
For twist-angle disorder, we first generate the correlated
$\delta\theta_R$ and then map $\theta_R=\theta+\delta\theta_R$ to the
onsite energies and NN hoppings.

\begin{figure*}[htbp]
    \centering
    \includegraphics[width=\linewidth]{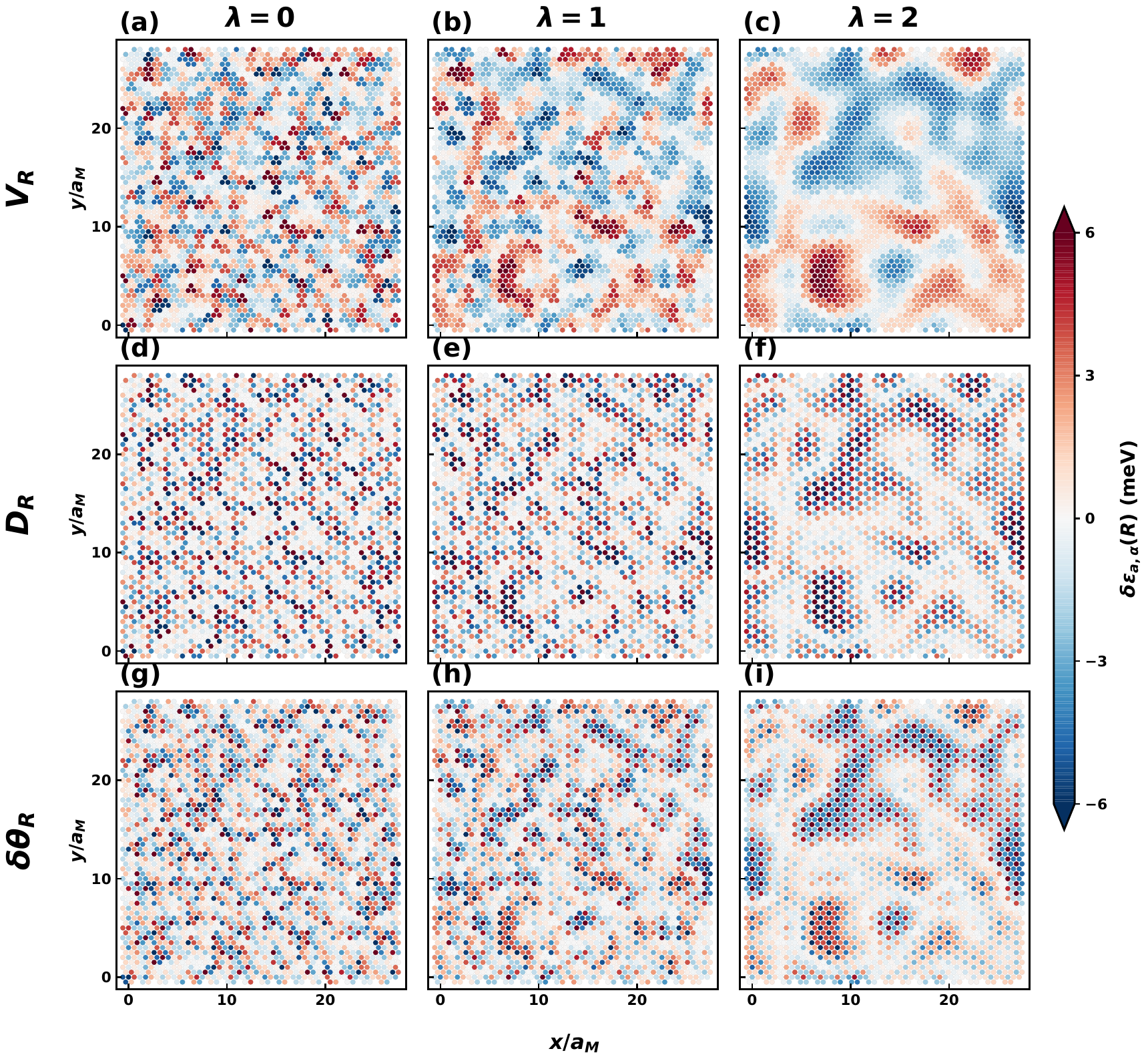}
    \caption{
Orbital-resolved onsite disorder in real space.
Rows show $V_R$ (a--c), $D_R$ (d--f), and $\delta\theta_R$ (g--i);
columns correspond to $\lambda=0$, $1$, and $2$.
Each dot represents a Wannier orbital, colored by its onsite energy
shift $\delta\epsilon_{a,\alpha}(R)=w_a(R)(M_a)_{\alpha\alpha}$.
All panels use the same underlying Gaussian realization and OBC sample
with $(N_x,N_y)=(16,28)$, at the same disorder strength
$\gamma_a=3\,\mathrm{meV}$.
The twist-angle panels use the full onsite-parameter mapping at
$\theta=2.45^\circ$, with $W_\theta\simeq1.0^\circ$.
The common color scale saturates beyond $\pm6\,\mathrm{meV}$, as indicated
by the colorbar extensions.
    }
    \label{fig:SM2_disorder-realization-maps}
\end{figure*}

Using the construction above, Fig.~\ref{fig:SM2_disorder-realization-maps}
illustrates the onsite energy shifts generated from one random Gaussian
disorder realization at $\gamma_a=3\,\mathrm{meV}$.
Increasing $\lambda$ produces broader, smoother spatial patterns on each
orbital sublattice.
Within each cell, $V_R$ shifts all three orbitals equally, whereas $D_R$
gives opposite MX/XM shifts and leaves MM unchanged.
For $\delta\theta_R$, the construction of angle-dependent onsite terms gives equal
MX/XM shifts and an opposite shift on MM.

\clearpage
\SMsection{Transport calculation results}
\label{sm:transport}

In this section, we present additional $\langle R_{xy}\rangle$ and
$\langle G_{2t}\rangle$ results over a wider parameter range, calculated using the six-terminal geometry and the two-terminal geometry, respectively, to examine general trends in
plateau behavior.
The results are organized into three subsections for potential disorder $V_R$, displacement-field
disorder $D_R$ and twist-angle disorder
$\theta_R$.
For $\theta_R$, we compare onsite disorder alone, nearest-neighbor (NN)
hopping disorder alone, and both contributions together to distinguish
their effects on transport.

\SMsubsection{Potential disorder $V_R$}

\begin{figure*}[htbp]
    \centering
    \includegraphics[width=\linewidth]{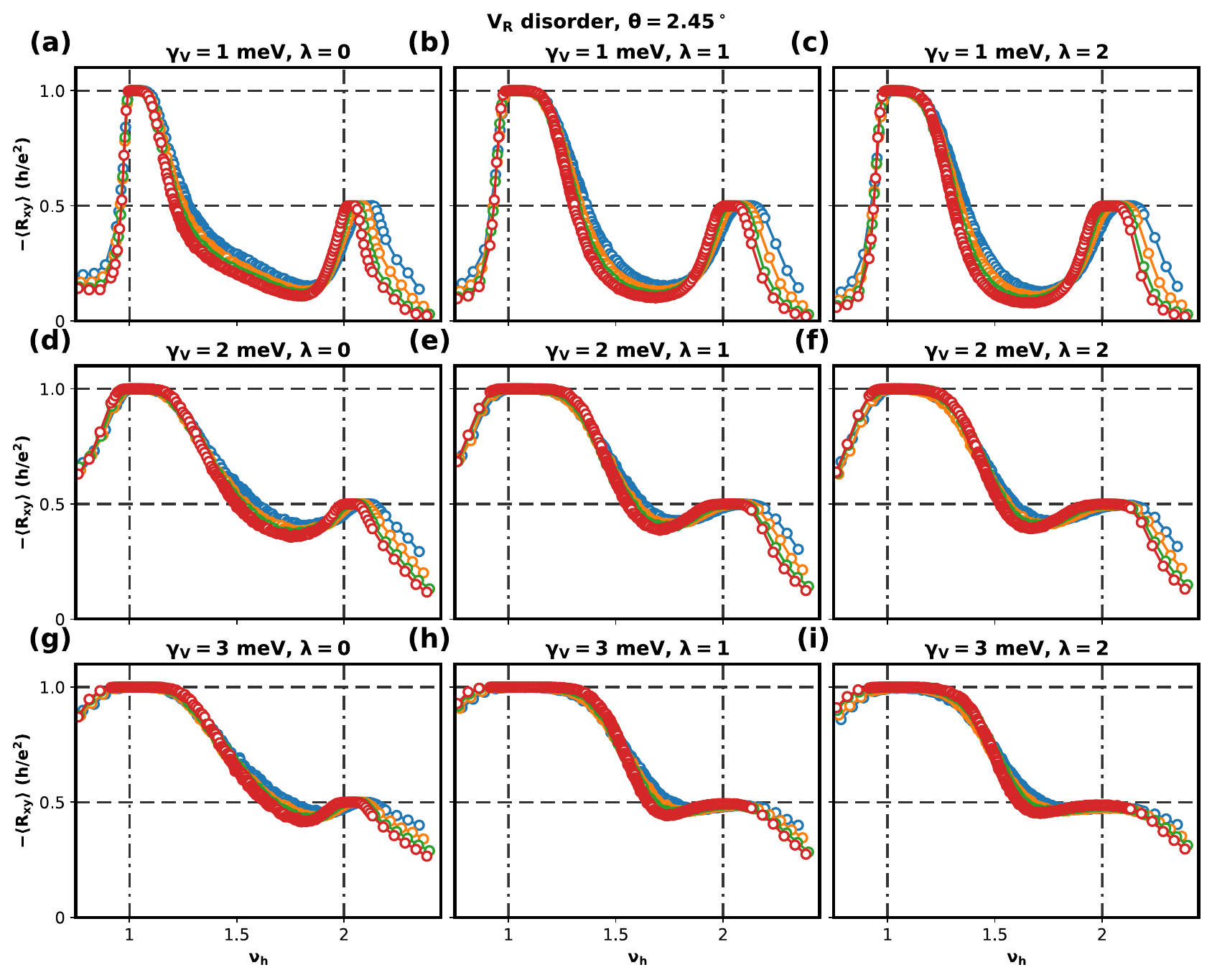}
    \caption{
Disorder-averaged Hall resistance $-\langle R_{xy}\rangle$ versus hole filling $\nu_h$
for potential disorder $V_R$ at $\theta=2.45^\circ$,
using the six-terminal geometry with open boundaries [Fig.~\ref{fig:Figure2}(a)].
Panels (a)--(c), (d)--(f), and (g)--(i) have
$\gamma_a=1$, $2$, and $3\,\mathrm{meV}$, respectively.
Columns from left to right correspond to $\lambda=0,1,2$.
Blue, orange, green, and red curves correspond to
$(N_x,N_y)=(16,28)$, $(24,40)$, $(36,60)$, and $(48,80)$, respectively.
Horizontal dashed lines mark $h/e^2$ and $h/(2e^2)$.
Vertical dash-dotted lines mark $\nu_h=1,2$.
    }
    \label{fig:SM3_V_R-disorder_six-terminal_Rxy_2.45deg}
\end{figure*}

\begin{figure*}[htbp]
    \centering
    \includegraphics[width=\linewidth]{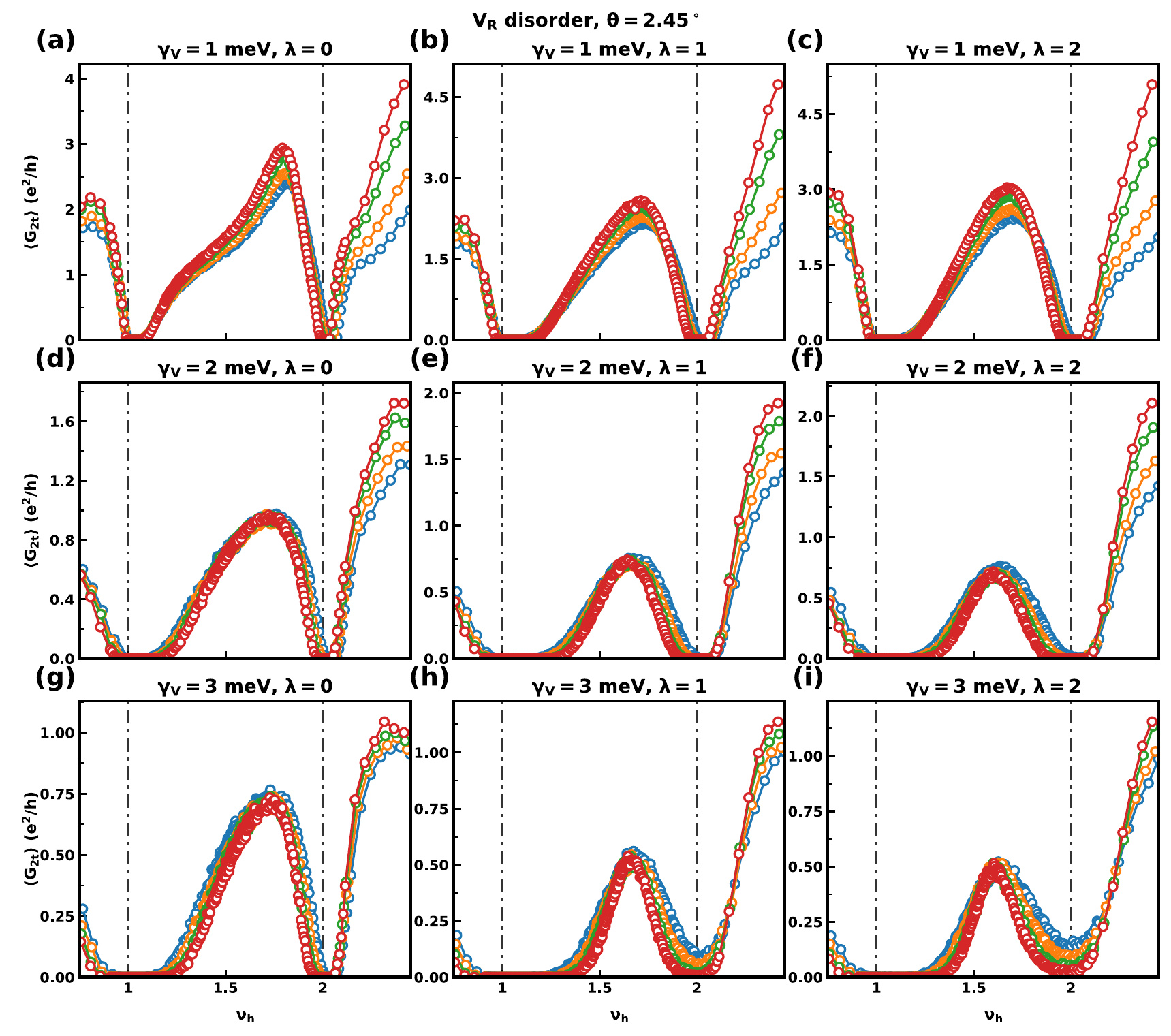}
    \caption{
Disorder-averaged two-terminal conductance $\langle G_{2t}\rangle$ versus hole filling $\nu_h$
for potential disorder $V_R$ at $\theta=2.45^\circ$.
The geometry has periodic boundaries along $y$,
and the conductance is corrected for the Sharvin contact resistance.
Panels (a)--(c), (d)--(f), and (g)--(i) have
$\gamma_a=1$, $2$, and $3\,\mathrm{meV}$, respectively.
Columns from left to right correspond to $\lambda=0,1,2$.
Blue, orange, green, and red curves correspond to
$(N_x,N_y)=(16,28)$, $(24,40)$, $(36,60)$, and $(48,80)$, respectively.
Vertical dash-dotted lines mark $\nu_h=1,2$.
    }
    \label{fig:SM3_V_R-disorder_two-terminal_G2t_2.45deg}
\end{figure*}

Figures~\ref{fig:SM3_V_R-disorder_six-terminal_Rxy_2.45deg} and
\ref{fig:SM3_V_R-disorder_two-terminal_G2t_2.45deg} show the Hall
resistance $-\langle R_{xy}\rangle$ and bulk conductance
$\langle G_{2t}\rangle$ under $V_R$ disorder at $\theta=2.45^\circ$.
We consider $\gamma_V=1,2,3\,\mathrm{meV}$ and $\lambda=0,1,2$,
with four sample sizes $N_x=16,24,36,48$.
Increasing $\gamma_V$ markedly broadens the plateaus at
$-\langle R_{xy}\rangle=h/e^2$ and $h/(2e^2)$, reflecting the strong
sensitivity to potential disorder.
For example, at $\lambda=2$ in the largest sample, the first plateau
width increases from approximately $\Delta\nu_h=0.2$ to $0.5$ as
$\gamma_V$ increases from $1$ to $3\,\mathrm{meV}$.
The intervening Hall dip weakens, and the transition between plateaus
becomes narrower and nearly direct.
Correspondingly, $\langle G_{2t}\rangle$ tends to zero in the plateau
regions, while its maximum between them falls from approximately
$3.0$ to $0.5\,e^2/h$.
At $\gamma_V=3\,\mathrm{meV}$, this peak is nearly size independent,
consistent with critical bulk transport at a Hall plateau transition.

\begin{figure*}[htbp]
   \centering
    \includegraphics[width=\linewidth]{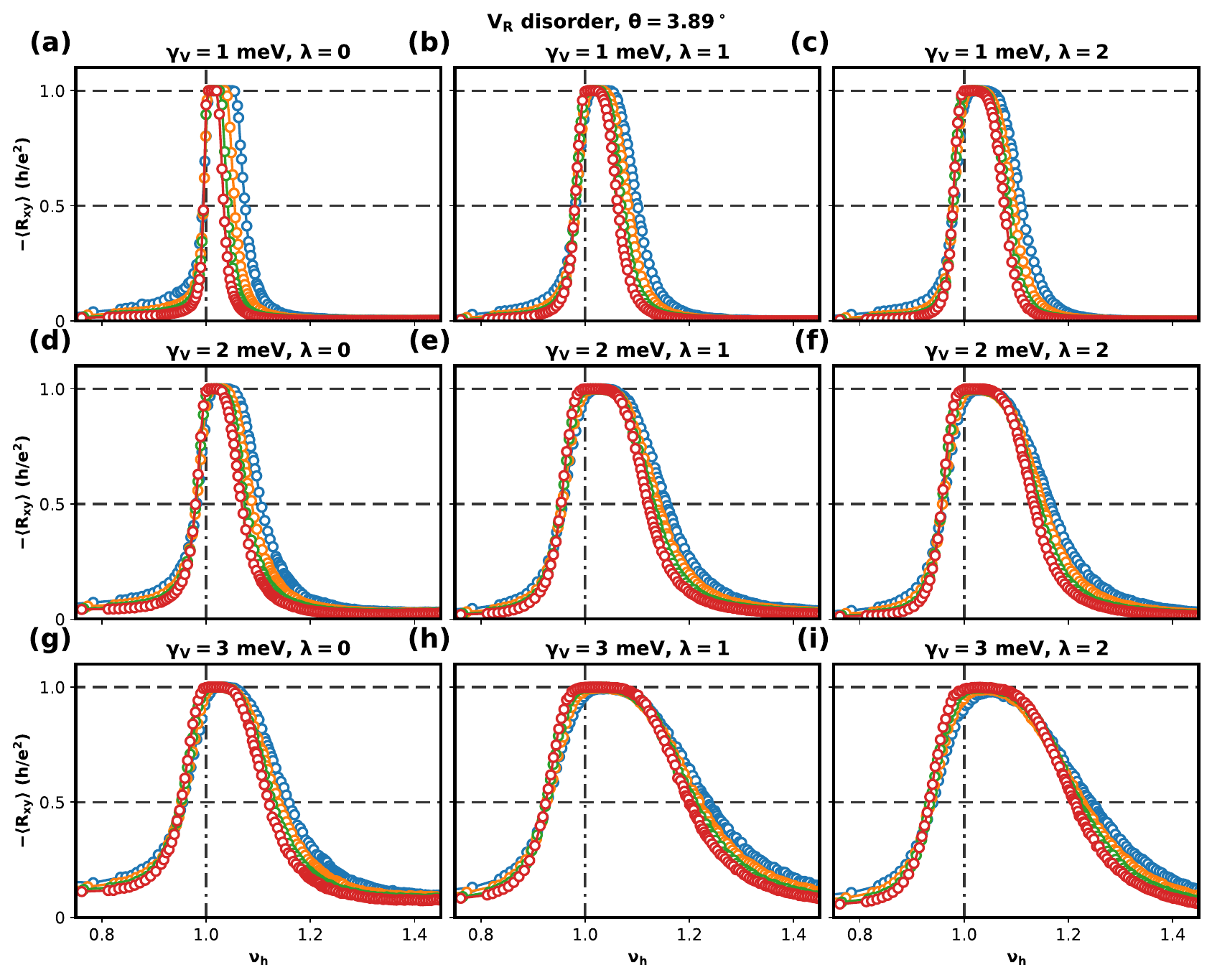}
    \caption{
Disorder-averaged Hall resistance $-\langle R_{xy}\rangle$ versus hole filling $\nu_h$
for potential disorder $V_R$ at $\theta=3.89^\circ$,
using the six-terminal geometry with open boundaries [Fig.~\ref{fig:Figure2}(a)].
Panels (a)--(c), (d)--(f), and (g)--(i) have
$\gamma_a=1$, $2$, and $3\,\mathrm{meV}$, respectively.
Columns from left to right correspond to $\lambda=0,1,2$.
Blue, orange, green, and red curves correspond to
$(N_x,N_y)=(16,28)$, $(24,40)$, $(36,60)$, and $(48,80)$, respectively.
Horizontal dashed lines mark $h/e^2$ and $h/(2e^2)$. The vertical dash-dotted line marks $\nu_h=1$.
    }
    \label{fig:SM3_V_R-disorder_six-terminal_Rxy_3.89deg}
\end{figure*}

\begin{figure*}[htbp]
    \centering
    \includegraphics[width=\linewidth]{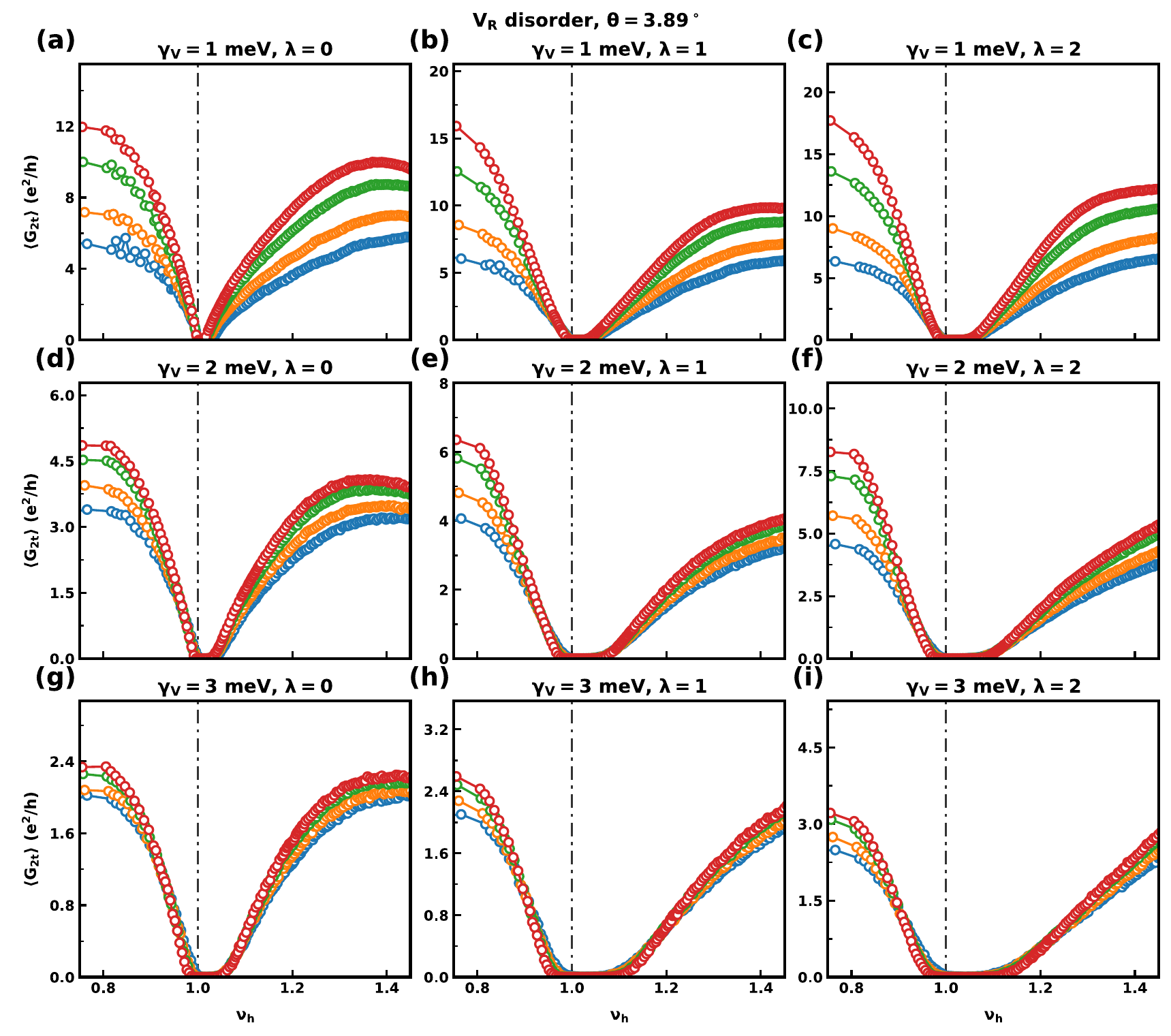}
    \caption{
Disorder-averaged two-terminal conductance $\langle G_{2t}\rangle$ versus hole filling $\nu_h$
for potential disorder $V_R$ at $\theta=3.89^\circ$.
The geometry has periodic boundaries along $y$,
and the conductance is corrected for the Sharvin contact resistance.
Panels (a)--(c), (d)--(f), and (g)--(i) have
$\gamma_a=1$, $2$, and $3\,\mathrm{meV}$, respectively.
Columns from left to right correspond to $\lambda=0,1,2$.
Blue, orange, green, and red curves correspond to
$(N_x,N_y)=(16,28)$, $(24,40)$, $(36,60)$, and $(48,80)$, respectively.
The vertical dash-dotted line marks $\nu_h=1$.
    }
    \label{fig:SM3_V_R-disorder_two-terminal_G2t_3.89deg}
\end{figure*}

Figures~\ref{fig:SM3_V_R-disorder_six-terminal_Rxy_3.89deg} and
\ref{fig:SM3_V_R-disorder_two-terminal_G2t_3.89deg} show the
corresponding results at $\theta=3.89^\circ$ for the same disorder
parameters and sample sizes.
Here, the two highest valence bands carry $C=-1$ and $+1$, with
zero total Chern number.
Consistent with this clean-band topology, the Hall response shows
a single plateau at $-\langle R_{xy}\rangle=h/e^2$ near $\nu_h=1$.
As at $2.45^\circ$, increasing $\gamma_V$ broadens the plateau and
suppresses bulk conduction.
The interval with nearly vanishing $\langle G_{2t}\rangle$ around
$\nu_h=1$ widens, while the conductance on either side decreases
substantially.
Outside this interval, the conductance still increases with sample
size, although this size dependence becomes weaker at stronger disorder.

At fixed $\gamma_V$, increasing $\lambda$ from $0$ to $2$ makes the
potential smoother and broadens the Hall plateaus at both twist angles
[Figs.~\ref{fig:SM3_V_R-disorder_six-terminal_Rxy_2.45deg}(g--i)
and \ref{fig:SM3_V_R-disorder_six-terminal_Rxy_3.89deg}(g--i)].
The corresponding intervals with $\langle G_{2t}\rangle\simeq0$
also widen, consistent with the Hall results
[Figs.~\ref{fig:SM3_V_R-disorder_two-terminal_G2t_2.45deg}(g--i)
and \ref{fig:SM3_V_R-disorder_two-terminal_G2t_3.89deg}(g--i)].

\clearpage
\SMsubsection{Displacement field disorder $D_R$}

\begin{figure*}[htbp]
    \centering
    \includegraphics[width=\linewidth]{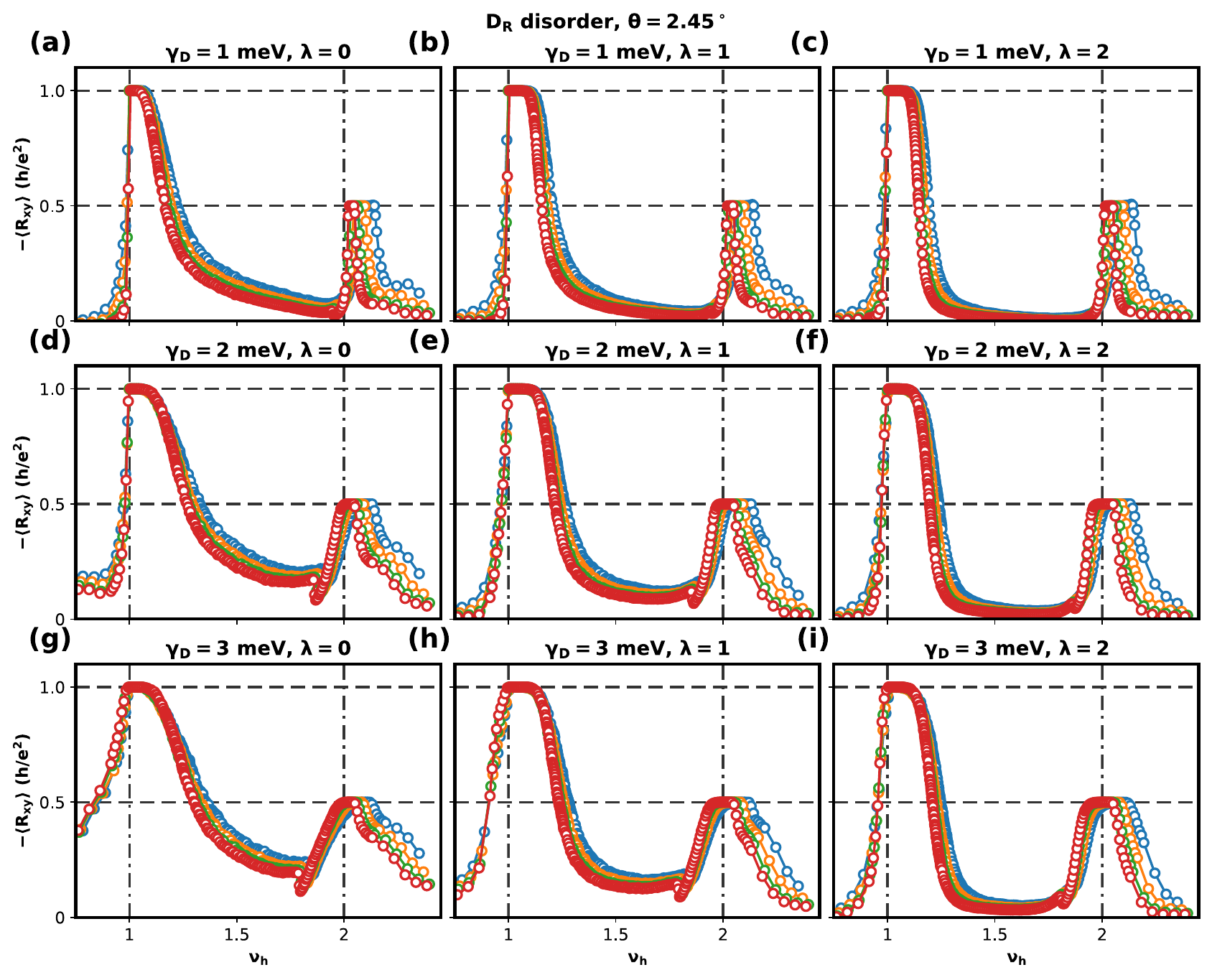}
    \caption{
Disorder-averaged Hall resistance $-\langle R_{xy}\rangle$ versus hole filling $\nu_h$
for displacement-field disorder $D_R$ at $\theta=2.45^\circ$,
using the six-terminal geometry with open boundaries [Fig.~\ref{fig:Figure2}(a)].
Panels (a)--(c), (d)--(f), and (g)--(i) have
$\gamma_a=1$, $2$, and $3\,\mathrm{meV}$, respectively.
Columns from left to right correspond to $\lambda=0,1,2$.
Blue, orange, green, and red curves correspond to
$(N_x,N_y)=(16,28)$, $(24,40)$, $(36,60)$, and $(48,80)$, respectively.
Horizontal dashed lines mark $h/e^2$ and $h/(2e^2)$.
Vertical dash-dotted lines mark $\nu_h=1,2$.
    }
    \label{fig:SM3_D_R-disorder_six-terminal_Rxy_2.45deg}
\end{figure*}

\begin{figure*}[htbp]
    \centering
    \includegraphics[width=\linewidth]{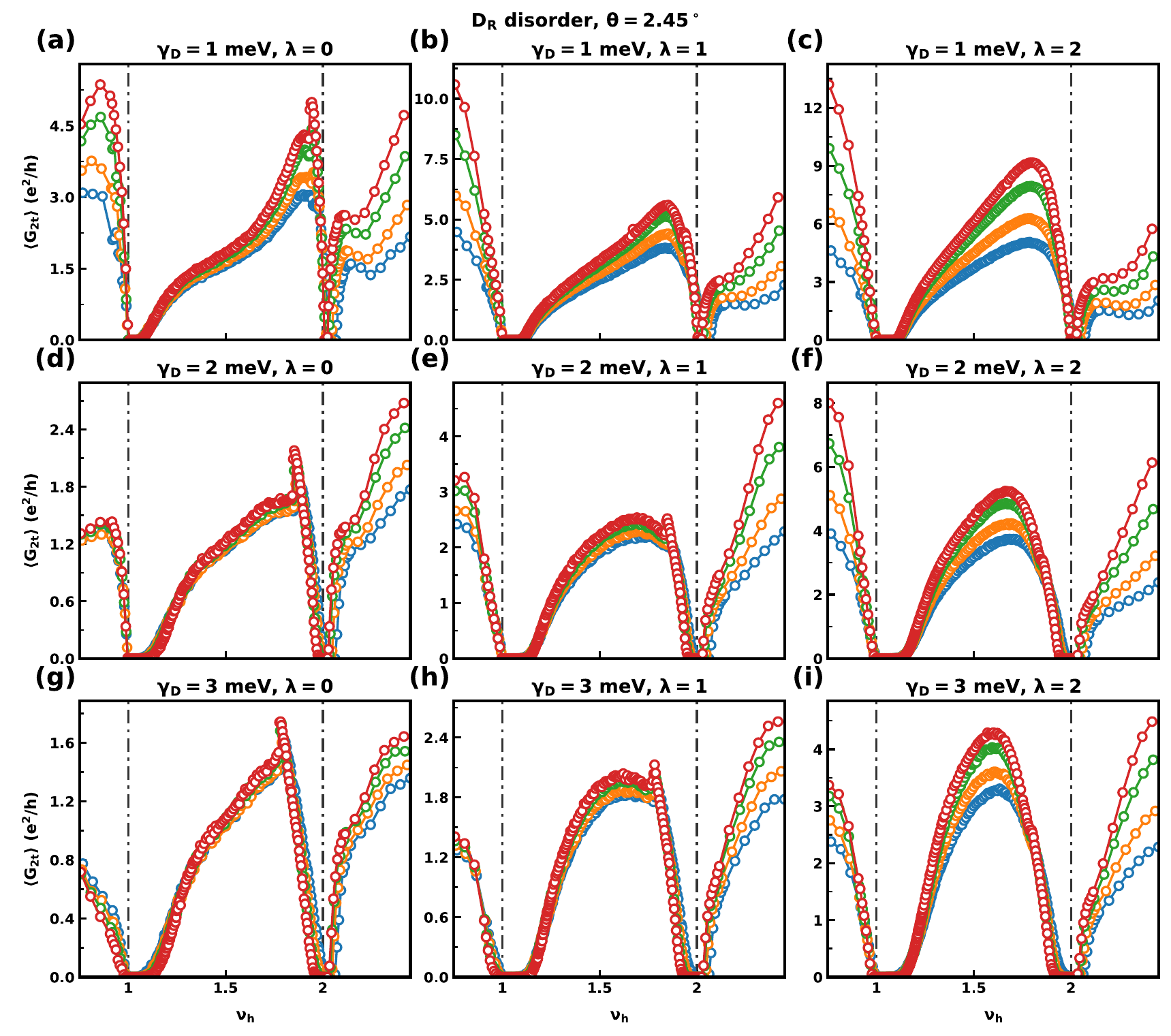}
    \caption{
Disorder-averaged two-terminal conductance $\langle G_{2t}\rangle$ versus hole filling $\nu_h$
for displacement-field disorder $D_R$ at $\theta=2.45^\circ$.
The geometry has periodic boundaries along $y$,
and the conductance is corrected for the Sharvin contact resistance.
Panels (a)--(c), (d)--(f), and (g)--(i) have
$\gamma_a=1$, $2$, and $3\,\mathrm{meV}$, respectively.
Columns from left to right correspond to $\lambda=0,1,2$.
Blue, orange, green, and red curves correspond to
$(N_x,N_y)=(16,28)$, $(24,40)$, $(36,60)$, and $(48,80)$, respectively.
Vertical dash-dotted lines mark $\nu_h=1,2$.
    }
    \label{fig:SM3_D_R-disorder_two-terminal_G2t_2.45deg}
\end{figure*}

Figures~\ref{fig:SM3_D_R-disorder_six-terminal_Rxy_2.45deg} and
\ref{fig:SM3_D_R-disorder_two-terminal_G2t_2.45deg} show the Hall
resistance $-\langle R_{xy}\rangle$ and bulk conductance
$\langle G_{2t}\rangle$ under $D_R$ disorder at $\theta=2.45^\circ$.
The disorder strengths, correlation lengths, and sample sizes are
the same as for $V_R$.
The plateaus at $-\langle R_{xy}\rangle=h/e^2$ and $h/(2e^2)$ remain
much narrower and broaden only modestly with increasing $\gamma_D$,
consistent with the weaker sensitivity to $D_R$ disorder.
For example, at $\lambda=2$ in the largest sample, the first plateau
width increases only from approximately $\Delta\nu_h=0.10$ to $0.15$
as $\gamma_D$ increases from $1$ to $3\,\mathrm{meV}$.
The plateaus remain separated by a broad transition region where
$-\langle R_{xy}\rangle$ dips toward zero.
Correspondingly, $\langle G_{2t}\rangle$ tends to zero in the plateau
regions, while its maximum between them decreases from approximately
$9.2$ to $4.3\,e^2/h$, remaining much larger than those of $V_R$ at the same disorder strength and correlation length.
At $\gamma_D=3\,\mathrm{meV}$ and $\lambda=2$, this maximum still
increases with sample size, indicating strong bulk conduction on
the simulated length scales.

\begin{figure*}[htbp]
    \centering
    \includegraphics[width=\linewidth]{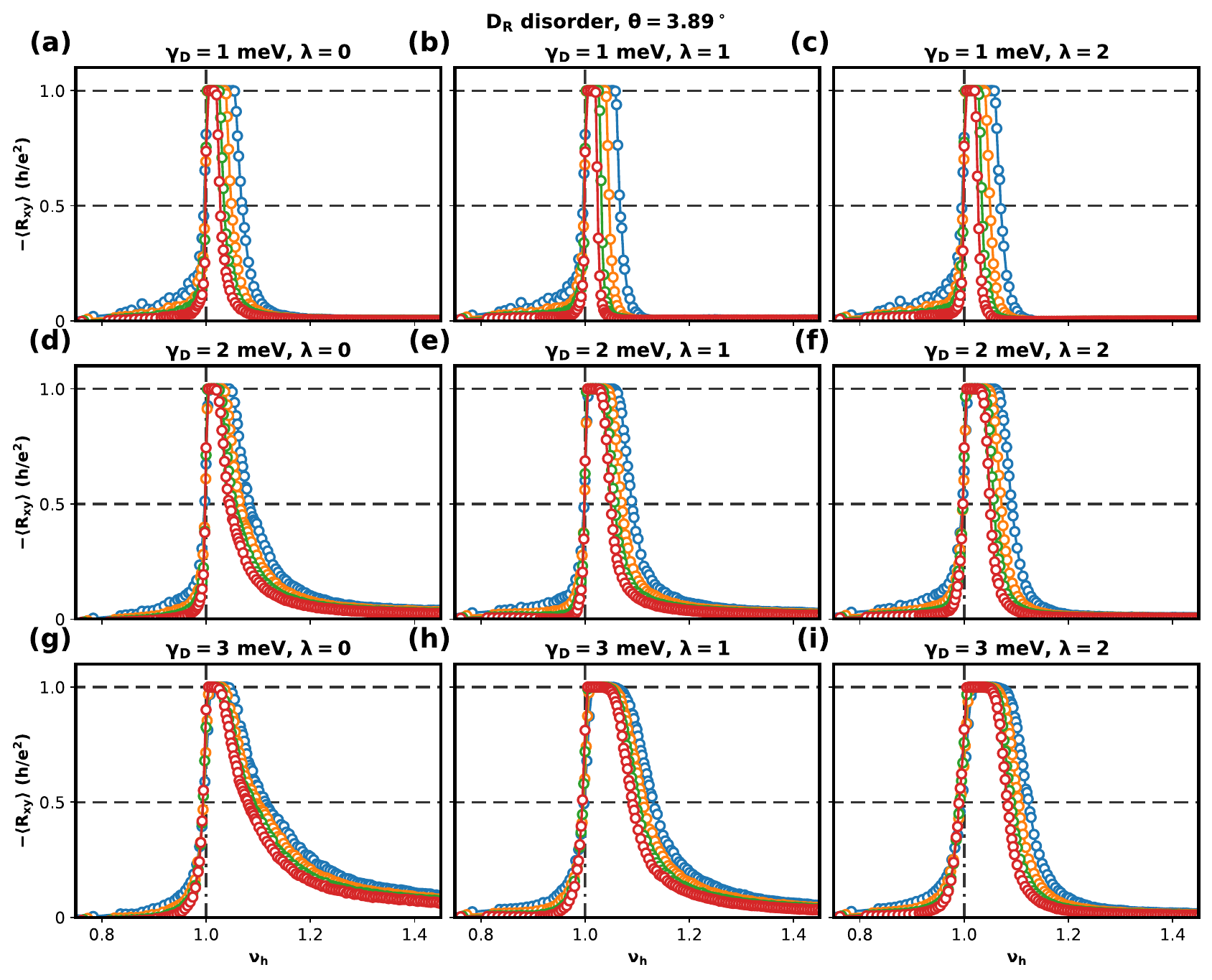}
    \caption{
Disorder-averaged Hall resistance $-\langle R_{xy}\rangle$ versus hole filling $\nu_h$
for displacement-field disorder $D_R$ at $\theta=3.89^\circ$,
using the six-terminal geometry with open boundaries [Fig.~\ref{fig:Figure2}(a)].
Panels (a)--(c), (d)--(f), and (g)--(i) have
$\gamma_a=1$, $2$, and $3\,\mathrm{meV}$, respectively.
Columns from left to right correspond to $\lambda=0,1,2$.
Blue, orange, green, and red curves correspond to
$(N_x,N_y)=(16,28)$, $(24,40)$, $(36,60)$, and $(48,80)$, respectively.
Horizontal dashed lines mark $h/e^2$ and $h/(2e^2)$.
The vertical dash-dotted line marks $\nu_h=1$.
    }
    \label{fig:SM3_D_R-disorder_six-terminal_Rxy_3.89deg}
\end{figure*}

\begin{figure*}[htbp]
    \centering
    \includegraphics[width=\linewidth]{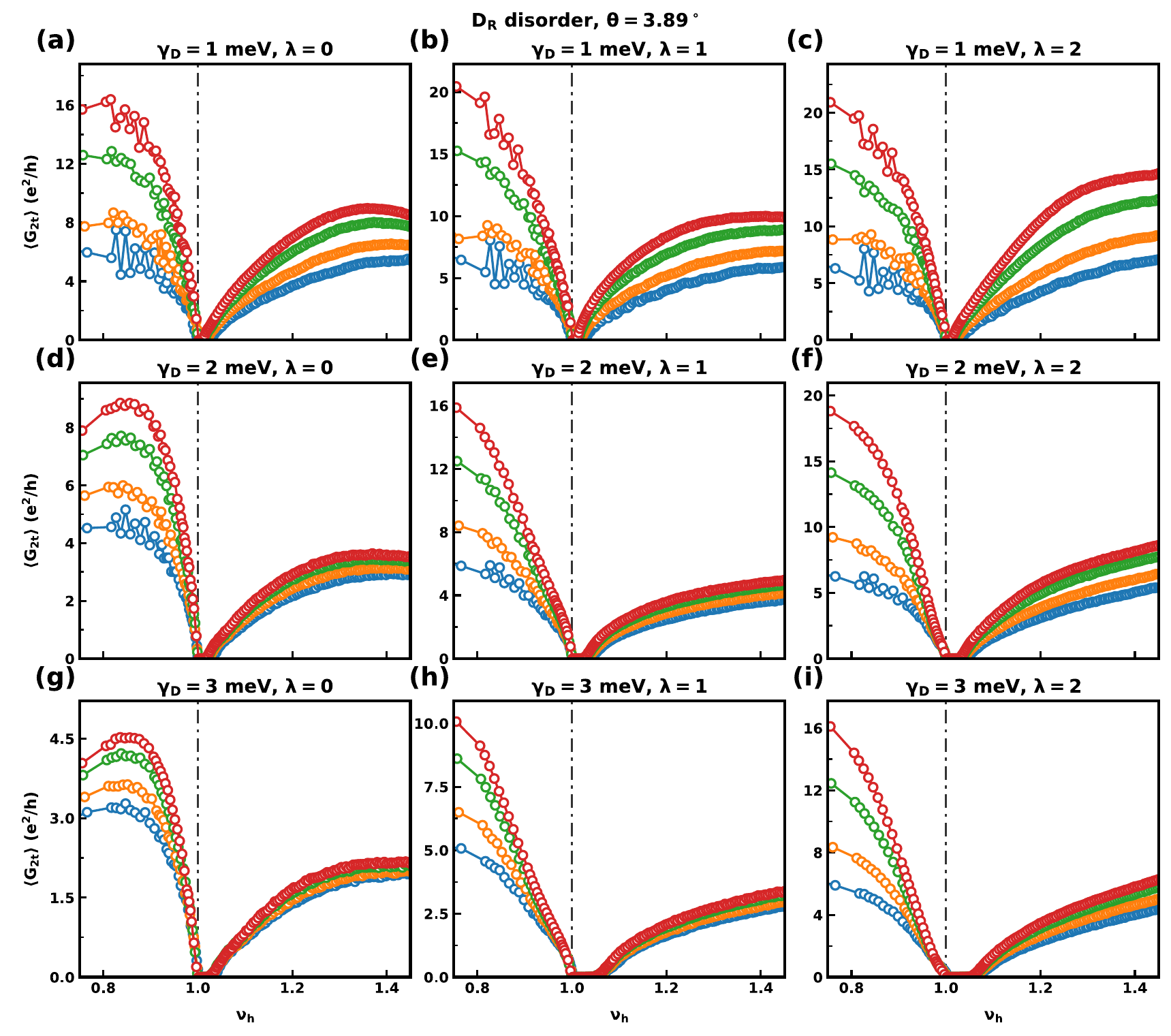}
    \caption{
Disorder-averaged two-terminal conductance $\langle G_{2t}\rangle$ versus hole filling $\nu_h$
for displacement-field disorder $D_R$ at $\theta=3.89^\circ$.
The geometry has periodic boundaries along $y$,
and the conductance is corrected for the Sharvin contact resistance.
Panels (a)--(c), (d)--(f), and (g)--(i) have
$\gamma_a=1$, $2$, and $3\,\mathrm{meV}$, respectively.
Columns from left to right correspond to $\lambda=0,1,2$.
Blue, orange, green, and red curves correspond to
$(N_x,N_y)=(16,28)$, $(24,40)$, $(36,60)$, and $(48,80)$, respectively.
The vertical dash-dotted line marks $\nu_h=1$.
    }
    \label{fig:SM3_D_R-disorder_two-terminal_G2t_3.89deg}
\end{figure*}

Similar behavior is found at $\theta=3.89^\circ$
[Figs.~\ref{fig:SM3_D_R-disorder_six-terminal_Rxy_3.89deg}
and \ref{fig:SM3_D_R-disorder_two-terminal_G2t_3.89deg}].
The single plateau at $-\langle R_{xy}\rangle=h/e^2$ near $\nu_h=1$
broadens only slightly with increasing $\gamma_D$ and remains
narrower than under $V_R$ disorder.
The bulk conductance is nearly zero in the plateau region and
decreases on either side as $\gamma_D$ increases.
Nevertheless, it retains a clear increase with sample size outside
the plateau, as at $2.45^\circ$.

At fixed $\gamma_D$, different from the behavior under $V_R$ disorder, increasing $\lambda$ changes the plateau widths
only slightly but enhances bulk conduction outside the plateaus.
At $2.45^\circ$, this is accompanied by a deeper Hall dip between
the two plateaus.
For example, at $\gamma_D=3\,\mathrm{meV}$ in the largest sample,
increasing $\lambda$ from $0$ to $2$ leaves the first plateau width
near $\Delta\nu_h=0.15$, while the maximum $\langle G_{2t}\rangle$
between plateaus rises from approximately $1.7$ to $4.3\,e^2/h$
[Figs.~\ref{fig:SM3_D_R-disorder_six-terminal_Rxy_2.45deg}(g--i)
and \ref{fig:SM3_D_R-disorder_two-terminal_G2t_2.45deg}(g--i)].
At $3.89^\circ$ and fixed $\gamma_D$, the Hall response outside the single plateau is
likewise further suppressed, while $\langle G_{2t}\rangle$ increases
on both sides with increasing $\lambda$
[Figs.~\ref{fig:SM3_D_R-disorder_six-terminal_Rxy_3.89deg}
and \ref{fig:SM3_D_R-disorder_two-terminal_G2t_3.89deg}]. The reason for enhanced Hall dip and bulk conduction under $D_R$ disorder is further discussed in SM.~\ref{sm:response}.

\clearpage
\SMsubsection{Twist angle disorder $\theta_R$}

To distinguish the onsite and hopping contributions (see Eq.~\eqref{eq:sm_full_twist_disorder}) to twist-angle
disorder, we compare three versions of
Eq.~\eqref{eq:sm_full_twist_disorder}: onsite disorder only, NN hopping
disorder only, and both contributions together.
The two contributions (onsite and NN hopping) are generated from the same local angle field.
All results use $\theta=2.45^\circ$, $W_\theta=0.2^\circ,0.6^\circ,1.0^\circ$,
$\lambda=0,1,2$, and the same sample sizes as above.
When the onsite term is included, these angle widths correspond to
$\gamma_\theta\simeq0.65,1.83,3.03\,\mathrm{meV}$, respectively,
as defined in Eq.~\eqref{eq:sm_twist_linear_strength}.

\begin{figure*}[htbp]
    \centering
    \includegraphics[width=\linewidth]{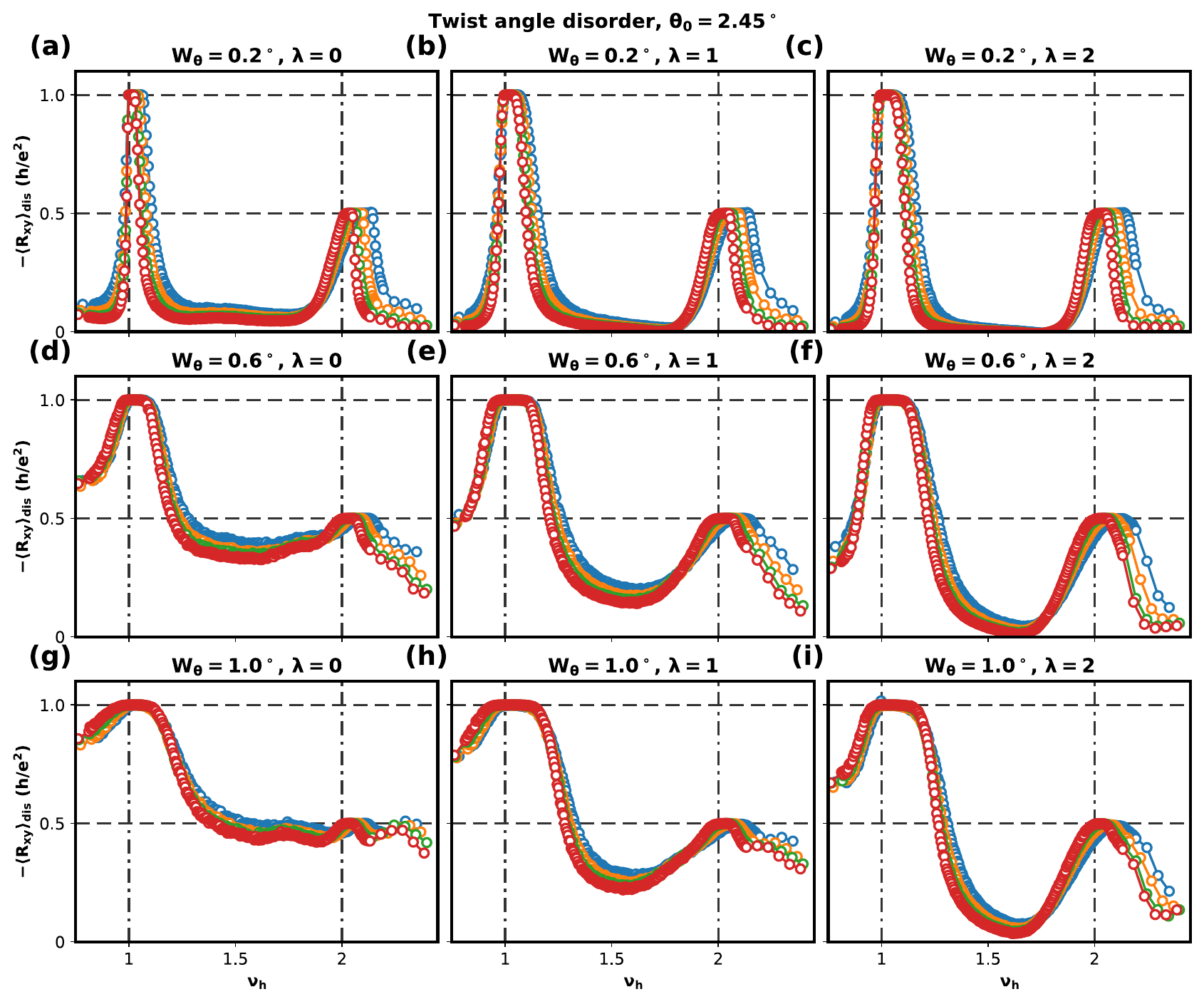}
    \caption{
Disorder-averaged Hall resistance $-\langle R_{xy}\rangle$ versus hole filling $\nu_h$
for twist-angle disorder $\theta_R$ at mean twist angle $\theta_0=2.45^\circ$,
using the six-terminal geometry with open boundaries [Fig.~\ref{fig:Figure2}(a)].
Both onsite and nearest-neighbor hopping variations are included.
Panels (a)--(c), (d)--(f), and (g)--(i) have twist-angle disorder strengths
$W_\theta=0.2^\circ$, $0.6^\circ$, and $1.0^\circ$, respectively.
Columns from left to right correspond to $\lambda=0,1,2$.
Blue, orange, green, and red curves correspond to
$(N_x,N_y)=(16,28)$, $(24,40)$, $(36,60)$, and $(48,80)$, respectively.
Horizontal dashed lines mark $h/e^2$ and $h/(2e^2)$.
Vertical dash-dotted lines mark $\nu_h=1,2$.
    }
    \label{fig:SM3_twist-angle-disorder_six-terminal_Rxy_2.45deg}
\end{figure*}

\begin{figure*}[htbp]
    \centering
    \includegraphics[width=\linewidth]{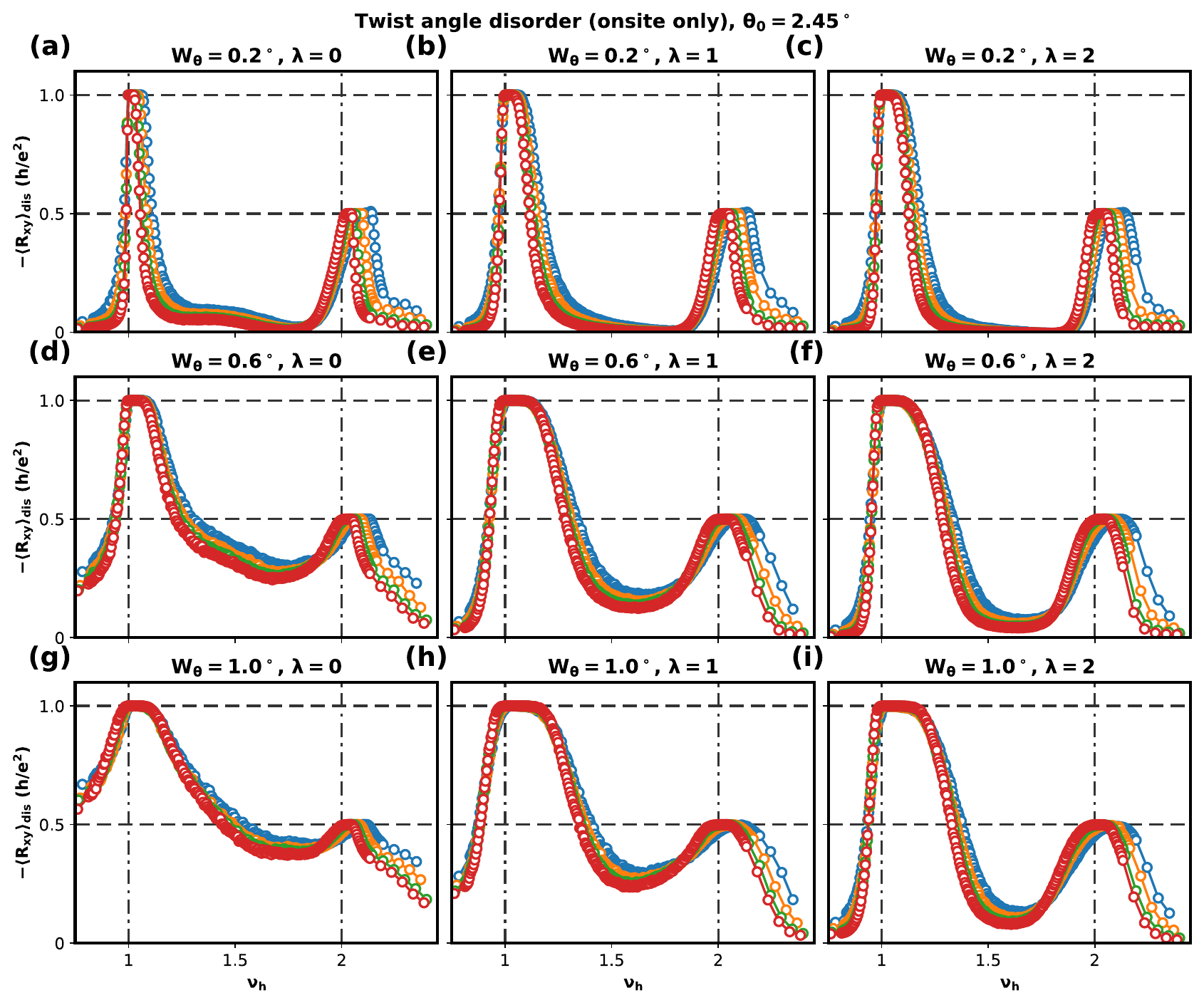}
    \caption{
Disorder-averaged Hall resistance $-\langle R_{xy}\rangle$ versus hole filling $\nu_h$
for twist-angle disorder $\theta_R$ at mean twist angle $\theta_0=2.45^\circ$,
using the six-terminal geometry with open boundaries [Fig.~\ref{fig:Figure2}(a)].
Only the twist-induced onsite variations are retained.
Panels (a)--(c), (d)--(f), and (g)--(i) have twist-angle disorder strengths
$W_\theta=0.2^\circ$, $0.6^\circ$, and $1.0^\circ$, respectively.
Columns from left to right correspond to $\lambda=0,1,2$.
Blue, orange, green, and red curves correspond to
$(N_x,N_y)=(16,28)$, $(24,40)$, $(36,60)$, and $(48,80)$, respectively.
Horizontal dashed lines mark $h/e^2$ and $h/(2e^2)$.
Vertical dash-dotted lines mark $\nu_h=1,2$.
    }
    \label{fig:SM3_twist-angle-disorder-onsite-only_six-terminal_Rxy_2.45deg}
\end{figure*}

The onsite-only and full twist-angle models show the same qualitative evolution
of the Hall plateaus
[Figs.~\ref{fig:SM3_twist-angle-disorder-onsite-only_six-terminal_Rxy_2.45deg}
and \ref{fig:SM3_twist-angle-disorder_six-terminal_Rxy_2.45deg}].
Increasing $\gamma_\theta$ broadens the plateaus, narrows the intervening
transition region, and weakens the Hall dip.
The maximum $\langle G_{2t}\rangle$ between the plateaus decreases markedly
in both models
[Figs.~\ref{fig:SM3_twist-angle-disorder-onsite-only_two-terminal_G2t_2.45deg}
and \ref{fig:SM3_twist-angle-disorder_two-terminal_G2t_2.45deg}] with increasing $\gamma_\theta$.
For $\lambda=0$, the conductance maximum increases with system size
at weak disorder but is nearly size independent at stronger disorder
over the system sizes studied.
At $W_\theta=1.0^\circ$, this maximum is approximately $0.9\,e^2/h$
in both models.

\begin{figure*}[htbp]
    \centering
    \includegraphics[width=\linewidth]{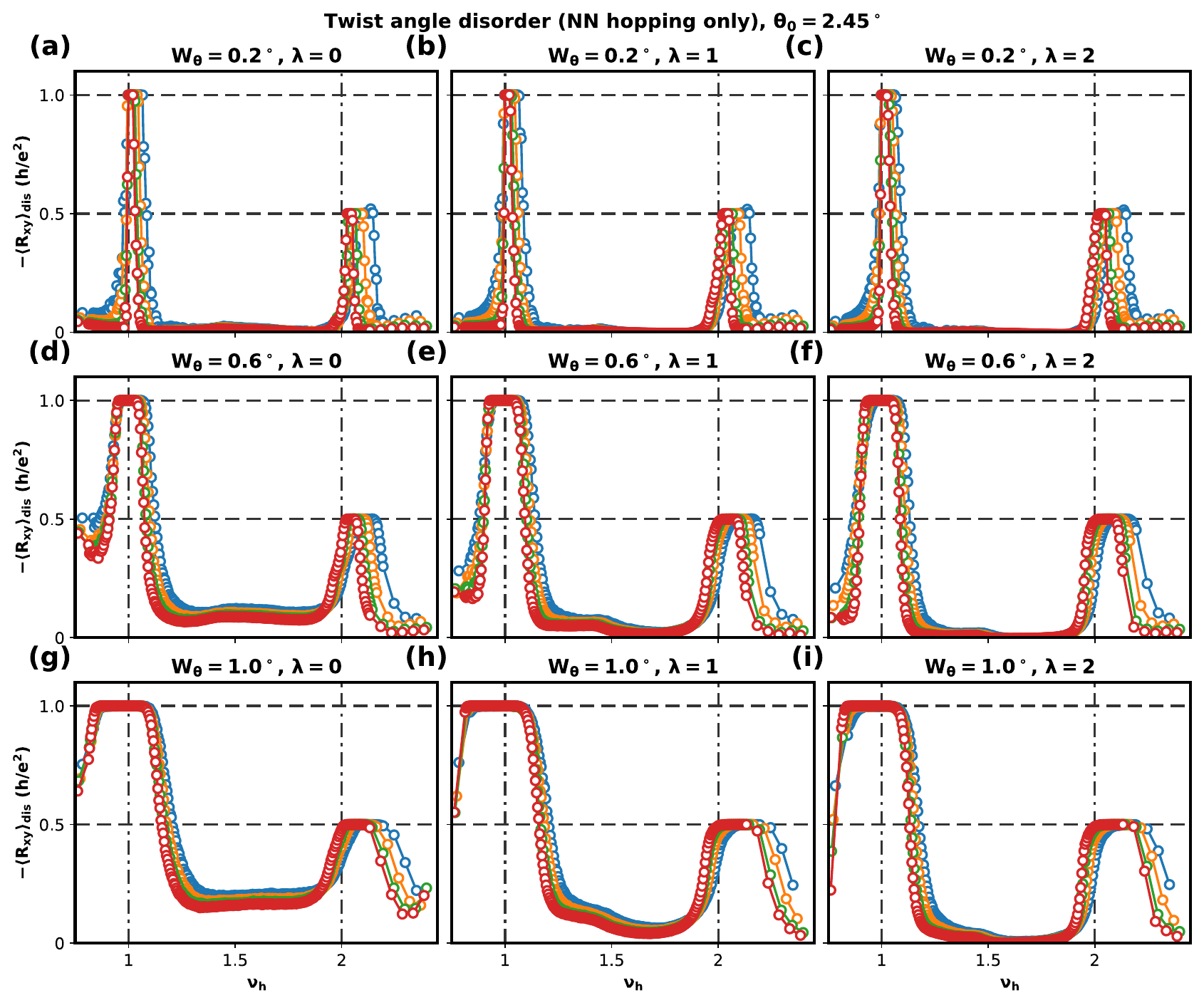}
    \caption{
Disorder-averaged Hall resistance $-\langle R_{xy}\rangle$ versus hole filling $\nu_h$
for twist-angle disorder $\theta_R$ at mean twist angle $\theta_0=2.45^\circ$,
using the six-terminal geometry with open boundaries [Fig.~\ref{fig:Figure2}(a)].
Only the twist-induced variations of nearest-neighbor hoppings are retained.
Panels (a)--(c), (d)--(f), and (g)--(i) have twist-angle disorder strengths
$W_\theta=0.2^\circ$, $0.6^\circ$, and $1.0^\circ$, respectively.
Columns from left to right correspond to $\lambda=0,1,2$.
Blue, orange, green, and red curves correspond to
$(N_x,N_y)=(16,28)$, $(24,40)$, $(36,60)$, and $(48,80)$, respectively.
Horizontal dashed lines mark $h/e^2$ and $h/(2e^2)$.
Vertical dash-dotted lines mark $\nu_h=1,2$.
    }
    \label{fig:SM3_twist-angle-disorder-NN-hopping-only_six-terminal_Rxy_2.45deg}
\end{figure*}

\begin{figure*}[htbp]
    \centering
    \includegraphics[width=\linewidth]{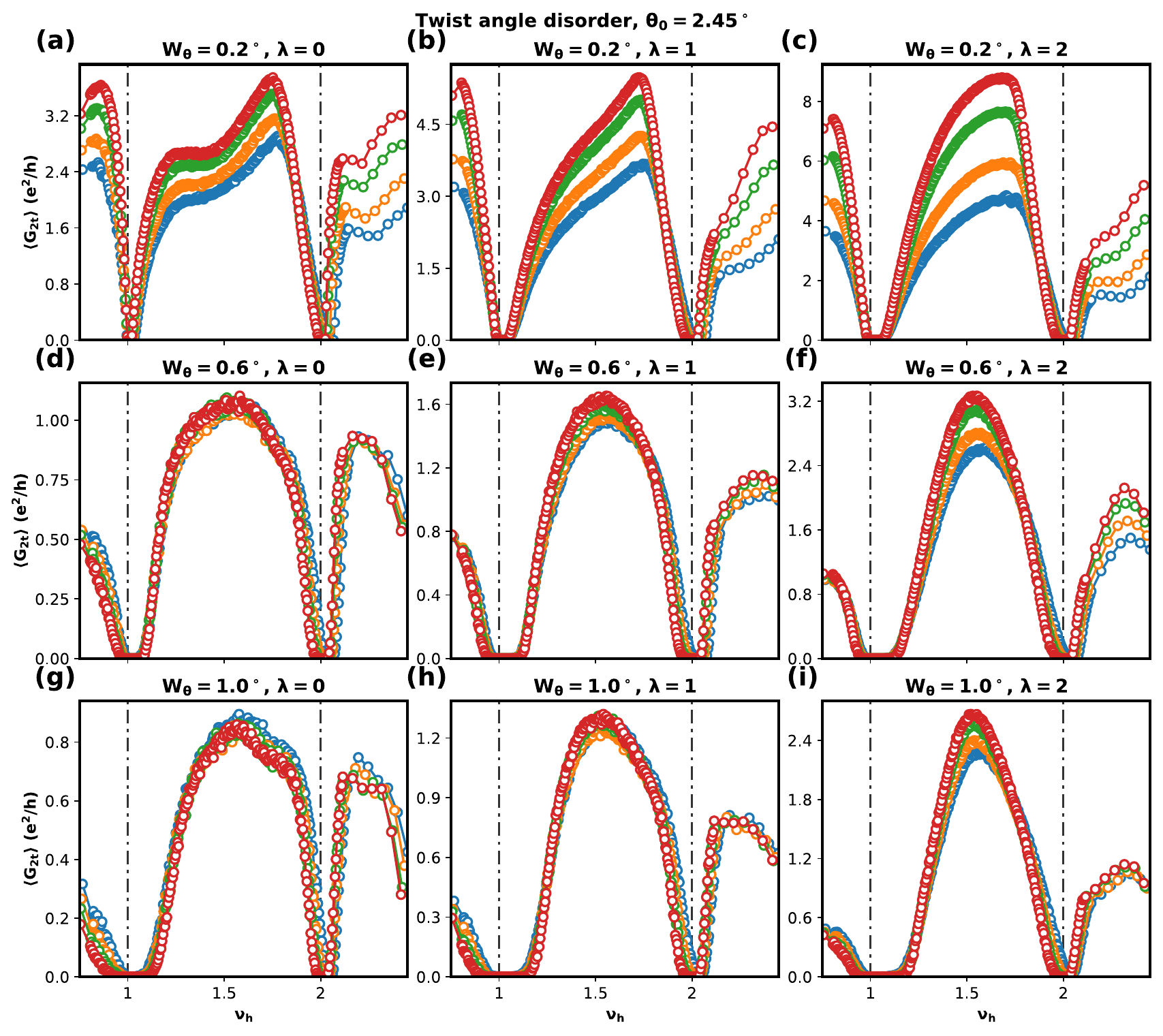}
    \caption{
Disorder-averaged two-terminal conductance $\langle G_{2t}\rangle$ versus hole filling $\nu_h$
for twist-angle disorder $\theta_R$ at mean twist angle $\theta_0=2.45^\circ$.
The geometry has periodic boundaries along $y$,
and the conductance is corrected for the Sharvin contact resistance.
Both onsite and nearest-neighbor hopping variations are included.
Panels (a)--(c), (d)--(f), and (g)--(i) have twist-angle disorder strengths
$W_\theta=0.2^\circ$, $0.6^\circ$, and $1.0^\circ$, respectively.
Columns from left to right correspond to $\lambda=0,1,2$.
Blue, orange, green, and red curves correspond to
$(N_x,N_y)=(16,28)$, $(24,40)$, $(36,60)$, and $(48,80)$, respectively.
Vertical dash-dotted lines mark $\nu_h=1,2$.
    }
    \label{fig:SM3_twist-angle-disorder_two-terminal_G2t_2.45deg}
\end{figure*}

\begin{figure*}[htbp]
    \centering
    \includegraphics[width=\linewidth]{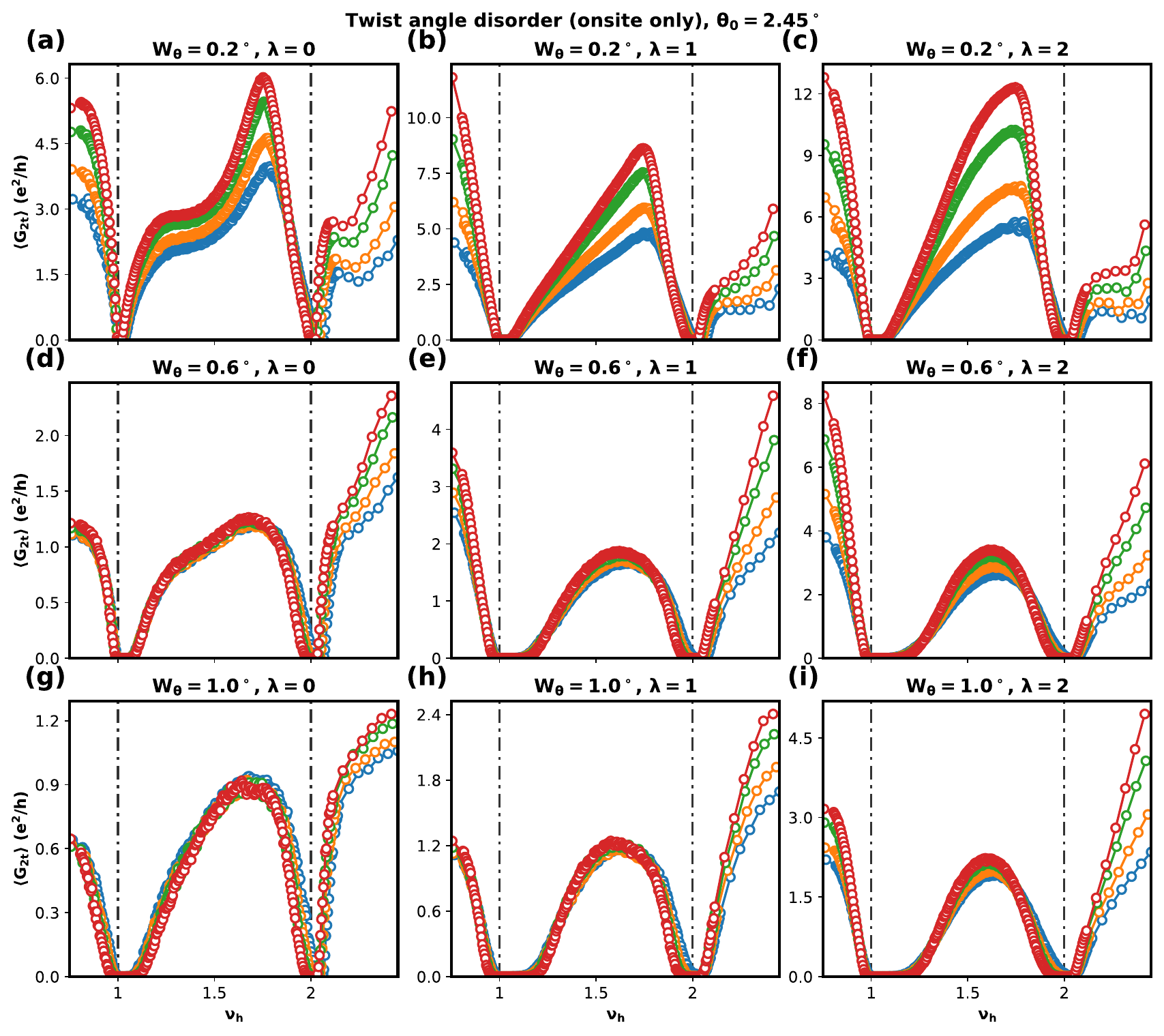}
    \caption{
Disorder-averaged two-terminal conductance $\langle G_{2t}\rangle$ versus hole filling $\nu_h$
for twist-angle disorder $\theta_R$ at mean twist angle $\theta_0=2.45^\circ$.
The geometry has periodic boundaries along $y$,
and the conductance is corrected for the Sharvin contact resistance.
Only the twist-induced onsite variations are retained.
Panels (a)--(c), (d)--(f), and (g)--(i) have twist-angle disorder strengths
$W_\theta=0.2^\circ$, $0.6^\circ$, and $1.0^\circ$, respectively.
Columns from left to right correspond to $\lambda=0,1,2$.
Blue, orange, green, and red curves correspond to
$(N_x,N_y)=(16,28)$, $(24,40)$, $(36,60)$, and $(48,80)$, respectively.
Vertical dash-dotted lines mark $\nu_h=1,2$.
    }
    \label{fig:SM3_twist-angle-disorder-onsite-only_two-terminal_G2t_2.45deg}
\end{figure*}

\begin{figure*}[htbp]
    \centering
    \includegraphics[width=\linewidth]{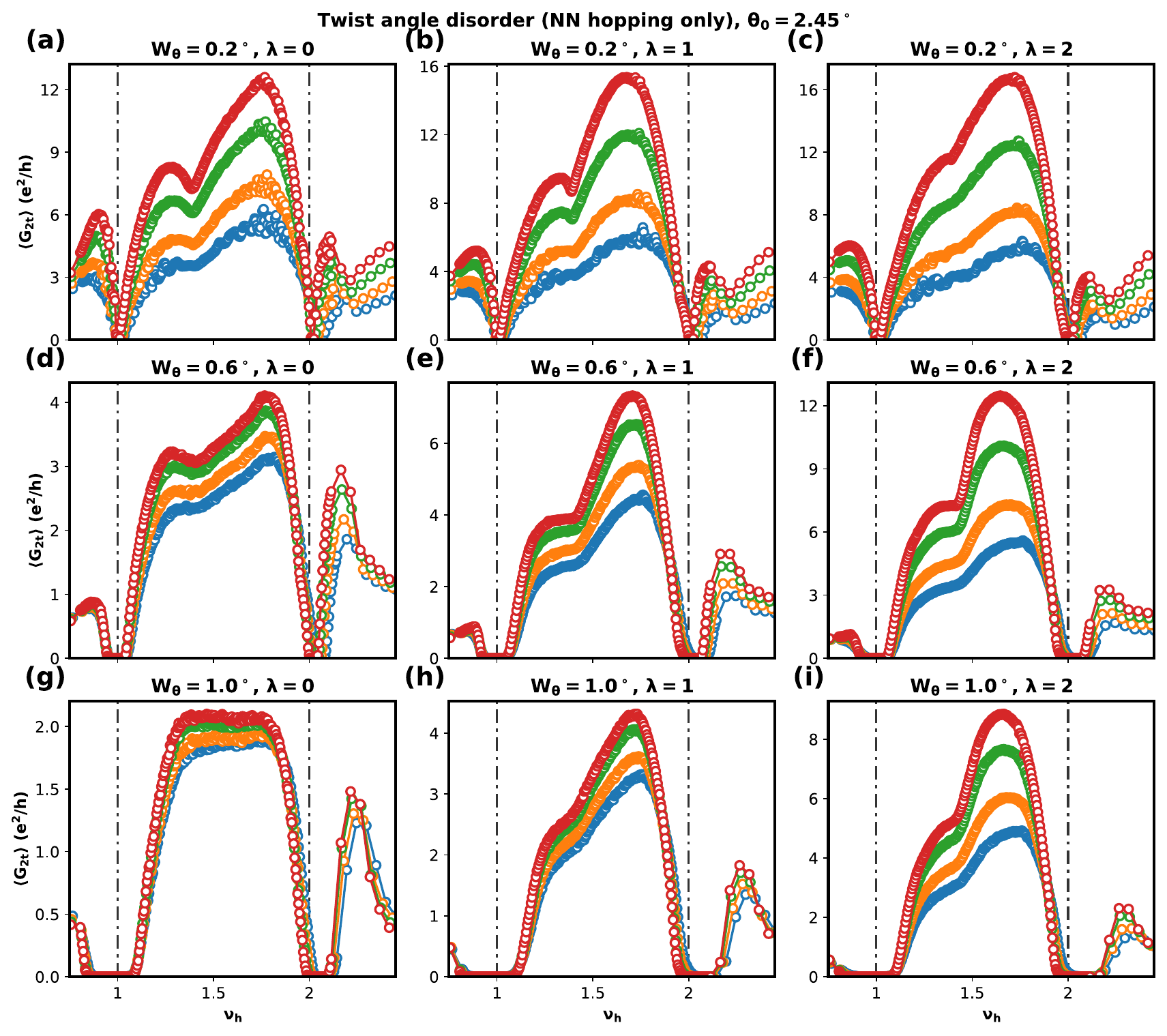}
    \caption{
Disorder-averaged two-terminal conductance $\langle G_{2t}\rangle$ versus hole filling $\nu_h$
for twist-angle disorder $\theta_R$ at mean twist angle $\theta_0=2.45^\circ$.
The geometry has periodic boundaries along $y$,
and the conductance is corrected for the Sharvin contact resistance.
Only the twist-induced variations of nearest-neighbor hoppings are retained.
Panels (a)--(c), (d)--(f), and (g)--(i) have twist-angle disorder strengths
$W_\theta=0.2^\circ$, $0.6^\circ$, and $1.0^\circ$, respectively.
Columns from left to right correspond to $\lambda=0,1,2$.
Blue, orange, green, and red curves correspond to
$(N_x,N_y)=(16,28)$, $(24,40)$, $(36,60)$, and $(48,80)$, respectively.
Vertical dash-dotted lines mark $\nu_h=1,2$.
    }
    \label{fig:SM3_twist-angle-disorder-NN-hopping-only_two-terminal_G2t_2.45deg}
\end{figure*}

At fixed disorder strength, increasing $\lambda$ changes the plateau
widths only modestly but substantially enhances conduction between them.
For example, at $W_\theta=1.0^\circ$ in the largest sample, increasing
$\lambda$ from $0$ to $2$ raises the maximum $\langle G_{2t}\rangle$
from approximately $0.9$ to $2.2\,e^2/h$ in the onsite-only model
and from $0.9$ to $2.7\,e^2/h$ in the combined model.
The Hall dip deepens, and the conductance develops a clearer increase
with sample size over the simulated
length scales.
These shared trends show that onsite part of the twist angle disorder dominates the
suppression of bulk conduction when both onsite and NN-hopping contributions are present.
With NN hopping disorder alone, the Hall response $\langle R_{xy}\rangle$ remains strongly
suppressed in the transition region between the plateaus, and the conductance maximum increases
with sample size throughout the parameter range shown
[Figs.~\ref{fig:SM3_twist-angle-disorder-NN-hopping-only_six-terminal_Rxy_2.45deg}
and \ref{fig:SM3_twist-angle-disorder-NN-hopping-only_two-terminal_G2t_2.45deg}].
Nevertheless, increasing the disorder strength still substantially
broadens the plateaus.
For example, at $\lambda=2$ in the largest sample, the first plateau
width increases from approximately $\Delta\nu_h=0.03$ to $0.27$ as
$W_\theta$ increases from $0.2^\circ$ to $1.0^\circ$.
Even at $W_\theta=1.0^\circ$, the maximum $\langle G_{2t}\rangle$
grows from approximately $4.9$ to $8.8\,e^2/h$ as $N_x$ increases
from $16$ to $48$.
This behavior suggests that hopping disorder tends to localize states
near the band edges, stabilizing Hall plateaus while preserving
strong bulk conduction in the band interior over the system sizes studied.

\clearpage

\SMsubsection{Finite-size scaling}

\begin{figure*}[htbp]
    \centering
    \includegraphics[width=\linewidth]{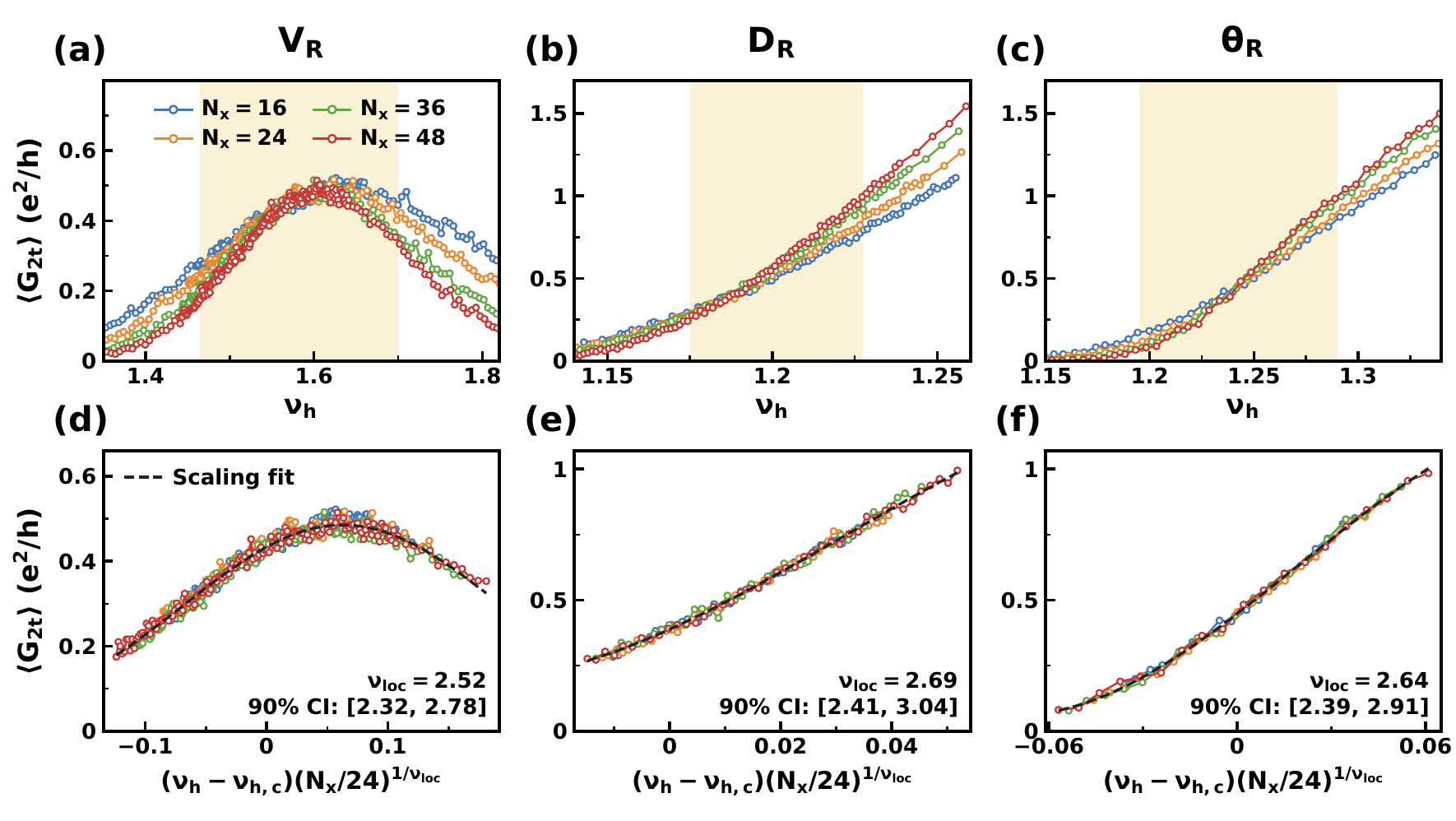}
    \caption{Finite-size scaling of the disorder-averaged bulk conductance.
(a--c) $\langle G_{2t}\rangle$ versus $\nu_h$ for $V_R$, $D_R$,
and $\theta_R$ disorder, respectively, corresponding to
Fig.~\ref{fig:Figure3}(d) of the main text and Fig.~\ref{fig:SM3_twist-angle-disorder-onsite-only_two-terminal_G2t_2.45deg}(i).
The strengths are $\gamma_V=\gamma_D=3\,\mathrm{meV}$ and
$W_\theta=1^\circ$ ($\gamma_\theta\simeq3$).
Colors denote $(N_x,N_y)=(16,28)$, $(24,40)$, $(36,60)$,
and $(48,80)$, as indicated in (a).
Yellow shading marks the fitting windows.
(d--f) The selected data plotted against the scaling variable $X$
defined in Eq.~\eqref{eq:sm_conductance_scaling}; black dashed curves show the fits.
Labels give the fitted $\nu_{\mathrm{loc}}$ and its $90\%$ bootstrap
confidence interval, evaluated from $400$ resamples with the fitting
window and functional form held fixed.
They measure window sensitivity, not statistical confidence intervals. The other parameters are
$\theta=2.45^\circ$, and $\lambda=2$
    }
    \label{fig:SM3_G2t_data-collapse_VR-DR-twist_2.45deg_lambda2}
\end{figure*}

In this part, we use finite-size scaling~\cite{Abrahams1979,FSS_Pichard_1981,TMM_MacKinnon_1981,Qi2026Anderson}
to examine the critical behavior of the bulk conductance $\langle G_{2t}\rangle$ in
Fig.~\ref{fig:Figure3}(d) of the main text for $V_R$ and $D_R$ disorder and Fig.~\ref{fig:SM3_twist-angle-disorder-onsite-only_two-terminal_G2t_2.45deg}(i) for $\theta_R$ disorder.
Our noninteracting single-valley model belongs to the unitary class
(class A), which in two dimensions supports delocalized 
states in the Hall or Chern plateau transition regions~\cite{Huckestein1995,EversMirlin2008}.
The strong sensitivity to $V_R$ favors bulk localization, leaving a
nearly size-independent conductance peak.
We therefore fit the region around this peak
[Fig.~\ref{fig:SM3_G2t_data-collapse_VR-DR-twist_2.45deg_lambda2}(a)].
For $D_R$ and $\theta_R$, weaker effective disorder leaves a broad
conducting region on the simulated length scales.
We instead analyze the crossings near its lower-filling edge
[Fig.~\ref{fig:SM3_G2t_data-collapse_VR-DR-twist_2.45deg_lambda2}(b,c)].

We fit the four sample sizes using the single-parameter scaling form
\begin{equation}
\frac{\langle G_{2t}\rangle}{e^2/h}
\simeq \mathcal F_a(X),\qquad
X=(\nu_h-\nu_{h,c})
\left(\frac{N_x}{24}\right)^{1/\nu_{\mathrm{loc}}},
\label{eq:sm_conductance_scaling}
\end{equation}
where $\nu_{h,c}$ and $\nu_{\mathrm{loc}}$ are the fitted critical
filling and localization-length exponent, respectively.
The reference size $24$ only sets the scale of $X$.
The fitting windows are shaded in the upper panels.
We represent $\ln\mathcal F_V$ by a quadratic polynomial in $X$
and $\mathcal F_D$ and $\mathcal F_\theta$ by cubic polynomials,
using weighted least squares with the standard errors of the mean.

The rescaled data collapse onto the fitted curves
[Fig.~\ref{fig:SM3_G2t_data-collapse_VR-DR-twist_2.45deg_lambda2}(d--f)],
with $\nu_{\mathrm{loc}}\simeq2.52$, $2.69$, and $2.64$ for
$V_R$, $D_R$, and $\theta_R$, respectively.
For these fixed windows and fitting forms, the corresponding
$90\%$ bootstrap intervals are $[2.32,2.78]$, $[2.41,3.04]$,
and $[2.39,2.91]$.
All three fitted exponents lie close to
$\nu_{\mathrm{loc}}\simeq2.6$, the characteristic value of the
integer quantum Hall universality class~\cite{Dresselhaus2022,Bera2024}.
supporting compatibility with this critical scaling despite the
different finite-size conductance profiles.

\clearpage
\SMsection{Charge response}
\label{sm:response}

In this section, we derive the intraband charge density susceptibility (Eq.~\eqref{eq:density_response_kernel}) and root-mean-square charge
response amplitude (Eq.~\eqref{eq:charge_response})  used in the main text.
The clean-band charge response provides a measure of the system's
sensitivity to each disorder channel within the perturbation theory
and provides a qualitative perspective on the transport results
in Sec.~\ref{sm:transport}.

\SMsubsection{Derivation of the charge density response}

The charge density is the total occupation of a moir\'e cell,
$\hat n_R=\sum_\alpha c_{R,\alpha}^\dagger c_{R,\alpha}$, whereas
the disorder couples to $\hat O_a(R)=c_R^\dagger M_a c_R$.
Here $\alpha=\mathrm{MX},\mathrm{XM},\mathrm{MM}$.
We consider $N$ periodic cells and use the unitary Fourier convention
$w_a(\bm q)=N^{-1/2}\sum_R w_a(R)e^{-i\bm q\cdot R}$,
with $N_k=N_q=N$.
The annihilation operator $d_{n\bm k}$ for a clean Bloch state in band $n$ is
\begin{equation}
\begin{aligned}
d_{n\bm k}
=\sum_\alpha u_{n\alpha}^*(\bm k)c_{\bm k,\alpha},\;
c_{\bm k,\alpha}
=\frac{1}{\sqrt N}\sum_R e^{-i\bm k\cdot R}c_{R,\alpha},
\end{aligned}
\label{eq:sm4_band_operator}
\end{equation}
where $u_{n\alpha}(\bm k)=\langle\alpha|u_{n\bm k}\rangle$
is the component of the Bloch eigenvector on Wannier orbital
$\alpha=\mathrm{MX},\mathrm{XM},\mathrm{MM}$.

Projecting both the density operator $\hat n_R$ and the disorder
coupling operator $\hat O_a(R)$ onto the target band $n$ gives
\begin{equation}
\begin{aligned}
\hat n_n(\bm q)
&=\frac{1}{\sqrt N}\sum_{\bm k}
\langle u_{n\bm k}|u_{n,\bm k+\bm q}\rangle
d_{n\bm k}^\dagger d_{n,\bm k+\bm q},\\
\hat O_{a,n}(-\bm q)
&=\frac{1}{\sqrt N}\sum_{\bm k}
F_a^{(n)}(\bm k,\bm q)
d_{n,\bm k+\bm q}^\dagger d_{n\bm k},
\end{aligned}
\label{eq:sm4_projected_operators}
\end{equation}
where $F_a^{(n)}(\bm k,\bm q)
=\langle u_{n,\bm k+\bm q}|M_a|u_{n\bm k}\rangle$.
%
%
The disorder-induced perturbation within the target band is
$H_{\mathrm{dis},a}^{(n)}
=\sum_{\bm q}w_a(\bm q)\hat O_{a,n}(-\bm q)$.

At fixed chemical potential $\mu$, static linear response gives
\begin{equation}
\begin{aligned}
\delta n_{a,n}(\bm q;\mu)
&=-\chi_{\rho,a}^{(n)}(\bm q;\mu)w_a(\bm q),\\
\chi_{\rho,a}^{(n)}(\bm q;\mu)
&=\int_0^\beta d\tau\,
\big\langle T_\tau\hat n_n(\bm q,\tau)
\hat O_{a,n}(-\bm q,0)\big\rangle_{0,c}.
\end{aligned}
\label{eq:sm4_kubo_response}
\end{equation}
Here $\delta n_{a,n}$ is the change in the equilibrium density,
$\beta=(k_BT)^{-1}$, $T_\tau$ denotes imaginary-time ordering,
and $\langle\cdots\rangle_{0,c}$ is the connected thermal average
in the clean system.
Both operators are fermion bilinears.
Evaluating their four-fermion correlation using Wick's theorem gives
\begin{equation}
\begin{aligned}
\chi_{\rho,a}^{(n)}(\bm q;\mu)
={}\frac{1}{N_k}\sum_{\bm k}
\frac{f_{n\bm k}-f_{n,\bm k+\bm q}}
{\varepsilon_{n,\bm k+\bm q}-\varepsilon_{n\bm k}}
\times
\langle u_{n\bm k}|u_{n,\bm k+\bm q}\rangle
F_a^{(n)}(\bm k,\bm q).
\end{aligned}
\label{eq:sm4_linear_response}
\end{equation}
Here $f_{n\bm k}=f(\varepsilon_{n\bm k}-\mu)$.
The Bloch-state overlap comes from summing the density over all
three orbitals, while $F_a^{(n)}$ describes the applied disorder channel.
Equation~\eqref{eq:sm4_linear_response} reproduces
Eq.~\eqref{eq:density_response_kernel} of the main text.

For zero-mean disorder, the induced density change also has zero
ensemble mean.
We therefore characterize its typical magnitude by the
root-mean-square charge response.
To perform the disorder average, we write
\begin{equation}
\overline{w_a(R)w_a(R')}
=\gamma_a^2 C_{a,\lambda}(R-R'),
\qquad C_{a,\lambda}(0)=1.
\label{eq:sm4_real_covariance}
\end{equation}
Fourier transformation gives
\begin{equation}
\begin{aligned}
\overline{w_a(\bm q)w_a(\bm q')^*}
&=\gamma_a^2 S_{a,\lambda}(\bm q)\delta_{\bm q,\bm q'},\\
S_{a,\lambda}(\bm q)
&=\sum_R C_{a,\lambda}(R)e^{-i\bm q\cdot R},
\qquad
\frac{1}{N_q}\sum_{\bm q}S_{a,\lambda}(\bm q)=1.
\end{aligned}
\label{eq:sm4_disorder_spectrum}
\end{equation}
Thus, $S_{a,\lambda}$ describes how the fixed disorder variance
is distributed among momentum transfers.
For onsite fields generated by the Gaussian filter
$F_\lambda(\bm q)$ in Eq.~\eqref{eq:sm_disorder_filter},
\begin{equation}
S_\lambda(\bm q)
=\frac{|F_\lambda(\bm q)|^2}
{N_q^{-1}\sum_{\bm q'}|F_\lambda(\bm q')|^2}
=\frac{e^{-\xi^2|\bm q|^2/2}}
{N_q^{-1}\sum_{\bm q'}e^{-\xi^2|\bm q'|^2/2}},
\qquad \xi=\lambda a_M.
\label{eq:sm4_gaussian_spectrum}
\end{equation}
This is the spectrum corresponding to the finite-grid covariance
in Eq.~\eqref{eq:sm_disorder_covariance}.
For $\lambda=0$, $S_0=1$; increasing $\lambda$ shifts weight toward
small $|\bm q|$ while preserving the local variance $\gamma_a^2$ in real space.
For a nonlinear twist-angle mapping, $S_{\theta,\lambda}$ is
obtained from the covariance of the mapped onsite fluctuations.

Using Eq.~\eqref{eq:sm4_kubo_response} and Parseval's identity,
the charge response amplitude becomes
\begin{equation}
\begin{aligned}
\Delta n_{a,n}(\mu;\lambda)
&=\left[\frac{1}{N}\sum_R
\overline{|\delta n_{a,n}(R;\mu)|^2}\right]^{1/2}=\left[\frac{1}{N_q}\sum_{\bm q}
\overline{|\delta n_{a,n}(\bm q;\mu)|^2}\right]^{1/2}\\
&=\left[\frac{1}{N_q}\sum_{\bm q}
|\chi_{\rho,a}^{(n)}(\bm q;\mu)|^2
\overline{|w_a(\bm q)|^2}\right]^{1/2}\\
&=\gamma_a\left[\frac{1}{N_q}\sum_{\bm q}
S_{a,\lambda}(\bm q)
|\chi_{\rho,a}^{(n)}(\bm q;\mu)|^2\right]^{1/2}.
\end{aligned}
\label{eq:sm4_charge_response}
\end{equation}
This gives Eq.~\eqref{eq:charge_response} of the main text for
both uncorrelated and spatially correlated disorder.

\SMsubsection{Charge density response in \tmt{} and connection to transport}

\begin{figure*}[htbp]
    \centering
    \includegraphics[width=\linewidth]{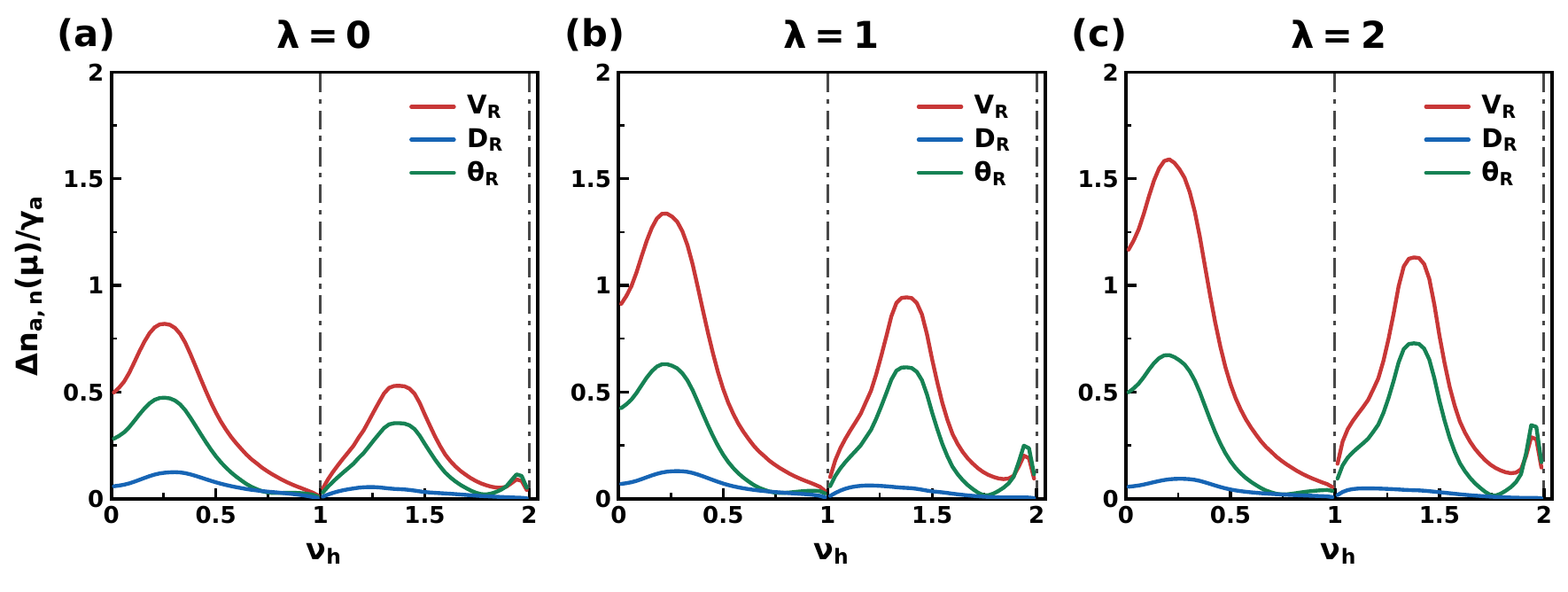}
    \caption{Charge response $\Delta n_{a,n}(\mu;\lambda)/\gamma_a$
versus hole filling $\nu_h$ at $\theta=2.45^\circ$.
(a--c) top and second highest $K$-valley valence band with twist-angle $\theta=2.45^\circ$.
Columns correspond to $\lambda=0,1,2$ from left to right, with
correlation length $\xi=\lambda a_M$.
Red, blue, and green curves denote $V_R$, $D_R$, and the onsite
component of $\theta_R$ disorder, respectively.
The amplitudes are normalized by the standard deviation of the onsite
fluctuations and have units $\mathrm{meV}^{-1}\mathrm{cell}^{-1}$.
All results use $N_k=N_q=101^2$ and a Fermi-function smoothing scale
of $0.04\,\mathrm{meV}$.
    }
    \label{fig:SM4_charge-response_lambda012_combined-bands}
\end{figure*}

Figure~\ref{fig:SM4_charge-response_lambda012_combined-bands} shows
$\Delta n_{a,n}(\mu;\lambda)/\gamma_a$ calculated from Eq.~\eqref{eq:sm4_charge_response} for the highest two valence
bands at $\theta=2.45^\circ$ and $\lambda=0,1,2$.
The chemical potential is expressed through the clean-band hole filling
$\nu_h$.
For every $\lambda$ shown, the main response maxima follow the same
order: $V_R$ is strongest, $D_R$ is weakest, and $\theta_R$ lies between
them.
This reflects the partial cancellation of opposite orbital contributions
for $D_R$ and $\theta_R$ compared with $V_R$ in the total-charge response.

Increasing $\lambda$ enhances the main $V_R$ response maxima from
approximately $0.82$ to $1.59$ in the top band and from $0.53$ to $1.13$
in the second band
[Fig.~\ref{fig:SM4_charge-response_lambda012_combined-bands}(a--c)].
The $D_R$ response remains much smaller, changes little between
$\lambda=0-2$.
The main $\theta_R$ maxima also increase, while remaining below those
of $V_R$.
These trends follow from the weighting in
Eq.~\eqref{eq:sm4_charge_response}: a smoother field emphasizes the
small-momentum charge response, which depends on the orbital channel.

Spatial smoothing changes the effective disorder strength, modifying the charge
response and disorder sensitivity through the orbital-dependent kernel
$\chi_{\rho,a}^{(n)}$.
It also suppresses large-momentum-transfer scattering, which can
reduce backscattering and enhance bulk conduction.

For $V_R$, the charge response increases substantially with $\lambda$,
indicating greater sensitivity to smooth potential disorder.
The Hall plateaus also show clear broadening as $\lambda$ increases
from $0$ to $2$ at $\theta=2.45^\circ$ and
$\gamma_V=3\,\mathrm{meV}$
[Fig.~\ref{fig:SM3_V_R-disorder_six-terminal_Rxy_2.45deg}(g--i)].
The plateau broadening suggests that
enhanced disorder sensitivity plays the dominant role, despite the
reduction in large-momentum scattering.

For $D_R$, the charge response remains weak and shows no comparable
enhancement with $\lambda$.
At the same angle, taking $\theta=2.45$, $\gamma_D=3\,\mathrm{meV}$ as an example, increasing $\lambda$ from $0$ to $2$ raises the conductance
maximum from approximately $1.7$ to $4.3\,e^2/h$
[Fig.~\ref{fig:SM3_D_R-disorder_two-terminal_G2t_2.45deg}(g--i)].
The Hall dip deepens while the plateau widths change only slightly
[Fig.~\ref{fig:SM3_D_R-disorder_six-terminal_Rxy_2.45deg}(g--i)].
These trends are consistent with reduced backscattering enhancing
bulk conduction while the disorder sensitivity remains weak.

For $\theta_R$ disorder, the main charge-response maxima lie between
those of $V_R$ and $D_R$, with a modest increase from
$\lambda=1$ to $2$
[Fig.~\ref{fig:SM4_charge-response_lambda012_combined-bands}(b,c)].
In the onsite-only model at $W_\theta=1^\circ$, the conductance
maximum between plateaus increases as $\lambda$ grows from $0$ to $2$ [Fig.~\ref{fig:SM3_twist-angle-disorder-onsite-only_two-terminal_G2t_2.45deg}(g--i)].
This suggests that reduced backscattering plays the dominant role
in enhancing bulk conduction, despite the increased disorder sensitivity.

\end{document}